%% file: main_double_column.tex
\documentclass[journal]{IEEEtran}
\usepackage{multirow}
\usepackage{amssymb}
\usepackage[cmex10]{amsmath}
\usepackage{graphicx}
\usepackage{stfloats}
\usepackage{xcolor}

\ifCLASSOPTIONcompsoc
\usepackage[caption=false,font=normalsize,labelfont=sf,textfont=sf]{subfig}
\else
\usepackage[caption=false,font=footnotesize]{subfig}
\fi

\title{Performance Analysis of Target Parameters Estimation Using Multiple Widely Separated Antenna Arrays}
\author{Peter ˜Khomchuk, ˜\IEEEmembership{Student ˜Member, ˜IEEE,
} Rick ˜S. ˜Blum, ˜\IEEEmembership{Fellow, ˜IEEE},
 Igal ˜Bilik, ˜\IEEEmembership{Member, ˜IEEE}
\thanks{This work was supported by the National Science Foundation under Grant No. ECCS-1405579.}
\thanks{}}

\begin{document}
%%  OPTIONAL -->   \ninept            <-- OPTIONAL, for nine pt only
%%}
%%\ninept
\maketitle
\input{psfig.sty}

\newcommand{\be}{\begin{equation}}
\newcommand{\ee}{\end{equation}}
\newcommand{\bea}{\begin{eqnarray}}
\newcommand{\eea}{\end{eqnarray}}
\newcommand{\beaa}{\begin{eqnarray*}}
\newcommand{\eeaa}{\end{eqnarray*}}
\newcommand{\ben}{\begin{enumerate}}
\newcommand{\een}{\end{enumerate}}
\newcommand{\bi}{\begin{itemize}}
\newcommand{\ei}{\end{itemize}}
\newcommand{\bs}{\begin{split}}
\newcommand{\es}{\end{split}}

\newcommand{\vecnorm}[1]{\left|\left|#1\right|\right|}

\newcommand{\bml}[1]{\mbox{\boldmath $ #1 $}}
\newcommand{\tbml}[2]{\tilde{\mbox{\boldmath$#1$}}_{\! #2}}
\newcommand{\hbml}[2]{\hat{\mbox{\boldmath$ #1 $}}_{\! #2}}
\newcommand{\bmll}[1]{\mbox{\boldmath $ #1 $}^{(\ell)}}

\newcommand{\one}{\frac{1}{n}}
\newcommand{\half}{\frac{1}{2}\:}
\newcommand{\Dhalf}{\mbox{$\frac{D}{2}$}}
\newcommand{\pomDhalf}{\mbox{$\left(1-\frac{D}{2}\right)$}}
\newcommand{\pxn}{\mbox{$P_{X^n}$}}
\newcommand{\pyn}{\mbox{$P_{Y^n}$}}
\newcommand{\pzn}{\mbox{$P_{Z^n}$}}
\newcommand{\wyxn}{\mbox{$W_{Y^n|X^n}$}}
\newcommand{\qxyn}{\mbox{$Q_{X^nY^n}$}}
\newcommand{\qyxn}{\mbox{$Q_{Y^n|X^n}$}}
\newcommand{\qxyzn}{\mbox{$Q_{X^nY^nZ^n}$}}
\newcommand{\pxyzn}{\mbox{$P_{X^nY^nZ^n}$}}
\newcommand{\pxyn}{\mbox{$P_{X^nY^n}$}}
\newcommand{\pxy}{\mbox{$\bml{P_{XY}}$}}
\newcommand{\qxy}{\mbox{$\bml{Q_{XY}}$}}
\newcommand{\px}{\mbox{$\bml{P_X}$}}
\newcommand{\py}{\mbox{$\bml{P_Y}$}}
\newcommand{\tp}{\mbox{$\tilde{P}$}}
\newcommand{\ptxx}{\mbox{$P_{\tilde{X}X}$}}
\newcommand{\eptxx}{\mbox{${\bf E}_{P_{\tilde{X}X}}$}}
\newcommand{\wyx}{\mbox{$\bml{W_{Y|X}}$}}
\newcommand{\w}{\mbox{$\bml{W}$}}
\newcommand{\qxn}{\mbox{$Q_{X^n}$}}
\newcommand{\qyn}{\mbox{$Q_{Y^n}$}}
\newcommand{\qx}{\mbox{$\bml{Q_X}$}}
\newcommand{\qy}{\mbox{$\bml{Q_Y}$}}
\newcommand{\vxn}{\mbox{$V_{X^n}$}}
\newcommand{\vyn}{\mbox{$V_{Y^n}$}}
\newcommand{\vxyn}{\mbox{$V_{X^nY^n}$}}
\newcommand{\vyxn}{\mbox{$V_{Y^n|X^n}$}}
\newcommand{\x}{\mbox{$\bml{X}$}}
\newcommand{\y}{\mbox{$\bml{Y}$}}
\newcommand{\z}{\mbox{$\bml{Z}$}}
\newcommand{\ixy}{\mbox{$\oo{\bml{I}}(\bml{X};\bml{Y})$}}
\newcommand{\iixy}{\mbox{$\uu{\bml{I}}(\bml{X};\bml{Y})$}}
\newcommand{\ixyzn}{\mbox{$i_{X^nY^n;Z^n}$}}
\newcommand{\ixyznl}{\mbox{$i_{X^nY^n;Z^n}(a^nb^n;c^n)$}}
\newcommand{\ixyznr}{\mbox{$i_{X^nY^n;Z^n}(X^nY^n;Z^n)$}}
\newcommand{\ixxyzn}{\mbox{$i_{X^n;Y^nZ^n}$}}
\newcommand{\ixxyznl}{\mbox{$i_{X^n;Y^nZ^n}(a^n;b^nc^n)$}}
\newcommand{\ixxyznr}{\mbox{$i_{X^n;Y^nZ^n}(X^n;Y^nZ^n)$}}
\newcommand{\iyzn}{\mbox{$i_{Y^n;Z^n}$}}
\newcommand{\iyznl}{\mbox{$i_{Y^n;Z^n}(b^n;c^n)$}}
\newcommand{\iyznr}{\mbox{$i_{Y^n;Z^n}(Y^n;Z^n)$}}
\newcommand{\ixzn}{\mbox{$i_{X^n;Z^n}$}}
\newcommand{\ixznl}{\mbox{$i_{X^n;Z^n}(a^n;c^n)$}}
\newcommand{\ixznr}{\mbox{$i_{X^n;Z^n}(X^n;Z^n)$}}
\newcommand{\vni}{\mbox{$V_{n,i}\left(\left\{y_j^n\right\},c^n\right)$}}
\newcommand{\rni}{\mbox{$R_{n,i}\left(\left\{y_j^n\right\},c^n\right)$}}
\newcommand{\umx}{\mbox{$U_{M_X}\left(\left\{y_j^n\right\},c^n\right)$}}
\newcommand{\tmx}{\mbox{$T_{M_X}\left(\left\{y_j^n\right\},c^n\right)$}}
\newcommand{\tvni}{\mbox{$\tilde{V}_{n,i}(b^n,c^n)$}}
\newcommand{\trni}{\mbox{$\tilde{R}_{n,i}(b^n,c^n)$}}
\newcommand{\tu}{\mbox{$\tilde{U}(b^n,c^n)$}}
\newcommand{\tllt}{\mbox{$\tilde{T}(b^n,c^n)$}}
\newcommand{\sr}{\mbox{$\oo{\bml{R}}$}}
\newcommand{\ir}{\mbox{$\uu{\bml{R}}$}}
\newcommand{\hy}{\mbox{$\oo{\bml{H}}(\bml{Y})$}}
\newcommand{\hx}{\mbox{$\oo{\bml{H}}(\bml{X})$}}
\newcommand{\uh}{\mbox{$\uu{\bml{H}}$}}
\newcommand{\oh}{\mbox{$\oo{\bml{H}}$}}
\newcommand{\er}{\mbox{$\bml{E}$}}
\newcommand{\ron}{\mbox{$\rho_n$}}
\newcommand{\ros}{\mbox{$\rho_s$}}
\newcommand{\ronb}{\mbox{$\bar{\rho}_n$}}
\newcommand{\rob}{\mbox{$\bar{\rho}$}}
\newcommand{\romax}{\mbox{$\rho_{max}$}}
\newcommand{\Pcirc}{\mbox{$\stackrel{\circ}{P}$}}
\newcommand{\Ecirc}{\mbox{$\stackrel{\circ}{E}$}}
\newcommand{\uu}{\underline}
\newcommand{\oo}{\overline}
\newcommand{\reals}{{\rm I\!R}}
\newcommand{\onei}{{\rm 1\!\!\!\:I}}
\newcommand{\dfn}{\stackrel{\triangle}{=}}
\def\ints{\mathop{\rm Z\kern -0.25em Z}\nolimits}
\def\complex{\mathop{|\kern -0.25em \rm C}\nolimits}
\newcommand{\realsd}{\reals^d}
\newcommand{\PP}{\mbox{${\cal P}$}}
\newcommand{\NN}{{\rm I\!\!\!\;N}}
\newcommand{\FF}{\mbox{${\cal F}$}}
\newcommand{\CC}{\mbox{${\cal C}$}}
\newcommand{\LL}{\mbox{${\cal L}$}}
\newcommand{\XX}{\mbox{${\cal X}$}}
\newcommand{\MM}{\mbox{${\cal M}$}}
\newcommand{\AAA}{\mbox{${\cal A}$}}
\newcommand{\AAB}{\mbox{$\bf{{\cal A}}$}}
\newcommand{\DD}{\mbox{${\cal D}$}}
\newcommand{\EE}{{\rm I\!\!\!\;E}}
\newcommand{\DI}{{\DD_{{\rm I}}}}
\newcommand{\Prob}{{\rm Prob\,}}
\newcommand{\la}{\lambda}
\newcommand{\Kalpha}{{K_\alpha}}
\newcommand{\Psialpha}{{\Psi_I(\alpha)}}
\newcommand{\PsiI}{{\Psi_I}}
\newcommand{\calCo}{{\cal{C}}^o}
\newcommand{\calB}{{\cal{B}}}
\newcommand{\Shat}{\hat{S}}
\newcommand{\mut}{\tilde{\mu}}
\newcommand{\boldx}{{\bf x}}
\newcommand{\boldX}{{\bf X}}
\newcommand{\boldy}{{\bf y}}
\newcommand{\boldz}{{\bf z}}
\newcommand{\boldp}{{\bf p}}
\newcommand{\uux}{\uu{x}}
\newcommand{\uuY}{\uu{Y}}

\def\muvec{{\mbox{\boldmath $\mu$}}}
\def\Gammavec{{\mbox{\boldmath $\Gamma$}}}
\def\btheta{{\mbox{\boldmath $\theta$}}}
\def\thetat{{\mbox{\scriptsize\boldmath $\theta$}}}

\newcommand{\avec}{{\bf{a}}}
\newcommand{\avect}{{\tilde{\bf{a}}}}
\newcommand{\bvec}{{\bf{b}}}
\newcommand{\cvec}{{\bf{c}}}
\newcommand{\dvec}{{\bf{d}}}
\newcommand{\evec}{{\bf{e}}}
\newcommand{\fvec}{{\bf{f}}}
\newcommand{\epsvec}{{\bf{\epsilon}}}
\newcommand{\pvec}{{\bf{p}}}
\newcommand{\qvec}{{\bf{q}}}
\newcommand{\Yvec}{{\bf{Y}}}
\newcommand{\yvec}{{\bf{y}}}
\newcommand{\uvec}{{\bf{u}}}
\newcommand{\wvec}{{\bf{w}}}
\newcommand{\wvect}{{\tilde{\bf{w}}}}
\newcommand{\xvec}{{\bf{x}}}
\newcommand{\zvec}{{\bf{z}}}
\newcommand{\mvec}{{\bf{m}}}
\newcommand{\nvec}{{\bf{n}}}
\newcommand{\rvec}{{\bf{r}}}
\newcommand{\Svec}{{\bf{S}}}
\newcommand{\Tvec}{{\bf{T}}}
\newcommand{\svec}{{\bf{s}}}
\newcommand{\vvec}{{\bf{v}}}
\newcommand{\gvec}{{\bf{g}}}
\newcommand{\gveca}{\gvec_{\alphavec}}
\newcommand{\uveca}{\uvec_{\alphavec}}
\newcommand{\hvec}{{\bf{h}}}
\newcommand{\ivec}{{\bf{i}}}
\newcommand{\kvec}{{\bf{k}}}
\newcommand{\etavec}{{\bf{\eta}}}
\newcommand{\onevec}{{\bf{1}}}
\newcommand{\zerovec}{{\bf{0}}}
\newcommand{\nuvec}{{\bf{\nu}}}
\newcommand{\alphavec}{{\boldsymbol{\alpha}}}
\newcommand{\psivec}{{\boldsymbol{\psi}}}
\newcommand{\varphivec}{{\boldsymbol{\varphi}}}
\newcommand{\thetavec}{\boldsymbol{\theta}}
\newcommand{\zetavec}{\boldsymbol{\zeta}}
\newcommand{\phivec}{\boldsymbol{\phi}}
\newcommand{\rhovec}{\boldsymbol{\rho}}
\newcommand{\omegavec}{\boldsymbol{\omega}}
\newcommand{\chivec}{\boldsymbol{\chi}}
\newcommand{\Phivec}{{\bf{\Phi}}}
\newcommand{\Thetavec}{{\bf{\Theta}}}
\newcommand{\deltakvec}{{\bf{\Delta k}}}
\newcommand{\Lambdamat}{{\bf{\Lambda}}}
\newcommand{\Deltamat}{{\bf{\Delta}}}
\newcommand{\invLambdamat}{\Lambdamat^{-1}}
\newcommand{\Gammamat}{{\bf{\Gamma}}}
\newcommand{\Amat}{{\bf{A}}}
\newcommand{\Bmat}{{\bf{B}}}
\newcommand{\Cmat}{{\bf{C}}}
\newcommand{\Dmat}{{\bf{D}}}
\newcommand{\Emat}{{\bf{E}}}
\newcommand{\Fmat}{{\bf{F}}}
\newcommand{\Gmat}{{\bf{G}}}
\newcommand{\Hmat}{{\bf{H}}}
\newcommand{\Jmat}{{\bf{J}}}
\newcommand{\Imat}{{\bf{I}}}
\newcommand{\Kmat}{{\bf{K}}}
\newcommand{\Lmat}{{\bf{L}}}
\newcommand{\Pmat}{{\bf{P}}}
\newcommand{\Pmatperp}{{\bf{P^{\bot}}}}
\newcommand{\Ptmatperp}{{\bf{P_2^{\bot}}}}
\newcommand{\Qmat}{{\bf{Q}}}
\newcommand{\invQmat}{\Qmat^{-1}}
\newcommand{\Smat}{{\bf{S}}}
\newcommand{\Tmat}{{\bf{T}}}
\newcommand{\Tmattilde}{\tilde{\bf{T}}}
\newcommand{\Tmatcheck}{\check{\bf{T}}}
\newcommand{\Tmatbar}{\bar{\bf{T}}}
\newcommand{\Rmat}{{\bf{R}}}
\newcommand{\Umat}{{\bf{U}}}
\newcommand{\Vmat}{{\bf{V}}}
\newcommand{\Wmat}{{\bf{W}}}
\newcommand{\Ymat}{{\bf{Y}}}
\newcommand{\Zmat}{{\bf{Z}}}
\newcommand{\Zeromat}{{\bf{0}}}

\newcommand{\cbl}{\left\lbrace}
\newcommand{\cbr}{\right\rbrace}

\newcommand{\Ry}{\Rmat_{\yvec}}
\newcommand{\Rz}{\Rmat_{\zvec}}
\newcommand{\RyInv}{\Rmat_{\yvec}^{-1}}
\newcommand{\Ryhat}{\hat{\Rmat}_{\yvec}}
\newcommand{\Rs}{\Rmat_{\svec}}
\newcommand{\Rn}{\Rmat_{\nvec}}
\newcommand{\Rninv}{\Rmat_{\nvec}^{-1}}
\newcommand{\Reta}{\Rmat_{\etavec}}
\newcommand{\Ralpha}{\Rmat_{\alphavec}}
\newcommand{\Ck}{\Cmat_{\kvec}}
\newcommand{\Cn}{\Cmat_{\nvec}}
\newcommand{\Cg}{\Cmat_{\gvec}}
\newcommand{\invRn}{\Rmat_{\nvec}^{-1}}
\newcommand{\Ical}{{\mathcal{I}}}
\newcommand{\Ncal}{{\mathcal{N}}}
\newcommand{\Jcal}{{\mathcal{J}}}
% This command splits the parentheses and braces
\newcommand{\LB}{\right. \\  \left.}
\newcommand{\Ccal}{{\mathcal{C}}}
\newcommand{\LDev}{\right.$ $\left.}
\newcommand{\define}{\stackrel{\triangle}{=}}
%%% For BOLD Greek Letters

\newcommand{\Psimat}{\mbox{\boldmath $\Psi$}}
\newcommand{\bzeta}{\mbox{\boldmath $\zeta$}}
\def\bzeta{{\mbox{\boldmath $\zeta$}}}
\def\btheta{{\mbox{\boldmath $\theta$}}}
\def\bgamma{{\mbox{\boldmath $\gamma$}}}
\def\Beta{{\mbox{\boldmath $\eta$}}}
\def\lam{{\mbox{\boldmath $\Gamma$}}}
\def\bomega{{\mbox{\boldmath $\omega$}}}
\def\bxi{{\mbox{\boldmath $\xi$}}}
\def\brho{{\mbox{\boldmath $\rho$}}}
\def\bmu{{\mbox{\boldmath $\mu$}}}
\def\bnu{{\mbox{\boldmath $\nu$}}}
\def\btau{{\mbox{\boldmath $\tau$}}}
\def\bphi{{\mbox{\boldmath $\phi$}}}
\def\bsigma{{\mbox{\boldmath $\Sigma$}}}
\def\bLambda{{\mbox{\boldmath $\Lambda$}}}
%%% For BOLD Greek Letters
\def\btheta{{\mbox{\boldmath $\theta$}}}
\def\bomega{{\mbox{\boldmath $\omega$}}}
\def\brho{{\mbox{\boldmath $\rho$}}}
\def\bmu{{\mbox{\boldmath $\mu$}}}
\def\bGamma{{\mbox{\boldmath $\Gamma$}}}
\def\bnu{{\mbox{\boldmath $\nu$}}}
\def\btau{{\mbox{\boldmath $\tau$}}}
\def\bphi{{\mbox{\boldmath $\phi$}}}
\def\bPhi{{\mbox{\boldmath $\Phi$}}}
\def\bxi{{\mbox{\boldmath $\xi$}}}
\def\bvarphi{{\mbox{\boldmath $\varphi$}}}
\def\bepsilon{{\mbox{\boldmath $\epsilon$}}}
\def\balpha{{\mbox{\boldmath $\alpha$}}}
\def\bvarepsilon{{\mbox{\boldmath $\varepsilon$}}}
\def\bXsi{{\mbox{\boldmath $\Xi$}}}
\def\betavec{{\mbox{\boldmath $\beta$}}}
\def\betavecsc{{\mbox{\boldmath \tiny $\beta$}}}
\def\xsivec{{\mbox{\boldmath $\xi$}}}
\def\xsivecsc{{\mbox{\boldmath \tiny $\xsivec$}}}
\def\alphavecsc{{\mbox{\boldmath \tiny $\alpha$}}}
\def\gammavec{{\mbox{\boldmath $\gamma$}}}
\def\etavecsc{{\mbox{\boldmath \tiny $\eta$}}}
\def\thetavecsc{{\mbox{\boldmath \tiny $\theta$}}}
\def\Ximat{{\mbox{\boldmath $\Xi$}}}
\def\xivec{{\mbox{\boldmath $\xi$}}}
\def\nuvec{{\mbox{\boldmath $\nu$}}}
\def\rhovec{{\mbox{\boldmath $\rho$}}}

\newcommand{\limn}{\lim_{n \rightarrow \infty}}
\newcommand{\limN}{\lim_{N \rightarrow \infty}}
\newcommand{\limr}{\lim_{r \rightarrow \infty}}
\newcommand{\limd}{\lim_{\delta \rightarrow \infty}}
\newcommand{\phit}{\phi_{\mbox{{\thetat}}}}
\newcommand{\phitk}{\phi_{\mbox{{\thetat}}}}

\newcommand{\AR}
 {\begin{array}[t]{c}
  \longrightarrow \\[-0.3cm]
  \scriptstyle {n\rightarrow \infty}x
  \end{array}}

\newcommand{\RAISE}{{\:\raisebox{.6ex}{$\scriptstyle{>}$}\raisebox{-.3ex}
{$\scriptstyle{\!\!\!\!\!<}\:$}}}

\newcommand{\ARROW}[1]
  {\begin{array}[t]{c}  \longrightarrow \\[-0.4cm] \textstyle{#1} \end{array} }

\newcommand{\pile}[2]
  {\left( \begin{array}{c}  {#1}\\[-0.2cm] {#2} \end{array} \right) }

\newcommand{\ffrac}[2]
  {\left( \frac{#1}{#2} \right)}

\def\squarebox#1{\hbox to #1{\hfill\vbox to #1{\vfill}}}
\newcommand{\qed}{\hspace*{\fill}
            \vbox{\hrule\hbox{\vrule\squarebox{.667em}\vrule}\hrule}\smallskip}

\newcommand{\ifff}{\mbox{\ if and only if\ }}

\newcommand{\ctg}{{\rm ctg}}
\newcommand{\limM}{\lim_{M \rightarrow \infty}}

\newcommand{\eps}{\epsilon}

\newcommand{\degree}{\scriptscriptstyle \circ }

\newcommand{\limsxupn}{\limsup_{n \rightarrow \infty}}
\newcommand{\liminfn}{\liminf_{n \rightarrow \infty}}

\newcommand{\st}{\stackrel}

\newcounter{MYtempeqncnt}

\input{paper_double_column.tex}
\end{document}

%% file: paper_double_column.tex
\begin{abstract}
Target parameter estimation performance is investigated for a radar employing a set of widely separated transmitting and receiving antenna arrays. Cases with multiple extended targets are considered under two signal model assumptions: stochastic and deterministic. The general expressions for the corresponding Cramer-Rao lower bound (CRLB) and the asymptotic properties of the maximum-likelihood (ML) estimator are derived for a radar with $M_t$ arrays of $L_t$ transmitting elements and $M_r$ arrays of $L_r$ receiving elements for both types of signal models.

It is shown that for an infinitely large product $M_tM_r$, and a finite $L_r$, the ML estimator is consistent and efficient under the stochastic model, while the deterministic model requires $M_tM_r$ to be finite and $L_r$ to be infinitely large in order to guarantee consistency and efficiency.

Monte Carlo simulations further investigate the estimation performance of the proposed radar configuration in practical scenarios with finite $M_tM_r$ and $L_r$, and a fixed total number of available receiving antenna elements, $M_r L_r$. The numerical results demonstrate that grouping receiving elements into properly sized arrays reduces the mean squared error (MSE) and decreases the threshold SNR. In the numerical examples considered, the preferred configurations employ $M_t M_r > 1$. In fact, when $M_t M_r$ becomes too small, due to the loss of the geometric gain, the estimation performance becomes strongly dependent on the particular scenario and can degrade significantly, while the CRLB may become a poor prediction of the MSE even for high SNR. This suggests it may be advantageous to employ approaches where neither $M_tM_r$ nor $L_r$ are too small.

\end{abstract}

\begin{IEEEkeywords}
Distributed arrays, array processing, multiple-input multiple-output (MIMO) radar, Cramer-Rao lower bound (CRLB), maximum likelihood (ML) estimate.
\end{IEEEkeywords}

\section{Introduction}
During the last decade, multiple-input multiple-output (MIMO) radars received significant attention from the research community \cite{DeMaio}-\cite{HaimBlum2}. MIMO radars employ multiple transmitted waveforms and jointly process signals received at the multiple receivers. Two MIMO radar concepts are commonly considered in the literature: MIMO radar with widely separated antennas \cite{DeMaio}-\cite{Aittomaki},  \cite{HaimBlum2}-\cite{Wei1} and MIMO radar with colocated antennas \cite{Bliss}-\cite{LiStoica0}. This work studies the multiple target parameter estimation performance of a radar configuration with widely separated antenna arrays, which combines the benefits from viewing the targets from different locations, called geometric diversity, with the benefits of employing standard coherent array processing.

Often, estimation performance is assessed by evaluation of bounds on the estimation errors. One of the most widely used bounds is the Cramer-Rao Lower Bound (CRLB). The CRLB for MIMO radar with widely separated antennas have been derived for target velocity estimation \cite{HaimBlum5}, coherent and noncoherent target location estimation \cite{HaimBlum6}, noncoherent joint location and velocity estimation \cite{HaimBlum8}, and multiple target parameter estimation \cite{Godrich1}, \cite{Wei1}. It is important to notice that all just referenced results consider radars with widely separated omnidirectional antennas and do not consider widely separated arrays.

This work derives the CRLB on the joint estimation of position and velocity of multiple targets with a set of widely separated antenna arrays. A similar configuration was proposed in \cite{Xu1}, where the idea of the MIMO radar with collocated antennas was extended to achieve both coherent processing and spatial diversity gains. While each antenna element in any closely-spaced array transmits a phase-shifted version of a common signal to allow coherent processing in the radar system considered in this paper, the radar system proposed in \cite{Xu1} consists of multiple widely spaced subarrays where the closely spaced antennas inside each transmitting subarray are assumed to transmit different orthogonal waveforms. The authors developed an iterative generalized-likelihood ratio test for target detection and parameter estimation, and compared performance of several spatial spectral estimators including the Capon and APES approaches. The CRLB was not considered in \cite{Xu1}. To our knowledge, the CRLB and the asymptotic properties of the ML estimator have not been discussed in the literature for the configuration studied in this paper. 

This paper derives the CRLB for two commonly used array processing signal models: stochastic and deterministic \cite{VanTrees}-\cite{Korso}. The stochastic model assumes the targets' reflectivities are random variables. If the reflectivities are normally distributed, which is the case considered in this paper, the stochastic model leads to the well known Swerling I and II target types \cite{Aittomaki}, \cite{Richards}. The deterministic model assumes the targets' reflectivities are deterministic unknowns and is often used when the assumption about the normality of the reflectivities is not realistic \cite{Yu}, \cite{Bekkerman}. Here the signal covariance matrix and the power of the additive white Gaussian noise are treated as unknown nuisance parameters under the stochastic model assumption. Similarly, the nuisance parameters for the deterministic model are the targets' reflectivities and the noise power. Notice that the previously derived bound in \cite{HaimBlum8} for MIMO radar with widely distributed antennas under the stochastic model does not consider nuisance parameters and treats the signal covariance matrix and the noise power as known quantities, while the bounds in \cite{HaimBlum5} and \cite{HaimBlum6} derived for the deterministic case consider the noise power to be known.

Under the deterministic signal model assumption, the CRLB depends on the particular realization of the targets' reflectivities which makes the application of such a bound complicated. In this work we consider an extended Miller-Chang bound (EMCB) \cite{Gini} which is calculated as the deterministic CRLB averaged over the different realizations of the targets' reflectivities. The EMCB provides a bound on the average variance of any scalar parameter of interest, where the average is taken using an assumed distribution for the parameter. This paper uses the same distribution assumed in the stochastic model. In the numerical results section the EMCB is evaluated for different configurations of the radar with widely separated antenna arrays and compared to the corresponding mean squared error (MSE).

The CRLB is a good prediction of the variance of the estimation error only in the asymptotic region when SNR is large or the number of taken data samples is large. The CRLB and the asymptotic properties of the corresponding ML estimator for conventional antenna arrays for both stochastic and deterministic models have been well studied in \cite{StoicaNehorai1} and \cite{StoicaNehorai2}. The asymptotic behavior of the ML estimator for MIMO radar with widely separated antennas has been considered only under the stochastic model assumption in a single target scenario \cite{HaimBlum8}. This work investigates a more general case of multiple distributed antenna arrays observing multiple targets. For a radar system with $M_t$ transmitting arrays of $L_r$ antenna elements and $M_r$ receiving arrays with $L_r$ antenna elements we study the asymptotic properties of the ML estimator in two scenarios: large $M_tM_r$ and large $L_r$.

The analysis shows that under the stochastic model assumption the ML estimates of the parameters of interest become consistent and efficient if $M_tM_r$ approaches infinity while $L_r$ is fixed to some constant value. On the other hand, having finite $M_tM_r$ and infinitely large $L_r$ does not provide consistency and efficiency. Under the deterministic model assumption the ML estimator is consistent but not efficient if $M_tM_r$ is asymptotically large and $L_r$ is constant, while finite $M_tM_r$ and infinitely large $L_r$ guarantee consistency and efficiency.

The asymptotic analysis provides an insight about the estimation performance of the proposed radar system with widely separated arrays but does not answer the question of an optimal allocation of a finite number of antennas into a finite number of distributed arrays. To further investigate this issue the paper provides a numerical comparison of the MSE and the corresponding CRLB and EMCB for a position estimation of two closely located targets using different configurations of the radar with distributed arrays and finite $M_tM_r$ and $L_r$.

Different configurations of the radar with widely separated arrays were compared while keeping the number of receiving antenna elements, $M_r L_r$, fixed. The numerical results show that in the scenarios with sufficiently large $M_t M_r$ and targets which were favorably positioned with respect to the distributed arrays, the MSE and the threshold SNR tended to decrease with increasing $L_r$ up to a point. When $M_tM_r$ was too small, the loss of geometric diversity (gains from different orientations between different arrays and a given target) became apparent. Equivalently, it became difficult to assure the targets were all well positioned with respect to the arrays. At this point, further decreases in $M_tM_r$ degraded performance. This indicates that it is often more beneficial to group receiving antenna elements into a small number of receiving arrays and there is a clear limit on the minimum number of receiving arrays which should be employed. However for a finite $M_t M_r$ and $L_r$ the MSE may not be well predicted by the CRLB even at the large SNR. The performance obtained did depend on the location of the targets with respect to the arrays and having a larger number of differently oriented arrays was extremely helpful in providing geometric diversity which tended to enhance the worst-case performance as targets were moved over some extensive region.

The obtained simulation results suggest that for a radar with a fixed finite number of receiving antenna elements the optimal number of transmit-to-receive array paths and the size of the arrays depends on the scenario, and it may be advantageous to use the radar configurations where neither $M_tM_r$ nor $L_r$ are too small. 

The main contributions of this paper are: a) new CRLBs for a radar with widely separated antenna arrays under stochastic and deterministic signal model assumptions; b) the derived CRLBs consider joint estimation of the parameters of multiple targets in the presence of the nuisance parameters; c) the study of the asymptotic properties of the ML estimator in the two extreme cases: the product $M_tM_r$ is large, and $L_r$ is large; d) the numerical study of the performance of ML estimator in the scenarios with a fixed finite number of receiving antenna elements $M_rL_r$.

It is worth pointing out that the obtained results can give the general trade-offs for the case of passive radar where the transmitters are from existing communication systems \cite{Griffiths}, \cite{Tan}. Assuming the passive system is able to estimate the transmitted signals with perfect accuracy, the derived CRLBs and the asymptotic ML results can be applied to predict the target parameter estimation performance in passive radar. The assumptions of perfectly estimated waveforms would result in a lower bounds on the MSE that a passive radar system can achieve.

The rest of the paper is organized as follows. Section II describes the received signal model for a radar with multiple widely spaced arrays, where each array employs coherent processing. Section III derives the likelihood functions and the CRLBs for both the stochastic and deterministic models. Section IV discusses the asymptotic properties of the ML estimator for both stochastic and deterministic signal models. The performance evaluation of the ML estimator using Monte-Carlo simulations is provided in Section V. Our conclusions are summarized in Section VI.

\section{Signal Model}
\label{sec:SignalModel}
Consider a radar system with $M_t$ transmitting and $M_r$ receiving arrays, which are arbitrarily distributed over the two-dimensional surveillance area. Each transmitting and receiving array consists of $L_t$ and $L_r$ antenna elements respectively. The center of the $k$th, $k = 1,2,\ldots,M_t$, transmitting array is located at $(x_{tk}, y_{tk})$, while the center of the $l$th, $l=1,2,\ldots,M_r$, receiving array is located at $(x_{rl}, y_{rl})$. Each array element of the $k$th transmitting array transmits a phase shifted version of the same waveform $\sqrt{\frac{E}{M_tL_t}}s_k(t)$, where $E$ is the total transmitted energy. Let $\wvec_k(\bar{\theta}_k)$ be an $L_t \times 1$ vector of beamforming coefficients which steers the $k$th transmitting array to point to the direction $\bar{\theta}_k$. The radar system observes $Q$ targets with coordinates $(x^q, y^q)$ and velocities $(v_{x}^q, v_{y}^q)$, where $q = 1, \ldots, Q$. The targets may act like non-point targets by exhibiting different reflections in different directions. Let $\avec_{tk}(\theta_{tk}^q)$ be an $L_t \times 1$ vector which represents propagation from the $k$th transmitting array toward the $q$th target located at the bearing angle $\theta_{tk}^q$.  Let $\avec_{rl}(\theta_{rl}^q)$ be the $L_r \times 1$ response vector of the $l$th receiving array to the plane wave arrived from the target $q$ at the angle $\theta_{rl}^q$. Notice that, the angles $\theta_{rl}^q$, $\theta_{tk}^q$ and $\bar{\theta}_{k}$ are defined with respect to a given specified direction. For linear arrays, this could be the direction normal to the corresponding receiving and transmitting arrays. In order to keep the results of this paper general for all array configurations, we don't assume any particular array geometry, unless stated otherwise. The targets are assumed to be located in the far-field of the arrays and narrowband signals are assumed. 

Assuming the observed $Q$ targets have constant velocities, the time delay for the signal traveling from the center of the $k$th transmitting array, reflected from the $q$th target and received at the center of the $l$th receiving array is determined by $\tau_{kl}^q = ( R_{tk}^q + R_{rl}^q ) / c$, where  $R_{tk}^q = \sqrt{(x_{tk} - x^q)^2 + (y_{tk} - y^q)^2}$ is the range from the center of the $k$th transmitting array to the $q$th target, $R_{rl}^q = \sqrt{(x_{rl} - x^q)^2 + (y_{rl} - y^q)^2}$ is the range from the $q$th target to the center of the $l$th receiving array, and $c$ is the wave propagation velocity. For $v_x^q, v_y^q \ll c$, the Doppler shift induced by the $q$th target on the signal transmitted by the $k$th transmitting array and received at the $l$th receiving array is defined by $\Omega_{kl}^q = v_{x}^q \Omega ( \cos\phi_{tk}^q + \cos\phi_{rl}^q)/c + v_{y}^q \Omega ( \sin\phi_{tk}^q + \sin\phi_{rl}^q)/c$, where $\phi_{tk}^q = \tan^{-1}((y_{tk} - y^q)/(x_{tk} - x^q))$, $\phi_{rl}^q = \tan^{-1}((y_{rl} - y^q)/(x_{rl} - x^q))$, and $\Omega$ is a carrier frequency. At every receiving array, the continuous-time baseband signal is sampled every $\triangle t$ seconds. Due to the sampling, the time delay and Doppler shift in the sampling domain are defined as $n_{kl}^q = \tau_{kl}^q / \triangle t$ and $\omega_{kl}^q = \Omega_{kl}^q \triangle t$ respectively. The sampled baseband signal at the $l$th receiving array, due to the transmission from the $k$th transmitting array and reflection from the $Q$ targets, can be modeled as the superposition of the $Q$ time delayed and Doppler-shifted versions of the transmitted signal $s_k(t)$ as \cite{DoganNehor1}
\be
\begin{split}
\rvec_{kl}\left[n\right] & = \sqrt{\frac{E}{M_tL_t}}  \sum_{q=1}^Q \alpha_{kl}^q \zeta_{kl}^q \avec_{rl}\left(\theta_{rl}^q\right)\\ 
\cdot & \left(\wvec_k^H\left(\bar{\theta}_k\right)\avec_{tk}\left(\theta_{tk}^q\right)\right) s_{k}\left[n-n_{kl}^q\right]e^{jn\omega_{kl}^q} + \evec_{kl}\left[n\right] \label{Model1}
\end{split}
\ee
where $\alpha_{kl}^q$ is the complex target reflectivity corresponding to the $(k,q,l)th$ ($k$th transmitting array, $q$th target, $l$th receiving array) path, $\zeta_{kl}^q = 1/(R_{tk}^q + R_{rl}^q)^2$ is a propagation loss coefficient, $s_k[n]$ is the sampled version of the continuous time signal $s_{k}(t)$, $\evec_{kl}[n]$ is an $L_r \times 1$ vector of the additive receiver noise, and $(\cdot)^H$ is a complex conjugate and transpose operation. To incorporate multiple temporal samples in the model (\ref{Model1}) we define the $N\times 1$ temporal steering vector for the $(k,q,l)th$ path as 
\bea
\bvec_{kl}\left(n_{kl}^q,\omega_{kl}^q\right) = \sqrt{\frac{E}{M_tL_t}} \left[ s_k[1-n_{kl}^q]e^{j\omega_{kl}^q1},\right.\nonumber \\  s_k[2-n_{kl}^q]e^{j\omega_{kl}^q2}, \ldots, \left. s_k[N-n_{kl}^q]e^{j\omega_{kl}^qN} \right]^T
\eea
where $(\cdot)^T$ denotes a matrix or a vector transpose operator. Considering $N$ temporal samples, the signal transmitted by the $k$th transmitting array and observed at the $l$th receiving array becomes 
\be
\label{Model2}
\rvec_{kl} = \sum_{q=1}^Q \alpha_{kl}^q \zeta_{kl}^q \avec_{rl}^q\otimes \left(\left(\wvec_k^H\avec_{tk}^q\right)\bvec_{kl}^q\right) + \evec_{kl}
\ee
where $\otimes$ denotes the Kroneker product, and for the simplicity of the further presentation the following shorthand notation is used $\avec_{tk}^q\triangleq\avec_{tk}(\theta_{tk}^q)$, $\avec_{rl}^q \triangleq \avec_{rl}(\theta_{rl}^q)$, $\wvec_k \triangleq  \wvec_k(\bar{\theta}_k)$, and $\bvec_{kl}^q \triangleq \bvec_{kl}(n_{kl}^q, \omega_{kl}^q)$.

%{\color{blue}{The proposed radar configuration with multiple distributed transmitting and receiving arrays provides the spatial diversity and allows to address the problem of the received signal fading due to the target's RCS fluctuation \cite{HaimBlum2}. Another possible solution to the receive signal fading is the transmission of the wideband waveforms. A direct introduction of the wideband waveforms to the model (\ref{Model2}) is challenging, since the received signals can no longer be represented as a superposition of the time-delayed and Doppler shifted versions of the transmitted waveforms, the beampatterns of the transmitting and receiving arrays are frequency dependent \cite{He}, and the waveform orthogonality is difficult to maintain over the large bandwidth due to the signal dispersion during the propagation. One of the ways to utilize a wide bandwidth in the proposed radar configuration would be to simultaneously transmit and receive multiple narrowband signals at different carrier frequencies. Such signaling scheme can be efficiently implemented using the Orthogonal Frequency Division Multiplexing (OFDM) \cite{Wu, Berger}}}.

Let the coherent processing interval (CPI) be $T_c = N \triangle t$. The targets' reflectivities $\alpha_{kl}^q$, $\forall k,q,l$ are assumed to remain constant during this interval. In addition, the narrowband assumption in (\ref{Model1}) implies that the propagation time of the signal across the array elements should be smaller than the reciprocal of the signal's bandwidth \cite{DoganNehor1}. 

In the signal model (\ref{Model2}), the bearing angles $\theta_{tk}^q$ and $\theta_{rl}^q$, the time delay $n_{kl}^q$, and the propagation loss coefficient $\zeta_{kl}^q$ are defined by the location of the $q$th target, while the Doppler shift $\omega_{kl}^q$, is defined by the both the $q$th target's location and velocity. Let $\psivec^q$ be a $P\times 1$ vector of parameters of interest for the $q$th target (e.g. if $\psivec^q = \left[x^q, y^q, v_x^q, v_y^q \right]^T$ then $P=4$), then the corresponding spatio-temporal steering vector for the $q$th target is
\be
\label{spatiotemporalsv}
\hvec_{kl}^q( \psivec^q ) = \zeta_{kl}^q\avec_{rl}^q\otimes \left(\left(\wvec_{tk}^H\avec_{tk}^q\right) \bvec_{kl}^q\right).
\ee

Now the received signal model in (\ref{Model2}) can be rewritten in the matrix form as 
\be
\label{Model3}
\rvec_{kl} = \Hmat_{kl}(\psivec)\alphavec_{kl} + \evec_{kl}
\ee
where $\alphavec_{kl} = \left[\alpha_{kl}^1,\alpha_{kl}^2,\cdots,\alpha_{kl}^Q\right]^T$ is a vector of tagets' reflectivities for the $kl$th transmit-to-receive array path, $\Hmat_{kl}(\psivec) = \left[\hvec_{kl}^1(\psivec^1),\hvec_{kl}^2(\psivec^2),\cdots,\hvec_{kl}^Q(\psivec^Q)\right]$ is a matrix of spatio-temporal steering vectors, and $\psivec =\left[ (\psivec^1)^T,\right.$ $\left.(\psivec^2)^T,\cdots,(\psivec^Q)^T \right]^T$ is a $PQ\times 1$ vector of unknown parameters for the $Q$ observed targets. Further this paper uses $\Hmat_{kl}$ to indicate $\Hmat_{kl}(\psivec)$. For the estimate $\hat{\psivec}$ of $\psivec$ we will use $\hat{\Hmat}_{kl}$ instead of $\Hmat_{kl}(\hat{\psivec})$.

The following assumptions allow for the signals transmitted by the $M_t$ transmitting arrays and received at the $l$th receiving array to be considered independently.

\textit{Assumption 1 (Orthogonal signals):} Assuming sufficiently large sample support $N$, the sampled waveforms transmitted by the $M_t$ transmitting arrays are orthogonal if 
\be
\label{orthog1}
\begin{split}
\bvec_{kl}^H\left(0,0\right)&\bvec_{k'l}\left(0,0\right)=\frac{E}{M_tL_r}\sum_{n=1}^{N}s_{k}[n]s_{k'}^*[n] \\ & = \left\lbrace
\begin{array}{cc}
\frac{E}{M_tL_r}\sum_{n=1}^{N}\vert s_i[n] \vert ^ 2 = \frac{E}{M_tL_r} & k=k' \\
0 & k \neq k'
\end{array} \right.
\end{split}
\ee
and the orthogonality is approximately maintained for the set of all possible time delays $n_{kl}^q$, $n_{k'l}^q$ and Doppler shifts $\omega_{kl}^q$, $\omega_{k'l}^q$ as per 
\be
\label{orthog2}
\begin{split}
\bvec_{kl}^H&\left(n_{kl}^q,\omega_{kl}^q\right)\bvec_{k'l}^H\left(n_{k'l}^q,\omega_{k'l}^q\right)\\
& =\frac{E}{M_tL_r}\sum_{n=1}^{N}s_k[n-n_{kl}^q]s_{k'}^*[n-n_{k'l}^q]e^{jn\left(\omega_{kl}^q-\omega_{k'l}^q\right)} \\
& \approx\left\lbrace
\begin{array}{cc}
\frac{E}{M_tL_r} & k=k' \\
0 & k \neq k'
\end{array} \right. .
\end{split}
\ee

The estimation of the parameters of interest in (\ref{Model3}) and the corresponding CRLB on those parameters depend on the assumptions made about the targets' reflectivities $\alphavec_{kl}$. Two typical approaches exist in the literature \cite{VanTrees}-\cite{Korso}. The first is known as the stochastic model which assumes randomly distributed $\alphavec_{kl}$ according to a zero-mean complex Gaussian distribution with unknown covariance matrix that needs to be estimated. Such a model is also referred to as unconditional since the distribution of the received data depends only on the statistics of the targets' reflectivities, and remains the same for different realizations of $\alphavec_{kl}$. The second approach is called deterministic and assumes $\alphavec_{kl}$ to be deterministic unknowns. Since the distribution of the received data in this case is conditioned on the particular values of the targets' reflectivities such a received signal model is also known as conditional. This work considers both signal models based on the following assumptions.

\textit{Assumption 2.1 (Stochastic signal model):} The stochastic signal model assumes $\alphavec_{kl}$ to be a zero-mean circularly-symmetric complex Gaussian random vector with a unknown $Q \times Q$ covariance matrix, $\Amat = E[\alphavec_{kl}\alphavec_{kl}^H]$, equal for all transmit-to-receive array paths. Since the transmitting and receiving arrays are widely separated, we assume that $\alphavec_{kl}$ and $\alphavec_{k'l'}$ are statistically independent if $k \ne k'$ or $l \neq l'$. 
Thus 
\be
\label{assump:21}
E[\alpha_{kl}^q\alpha_{k'l'}^{q'}] = \left\lbrace
\begin{array}{cc}
[\Amat]_{qq'} & k=k', l=l' \\
0 & otherwise
\end{array} \right. .
\ee

\textit{Assumption 2.2 (Deterministic signal model):} Vectors of targets' reflectivities $\alphavec_{kl}$ are assumed to be deterministic and unknown.

\textit{Assumption 3 (Uncorrelated spatiotemporal noise):} The noise in (\ref{Model3}) is spatially and temporally white zero-mean complex Gaussian random vector uncorrelated for different transmit-to-receive array paths
\be
\label{assump:3}
E[\evec_{kl}\evec_{k'l'}] = \left\lbrace
\begin{array}{cc}
\sigma^2\Imat_{L_rN} & k=k', l=l'\\
\Zeromat & otherwise
\end{array} \right.
\ee
where $\sigma^2$ is an unknown noise power, which is treated as a nuisance parameter during the target parameter estimation and $\Imat_{L_rN}$ is an $L_rN\times L_rN$ identity matrix.

The rest of the paper develops the CRLBs and studies the asymptotic properties of the ML estimator for stochastic and deterministic signal models based on the Assumptions 2.1 and 2.2 respectively.

\section{CRLB}
The covariance matrix of any unbiased estimator $\hat{\gammavec}$ of the vector of unknown parameters $\gammavec$ satisfies the following inequality \cite{Kay1}
\be
Cov[\hat{\gammavec}] \succeq  \Ical^{-1}(\gammavec)
\ee
where
\be
\Ical(\gammavec) = E\left[\frac{\partial}{\partial \gammavec } \Lambda(\rvec; \gammavec)\left(\frac{\partial}{\partial \gammavec } \Lambda(\rvec; \gammavec)\right)^H\right]
\ee
is a Fisher information matrix (FIM), $\Lambda(\rvec;\gammavec)$ is a likelihood function, $\rvec$ is a collected data vector, and the symbol $\succeq$ indicates that the difference $Cov[\hat{\gammavec}] - \Ical^{-1}(\gammavec)$ is a positive semidefinite matrix. The CRLB for the vector of unknown parameters $\gammavec$ is defined as an inverse of the FIM
\be
\Cmat_{CRLB}(\gammavec) = \Ical^{-1}(\gammavec).
\ee
This section presents the CRLBs for the stochastic and the deterministic signal models for radar with multiple widely separated arrays.

\subsection{Stochastic Model}
\label{sec:stochmodel}
According to the the stochastic model in Assumption 2.1, the vector of targets' reflectivities $\alphavec_{kl}$ at the $kl$th transmit-to-receive array path is a sample from the complex Gaussian random process with zero-mean and $Q\times Q$ non-singular covariance matrix $\Amat = E\left[\alphavec_{kl}\alphavec_{kl}^H\right]$. Since the targets' reflectivities and the additive noise are mutually independent, the received signal $\rvec_{kl}$ in (\ref{Model3}) is a zero-mean complex Gaussian with a covariance matrix
\be
\label{CovarMat}
\Rmat_{kl} = E\left[ \rvec_{kl}\rvec_{kl}^H \right] = \Hmat_{kl}\Amat\Hmat_{kl}^H + \sigma^2\Imat_{L_rN}.
\ee
Under Assumptions 1, 2.1 and 3 the likelihood function becomes a product of individual likelihood functions for each transmit-to-receive array path
\be
\label{likelihood_stoch}
\begin{split}
\Lambda_s (\rvec_{11},\rvec_{12},&\ldots,\rvec_{M_tM_r}; \psivec, \Amat, \sigma^2) \\ & = \prod_{l=1}^{M_r}\prod_{k=1}^{M_t}\frac{1}{\pi^{NL_r}\left|\Rmat_{kl}\right|}e^{-\rvec_{kl}^H\Rmat_{kl}^{-1}\rvec_{kl}}
\end{split}
\ee
For the stochastic model the unknown parameters can be gathered in a single vector
\be
\label{prms_stoch}
\gammavec = \left[\psivec, \rhovec, \sigma^2 \right]^T
\ee
where $\rhovec$ is the $Q^2 \times 1$ vector containing real and imaginary parts of the elements in  $\Amat$. 

The $ij$th element of the FIM for the vector of unknown parameters $\gammavec$ in (\ref{prms_stoch}) is \cite{Kay1}
\be
\label{FIMstoch}
[\Ical(\gammavec)]_{ij} = \sum_{l=1}^{M_r}\sum_{k=1}^{M_t}Tr\cbl\frac{d\Rmat_{kl}}{d\gamma_i}\Rmat_{kl}^{-1}\frac{d\Rmat_{kl}}{d\gamma_j}\Rmat_{kl}^{-1}\cbr.
\ee
The FIM as defined in (\ref{FIMstoch}) contains the information about unknown parameters of interest $\psivec$ as well as the information about the nuisance parameters, the elements of the unknown signal covariance matrix $\Amat$ stored in $\rhovec$ and the noise variance $\sigma^2$. Let $\cvec^q$ be a $q$th column of the matrix $\Amat$, and $\dvec_{kl}^{qp} = \frac{d}{d(\psivec^q)_p}\hvec_{kl}^q$ be a derivative of the spatio-temporal steering vecotor $\hvec_{kl}^q$ in (\ref{spatiotemporalsv}) with respect to the $p$th, $p=1,2,\ldots P$, element of the vector $\psivec^q$. The following expression for the CRLB on the estimation errors of the targets' parameters of interest only is derived from the FIM in (\ref{FIMstoch}) in Appedinx \ref{AppendixA} 
\be
\label{FIMstochFianl}
\begin{split}
&\Cmat_s^{-1}(\psivec) = \sum_{l=1}^{M_r}\sum_{k=1}^{M_t} \Gmat_{kl}^H\Gmat_{kl}\\
&- \Gmat_{kl}^H\Fmat_{kl}\left(\sum_{l'=1}^{M_r}\sum_{k'=1}^{M_t}\Fmat_{k'l'}^H\Fmat_{k'l'}\right)^{-1}
\left(\sum_{l'=1}^{M_r}\sum_{k'=1}^{M_t} \Fmat_{k'l'}^H\Gmat_{k'l'}\right)
\end{split}
\ee
where
\be
\Fmat_{kl} = \begin{bmatrix}\Rmat_{kl}^{-T/2}\Hmat_{kl}^C\otimes\Rmat_{kl}^{-1/2}\Hmat_{kl} & vec\left(\Rmat_{kl}^{-1}\right) \end{bmatrix}
\ee
and the $m$th, $m=P(q-1)+p$, column of matrix $\Gmat_{kl}$ is defined as follow
\be
\begin{split}
\left[\Gmat_{kl}\right]_m & = vec\left(\Rmat_{kl}^{-1/2}\left(\Hmat_{kl}\cvec^q\left(\dvec_{kl}^{qp}\right)^H \right.\right. \\ & + \left.\left. \dvec_{kl}^{qp}\left(\cvec^q\right)^H\Hmat_{kl}^H\right)\Rmat_{kl}^{-1/2}\right).
\end{split}
\ee
Here $vec(\cdot)$ is a vectorization operator that stacks columns of a matrix on top of each other, and $(\cdot)^C$ denotes a complex conjugate operation. Notice that the derived bound depends on the time delays $n_{kl}^q$, the bearing angles $\theta_{tk}^q$ and $\theta_{rl}^q$, the Doppler shift $\omega_{kl}^q$, and the propagation loss coefficients $\zeta_{kl}^q$  through the terms $\dvec_{kl}^{qp}$ which can be expanded using the differentiation chain rule as follows
\bea
\label{chainrule}
\dvec_{kl}^{qp} & = & \frac{\partial \hvec_{kl}^q}{\partial (\psivec^q)_p} = \frac{\partial \hvec_{kl}^q}{\partial n_{kl}^q}\frac{\partial n_{kl}^q}{\partial (\psivec^q)_p} + \frac{\partial \hvec_{kl}^q}{\partial \theta_{tk}^q}\frac{\partial \theta_{tk}^q}{\partial (\psivec^q)_p} \\
 &+& \frac{\partial \hvec_{kl}^q}{\partial \theta_{rl}^q}\frac{\partial \theta_{rl}^q}{\partial (\psivec^q)_p} + \frac{\partial \hvec_{kl}^q}{\partial \omega_{kl}^q}\frac{\partial \omega_{kl}^q}{\partial (\psivec^q)_p} + \frac{\partial \hvec_{kl}^q}{\partial \zeta_{kl}^q}\frac{\partial \zeta_{kl}^q}{\partial (\psivec^q)_p}. \nonumber
\eea
To our knowledge the general expression for the CRLB in (\ref{FIMstochFianl}) cannot be significantly simplified since it requires an inverse of a sum of matrices. 

One can observe from (\ref{CovarMat}) that the covariance matrix $\Rmat_{kl}$ of the received radar echoes is different for each transmit-to-receive array path. Thus, the radar echoes, $\rvec_{kl}$, are independent but not identically distributed (i.n.i.d.) random vectors. The result in (\ref{FIMstochFianl}) is a summation of $M_tM_r$ terms, such that each transmit-to-receive array path contributes information about the targets' parameters of interest.

\subsection{Deterministic Model}
\label{subsec:DetModel}
Assumption 2.2 of the deterministic signal model treats the vector of targets' reflectivities $\alphavec_{kl}$ as a deterministic unknown nuisance parameter. Both the real and imaginary part of $\alphavec_{kl}$ has to be estimated for each $k$ and $l$ jointly with the elements of $\psivec$, and unknown noise power $\sigma^2$. Thus the vector of unknown parameters for the deterministic model is
\be
\gammavec = \left[\psivec, \alphavec, \sigma^2 \right]^T
\ee
where $\alphavec = \left[Re\cbl\alphavec_{11}^T\cbr, Im\cbl\alphavec_{11}^T\cbr, Re\cbl\alphavec_{12}^T\cbr, Im\cbl\alphavec_{12}^T\cbr,\right.$ $\left.\ldots, Re\cbl\alphavec_{M_tM_r}^T\cbr, Im\cbl\alphavec_{M_tM_r}^T\cbr\right]^T $ is the $2QM_tM_r \times 1$ vector that contains the real and imaginary parts of the unknown targets' reflectivities. In this case the received data at the $kl$th path (\ref{Model3}) has the complex Gaussian distribution with the mean vector $\muvec_{kl} = \Hmat_{kl}\alphavec_{kl}$ and the covariance $\Rmat_{kl} = \sigma^2\Imat_{L_rN}$. The corresponding likelihood function is
\be
\label{Likelihood_det}
\begin{split}
\Lambda_d(&\rvec_{11},\rvec_{12},\ldots,\rvec_{M_tM_r};\psivec, \alphavec, \sigma^2)\\
 =& \prod_{l=1}^{M_r}\prod_{k=1}^{M_t}\frac{1}{\pi^{NL_r}\left|\sigma^2\Imat\right|}e^{-\sigma^{-2}\left(\rvec_{kl}-\Hmat_{kl}\alphavec_{kl}\right)^H\left(\rvec_{kl}-\Hmat_{kl}\alphavec_{kl}\right)}
\end{split}
\ee
For the described signal model, the $ij$th element of FIM has a following general form \cite{Kay1}
\be
\label{FIMdet}
\begin{split}
[\Ical(\gammavec)]_{ij} = \sum_{l=1}^{M_r}\sum_{k=1}^{M_t}&\left[Tr\cbl\Rmat_{kl}^{-1} \frac{d\Rmat_{kl}}{d\gamma_i} \Rmat_{kl}^{-1}\frac{d\Rmat_{kl}}{d\gamma_j} \cbr \right. \\
+ & \left. 2Re\cbl \frac{d\muvec_{kl}^H}{d\gamma_i} \Rmat_{kl}^{-1} \frac{d\muvec_{kl}}{d\gamma_j} \cbr\right]
\end{split}
\ee
The FIM in ($\ref{FIMdet}$) is defined for the vector of unknown parameters $\gammavec$ which contains both the parameters of interest $\psivec$ and the nuisance parameters $\alphavec$ and $\sigma$. The corresponding CRLB for the vector of targets' parameters of interests only is derived in Appendix \ref{AppendixB}, and can be written as
\bea
\label{FIMdetFinal}
\Cmat_d^{-1}\left(\psivec \right) & = & \frac{2}{\sigma^2}Re\cbl \sum_{l=1}^{M_r}\sum_{k=1}^{M_t} \left( \Dmat_{kl}^H \Pi_{\Hmat_{kl}}^{\perp} \Dmat_{kl} \right) \right. \\ & \odot & \left. \left(\alphavec_{kl}\alphavec_{kl}^H 
\otimes \onevec_{P\times P}\right)^T\cbr \nonumber
\eea
where
\be
\label{projection_matrix}
\Pi_{\Hmat_{kl}}^{\perp} = \Imat_{L_rN} - \Hmat_{kl}\left(\Hmat_{kl}^H\Hmat_{kl}\right)^{-1}\Hmat_{kl}^H
\ee
is a projection matrix on a subspace orthogonal to a null space of $\Hmat_{kl}^H$, $\Dmat_{kl} = \left[ \dvec_{kl}^{11}, \dvec_{kl}^{12},\ldots, \dvec_{kl}^{QP}\right]$, $\onevec_{P\times P}$ is the $P\times P$ all-ones matrix, and $\otimes$ denotes the Hadamard matrix product. The derived CRLB depends on the time delays, the bearing angles, the Doppler shifts, and the propagation loss coefficients through the columns of the matrices $\Dmat_{kl}$ that can be explicitly written using the differentiation chain rule as shown in (\ref{chainrule}). The resultant CRLB has a similar form with the CRLB for the single array case in \cite{StoicaNehorai2}, where instead of the summation over the multiple snapshots taken at the same array, the expression in (\ref{FIMdetFinal}) has a summation over the different transmit-to-receive array paths. Notice that similarly to the stochastic model case discussed in Section \ref{sec:stochmodel}, the radar echoes received over different transmit-to-receiver array paths have different distributions, thus the likelihood function in (\ref{FIMdetFinal}) is a summation of i.n.i.d. terms.

The performance comparison of different radar systems with multiple distributed arrays requires evaluation of (\ref{FIMdetFinal}) for each different set of targets' reflectivities. Making additional assumptions about the nuisance parameters allows us to simplify the performance evaluation by obtaining a single value of the bound instead of a set of values. Multiple ways to remove the dependence of the bound on the nuisance parameters have been proposed in the literature. For example, the hybrid CRLB assumes nuisance parameters to be random with known prior distribution \cite{VanTrees, Rockah}. Other methods, known as the modified Cramer-Rao bound, the Miller-Chang bound (MCB) and the extended MCB (EMCB), are discussed and compared in \cite{Gini}. These bounds characterize the estimation performance averaged over the different values of the vector of nuisance parameters. The EMCB is shown to be the tightest among the discussed Cramer-Rao like bounds. The EMCB is calculated by first deriving the CRLB for the joint estimation of the vector of prarameters of interest $\psivec$ and the nuisance parameters $\alphavec$, and then averaging the result over $\alphavec$ assuming it has some known probability distribution. The resulting expression is a bound on the expected value of the variance of the estimator $\hat{\psivec}$ taken with respect to the targets' reflectivities $\alphavec$
\be
\label{emcb}
E_{\alphavec}[var[\psivec]] \succcurlyeq \Cmat_{EMCB}(\psivec) = E_{\alphavec} \left[\Cmat_d(\psivec)\right].
\ee
The EMCB is used in this work for performance evaluation of the ML estimator for the deterministic model, since it can be evaluated using Monte-Carlo simulations.

\section{Maximum Likelihood Estimation}
\label{sec:ML}
This section investigates the asymptotic properties of the ML estimator for both signal models: deterministic and stochastic as defined in Assumption 2.1 and 2.2, respectively. Based on the standard theory, the ML estimator is known to be consistent and efficient if the number of observations approaches infinity \cite{Kay1} while the number of nuisance parameters stays fixed. Since the received signal model in (\ref{Model3}) assumes each receiving array takes only a single snapshot, in the stochastic (\ref{likelihood_stoch}) and the deterministic (\ref{Likelihood_det}) likelihood functions the number of observations is equal to the number of transmit-to-receive array pairs, $M_tM_r$. Under the stochastic model assumption the number of nuisance parameters stays fixed as $M_tM_r$ grows, while for the deterministic model it increases. Hence, as $M_tM_r$ approaches infinity the standard theory can only be applied to the stochasitc model. In the following section we verify the asymptotic results for the stochastic model when $M_tM_r$ is large, and investigate the corresponding asymptotic properties for the deterministic model. In addition, we consider the asymptotic properties of ML in the scenario with fixed $M_tM_r$ and infinitely large $L_r$ for both models.

\subsection{Stochastic ML Estimator}
\label{subsec:StoML}
The ML estimates of the parameters are found by maximizing the corresponding log-likelihood function. Under the stochastic signal model assumption, the maximization of the log of the likelihood function in (\ref{likelihood_stoch}) is equivalent to the minimization of the following function
\be
\label{StochLogLike}
\begin{split}
\LL_s&\left(\psivec, \Amat, \sigma^2\right) \\
 = &\frac{1}{M_rM_tL_r}\sum_{l=1}^{M_r}\sum_{k=1}^{M_t}\left[ \log\left|\Rmat_{kl}\right| + Tr \cbl\Rmat_{kl}^{-1} \rvec_{kl} \rvec_{kl}^H\cbr\right].
\end{split}
\ee
To our knowledge the variables $\psivec$, $\Amat$, and $\sigma^2$ are not separable, and in general it is impossible to find closed form solutions that minimize the function in (\ref{StochLogLike}). On the other hand, some insight about the ML estimates of these parameters can be obtained by considering the asymptotic performance of the ML estimator when $M_tM_r$ and $L_r$ are large.

\subsubsection{Large $M_tM_r$}
\label{LargeMMStoch}
Consider the function in (\ref{StochLogLike}) evaluated at some given values of $\hat{\psivec}$, $\hat{\Amat}$, and $\hat{\sigma}^2$
\be
\begin{split}
\label{StochLogLikeEst}
\LL_s&\left(\hat{\psivec}, \hat{\Amat}, \hat{\sigma}^2\right) \\
=& \frac{1}{M_rM_tL_r}\sum_{l=1}^{M_r}\sum_{k=1}^{M_t}\left[ \log\left|\hat{\Rmat}_{kl}\right| + Tr \cbl\hat{\Rmat}_{kl}^{-1} \rvec_{kl} \rvec_{kl}^H\cbr\right]
\end{split}
\ee
where $\hat{\Rmat}_{kl} = \hat{\Hmat}_{kl}\hat{\Amat}\hat{\Hmat}_{kl}^H + \hat{\sigma}^2\Imat_{L_rN}$ is a positive definite matrix. In order to show that the ML estimates obtained by minimizing (\ref{StochLogLike}) are consistent we first show that the function in (\ref{StochLogLikeEst}) converges to the expected value when $M_tM_r \rightarrow \infty$, and then that the resulting expected value achieves its minimum at the true values of the parameters. 

Since the received signals, $\rvec_{kl}$, at different transmit-to-receive array paths have different covariance matrices $\Rmat_{kl}$ (\ref{CovarMat}), the function in (\ref{StochLogLikeEst}) is the sum of i.n.i.d. random variables. According to Kolmogorov's strong law of large numbers, the sample mean of i.n.i.d random variables converges to its expected value, as long as the variance of each individual summand is bounded \cite{Papoulis}. Appendix \ref{AppendixCA} demonstrates that the variance of the $kl$th term in $\LL_s\left(\hat{\psivec}, \hat{\Amat}, \hat{\sigma}^2\right)$ is always finite. Thus, Kolmogorov's strong law of large numbers can be applied to (\ref{StochLogLikeEst}) by letting $M_tM_r\rightarrow\infty$
\bea
\begin{split}
&\LL_s\left(\hat{\psivec}, \hat{\Amat}, \hat{\sigma}^2\right)\xrightarrow{a.s.} E\left[ \frac{1}{M_rM_tL_r}\sum_{l=1}^{M_r}\sum_{k=1}^{M_t} \left[\log\left|\hat{\Rmat}_{kl}\right| \right.\right. \\
&\qquad\qquad\qquad\qquad\qquad\qquad + \left.\left.Tr \cbl\hat{\Rmat}_{kl}^{-1} \rvec_{kl} \rvec_{kl}^H\cbr\right] \vphantom{\sum_{l=1}^{M_r}}\right]\\
& = \frac{1}{M_rM_tL_r}\sum_{l=1}^{M_r}\sum_{k=1}^{M_t} \left[\log\left|\hat{\Rmat}_{kl}\right| + Tr \cbl \hat{\Rmat}_{kl}^{-1} E\left[\rvec_{kl} \rvec_{kl}^H\right]\cbr \right] \\
& = \frac{1}{M_rM_tL_r}\sum_{l=1}^{M_r}\sum_{k=1}^{M_t} \left[\log\left|\hat{\Rmat}_{kl}\right| + Tr \cbl\hat{\Rmat}_{kl}^{-1} \Rmat_{kl} \cbr\right] \label{mle_stoch_app_asymptotic}
\end{split}
\eea
To show that the ML estimates are consistent, $\LL_s\left(\hat{\psivec}, \hat{\Amat}, \hat{\sigma}^2\right)$ in (\ref{mle_stoch_app_asymptotic}) is shown to be minimized by the true values of the parameters. The lower bound on (\ref{mle_stoch_app_asymptotic}) can be obtained by applying the inequality (\ref{StoicasInequality}) stated in Appendix \ref{Inequalities}
\be
\label{InequalityStoica}
\begin{split}
\frac{1}{M_rM_tL_r}&\sum_{l=1}^{M_r}\sum_{k=1}^{M_t} \log\left|\hat{\Rmat}_{kl}\right| + Tr \cbl\hat{\Rmat}_{kl}^{-1} \Rmat_{kl} \cbr \\
\geq &  N + \frac{1}{M_tM_rL_r}\sum_{l=1}^{M_r}\sum_{k=1}^{M_t} \log\det{\Rmat_{kl}} = const
\end{split}
\ee
while the equality in (\ref{InequalityStoica}) holds only if $\hat{\Rmat}_{kl}=\Rmat_{kl}$, $\forall k,l$. Thus, for $M_tM_r\rightarrow \infty$, the function in (\ref{StochLogLike}) is minimized when the estimates are equal to the true values of parameters: $\hat{\psivec} = \psivec$, $\hat{\Amat} = \Amat$, and $\hat{\sigma}^2 = \sigma^2$. Therefore, the ML estimator is consistent for $M_tM_r \rightarrow \infty$. In addition, according to the standard theory, since the number of nuisance parameters is fixed, the ML estimator for targets' parameters of interest is also efficient.

\subsubsection{Large $L_r$}
\label{LargeLrStoch}
This subsection considers the asymptotic behavior of the likelihood function in (\ref{StochLogLikeEst}) when the size of the receiving arrays, $L_r$, approaches infinity and the product $M_tM_r$ is finite. Similar to the case with an infinite number of transmit-to-receive array paths, $M_tM_r$, considered in the previous subsection, we first verify that (\ref{StochLogLikeEst}) converges to its expected value when $L_r\rightarrow \infty$. Appendix \ref{AppStochLr} shows that as $L_r$ approaches infinity, (\ref{StochLogLikeEst}) converges to
\be
\label{LargeLrStochLL}
\begin{split}
\LL_s\left(\hat{\psivec}, \hat{\Amat}, \hat{\sigma}^2\right)& \xrightarrow{a.s.} \frac{1}{M_rM_tL_r}\sum_{l=1}^{M_r}\sum_{k=1}^{M_t}\left[\log\left|\hat{\Rmat}_{kl}\right| \right. \\
 +& \left. Tr\cbl\hat{\Rmat}_{kl}^{-1}\left(\Hmat_{kl}\alphavec_{kl}\alphavec_{kl}^H\Hmat_{kl}^H + \sigma^2\Imat\right)\cbr\right]
\end{split}
\ee
which is not equal to the expected value in (\ref{mle_stoch_app_asymptotic}), provided $M_t$ and $M_r$ are finite. As it can be seen from (\ref{LargeLrStochLL}), the summation also requires $M_tM_r$ to be infinitely large in order for the trace term to converge to the expected value in (\ref{mle_stoch_app_asymptotic}). The consistent estimates of the targets' parameters of interest (locations and velocities) require consistent estimates of the corresponding bearing angles, time delays and Doppler shifts. The infinite receiving arrays can provide ideal estimates of the bearing angles, however for the time delays and Doppler shifts to be ideally estimated the number of transmit-to-receive array pairs must be much greater than the number of observed targets, $M_tM_r\gg Q$. Therefore, the ML estimates under the stochastic model assumption and infinitely large receiving arrays are inconsistent if $M_tM_r$ is finite.

\subsection{Deterministic ML Estimator}
\label{subsec:DetML}
Under the deterministic model assumption, the estimates of the unknown parameters can be found by the maximization of the log of the likelihood function in (\ref{Likelihood_det}) which is equivalent to the minimization of the function
\be
\label{lldet}
\begin{split}
\LL_d(\psivec, \alphavec, \sigma^2) = \frac{1}{M_tM_rL_r}&\sum_{l=1}^{M_r}\sum_{k=1}^{M_t} \left[L_rN\log\left(\pi\sigma^2\right) \right. \\
-& \left.\sigma^{-2} \Vert\rvec_{kl} - \Hmat_{kl}\alphavec_{kl}\Vert^2\right]
\end{split}
\ee
The estimate of the vectors of targets' reflectivities can be obtained by differentiating (\ref{lldet}) with respect to $\alphavec_{kl}$, and equating the resultant derivatives to zero while assuming the two other parameters $\sigma^2$ and $\psivec$ to be equal to their corresponding estimates $\hat{\sigma}^2$ and $\hat{\psivec}$. This leads to a necessary condition which expresses $\hat{\alphavec}_{kl}$ as a function of $\hat{\psivec}$ as
\be
\label{DetAlphaHat}
\hat{\alphavec}_{kl}(\hat{\psivec}) = \left(\hat{\Hmat}_{kl}^H\hat{\Hmat}_{kl}\right)^{-1}\hat{\Hmat}_{kl}^H\rvec_{kl}.
\ee
These steps also provide the estimate of the noise power as a function of $\hat{\psivec}$ as
\bea
\label{DetSigmaHat}
\hat{\sigma}^2(\hat{\psivec}) = \frac{1}{L_rNM_tM_r} \sum_{l=1}^{M_r} \sum_{k=1}^{M_t} Tr\cbl\Pi_{\hat{\Hmat}_{kl}}^{\perp} \rvec_{kl} \rvec_{kl}^H\cbr
\eea
where $\Pi_{\hat{\Hmat}_{kl}}^{\perp} = \Imat_{L_rN} - \hat{\Hmat}_{kl} \left(\hat{\Hmat}_{kl}^H \hat{\Hmat}_{kl}\right)^{-1} \hat{\Hmat}_{kl}^H$ is a projection matrix $\Pi_{\Hmat_{kl}}^{\perp}$ evaluated at the estimate $\hat{\psivec}$. Substituting $\hat{\sigma}^2(\hat{\psivec})$ and $\hat{\alphavec}_{kl}(\hat{\psivec})$ back into (\ref{lldet}) instead of the corresponding variables, the estimate of $\psivec$ can be found as the minimizer of the function
\be
\label{MLdet}
\begin{split}
\hat{\psivec} = & \arg\min_{\psivec}\hat{\sigma}^2(\psivec) = \arg\min_{\psivec} F\left(\psivec\right) \\
=& \arg\min_{\psivec} \frac{1}{L_rNM_tM_r}\sum_{l=1}^{M_r} \sum_{k=1}^{M_t} Tr\cbl\Pi_{\Hmat_{kl}}^{\perp} \rvec_{kl}\rvec_{kl}^H\cbr
\end{split}
\ee
Thus, the ML estimate of the vector of parameters of interest $\psivec$ is a vector $\hat{\psivec}$ which minimizes the estimate of the noise power $\hat{\sigma}^2(\hat{\psivec})$ in (\ref{DetSigmaHat}). The likelihood function in (\ref{MLdet}) is obtained form (\ref{lldet}) by removing the undesirable dependence on the nuisance parameters $\alphavec$ and $\sigma^2$. Such a likelihood function is known as the concentrated \cite{StoicaNehorai3} or profile \cite{Murphy} likelihood. Notice, since Assumption 2.2 considers the targets' reflectivities as deterministic unknowns, the number of unknown variables that have to be estimated grows with the number of antenna arrays, $M_tM_r$.

\subsubsection{Large $M_tM_r$}
\label{DetLargeMM}
Insight about the performance of the ML estimator in (\ref{MLdet}) can be obtained by studying an asymptotic case when the number of transmit-to-receive array pairs, $M_tM_r$, is large. Consider (\ref{MLdet}) evaluated at some given value of $\hat{\psivec}$
\be
\label{MLdetEst}
F\left(\hat{\psivec}\right) = \frac{1}{L_rNM_tM_r}\sum_{l=1}^{M_r} \sum_{k=1}^{M_t} Tr\cbl\Pi_{\hat{\Hmat}_{kl}}^{\perp} \rvec_{kl}\rvec_{kl}^H\cbr.
\ee
Similar to Section \ref{LargeMMStoch} the consistency of $\hat{\psivec}$ can be shown by first demonstrating that (\ref{MLdetEst}) converges to the expected value when $M_tM_r \rightarrow \infty$, and then showing that the obtained expected value is minimized by the true value of the vector of parameter $\psivec$.

Under Assumption 2.2 of the deterministic signal model, the radar echoes $\rvec_{kl}$ received at different transmit-to-receive array paths have different means, $\muvec_{kl} = \Hmat_{kl}\alphavec_{kl}$. Hence, (\ref{MLdetEst}) is a summation of i.n.i.d. random variables. According to Kolmogorov's strong low of large numbers, this summation converges to the expected value, if the variance of each summand is finite. Appendix \ref{AppendixDA} shows that for targets' reflectivities with bounded magnitude, $\vert\alpha_{kl}^q\vert^2 \leq \vert\alpha_{max}\vert^2$, $\forall k,q,l$, the variance of the $kl$th term in (\ref{MLdetEst}) is bounded. Thus, when the number of transmit-to-receive array paths, $M_tM_r$, is approaching infinity, the function in (\ref{MLdetEst}) converges to the expected value
\be
\begin{split}
F(\hat{\psivec}) &\xrightarrow{a.s.} E\left[\frac{1}{L_rNM_tM_r}\sum_{l=1}^{M_r} \sum_{k=1}^{M_t} Tr\cbl\Pi_{\hat{\Hmat}_{kl}}^{\perp} \rvec_{kl}\rvec_{kl}^H\cbr\right] \\
= & \frac{1}{L_rNM_tM_r}\sum_{l=1}^{M_r} \sum_{k=1}^{M_t} Tr\cbl\Pi_{\hat{\Hmat}_{kl}}^{\perp} E\left[\vphantom{\left(\right)^H} \left(\Hmat_{kl}\alphavec_{kl} + \evec_{kl}\right) \right.\right.\\
\quad & \qquad\qquad\qquad\qquad\qquad \cdot  \left.\left.\left(\Hmat_{kl}\alphavec_{kl} + \evec_{kl}\right)^H\right]\cbr\\
= & \frac{1}{L_rNM_tM_r}\sum_{l=1}^{M_r} \sum_{k=1}^{M_t}\left[Tr\cbl \Pi_{\hat{\Hmat}_{kl}}^{\perp} \Hmat_{kl}\alphavec_{kl}\alphavec_{kl}^H\Hmat_{kl}^H \cbr \right. \\
\quad & \qquad\qquad\qquad\qquad\qquad + \left.\sigma^2 Tr\cbl \Pi_{\hat{\Hmat}_{kl}}^{\perp}\cbr\right] \label{AsympDet}
\end{split}
\ee
where the last identity follows since the noise vector $\evec_{kl}$ is zero-mean with the covariance $\sigma^2\Imat_{L_rN}$.

Since the matrix inside the first trace operator in (\ref{AsympDet}) is positive semidefinite, and using the fact that the trace of the orthogonal projection matrix is equal to its rank, $Tr\cbl \Pi_{\hat{\Hmat}_{kl}}^{\perp}\cbr = L_rN-Q$, a bound on the expected value of (\ref{MLdetEst}) follows as
\be
\label{AsympDet1}
\begin{split}
F(\hat{\psivec}) &\xrightarrow{a.s.} \sigma^2\frac{L_rN - Q}{L_rN} \\
+& \frac{1}{L_rNM_tM_r}\sum_{l=1}^{M_r} \sum_{k=1}^{M_t}Tr\cbl \Pi_{\hat{\Hmat}_{kl}}^{\perp} \Hmat_{kl}\alphavec_{kl}\alphavec_{kl}^H\Hmat_{kl}^H \cbr \\
&\qquad\qquad\qquad\qquad\qquad\geq \sigma^2\frac{L_rN - Q}{L_rN}
\end{split}
\ee
Further $Tr\cbl \Pi_{\hat{\Hmat}_{kl}}^{\perp} \Hmat_{kl}\alphavec_{kl}\alphavec_{kl}^H\Hmat_{kl}^H \cbr$ in (\ref{AsympDet1}) is equal to zero if and only if $\hat{\Hmat}_{kl} = \Hmat_{kl}$, thus the equality in (\ref{AsympDet1}) holds only if the estimate is equal to the true value of the parameters vector. Therefore, the estimate of the targets' parameters vector $\hat{\psivec}$ is consistent.

However the estimates of the vectors of targets' reflectivities $\alphavec_{kl}$ are inconsistent. This can be seen from (\ref{DetAlphaHat}) by letting $\hat{\Hmat}_{kl}=\Hmat_{kl}$
\be
\label{DetAlphaHatAsymp}
\hat{\alphavec}_{kl}(\psivec) = \alphavec_{kl} + (\Hmat_{kl}^H\Hmat_{kl})^{-1} \Hmat_{kl}^H\evec_{kl}
\ee
Additionally, since $\sigma^2(\psivec) = F(\psivec)$ and following the asymptoic result in (\ref{AsympDet1}), the estimate of the noise power in (\ref{DetSigmaHat}) converges to
\be
\label{DetSigmaHatAsymp}
\hat{\sigma^2}(\psivec) \rightarrow \frac{L_rN-Q}{L_rN} \sigma^2.
\ee
Thus, infinitely large product $M_tM_r$ does not provide a consistent estimate of the noise power. The biased estimates of the targets' reflectivities and the noise power for large $M_tM_r$ result in the ML estimate of $\psivec$ being consistent but not efficient.

\subsubsection{Large $L_r$}
\label{sec:LargeLrDet}
This section proves consistency and efficiency of the estimates of $\psivec$, $\alphavec$ and $\sigma^2$, for a finite number of transmit-to-receive array paths, $M_tM_r$, and infinitely large size of the receiving arrays $L_r$.

Appendix \ref{AppendixDB} proves that as $L_r\rightarrow \infty$, the function $F(\hat{\psivec})$ in (\ref{MLdetEst}) converges to the same expected value as in (\ref{AsympDet1}) which is minimized when $\hat{\psivec}=\psivec$. Hence, the estimates of $\psivec$ in (\ref{MLdet}) are consistent if $L_r\rightarrow \infty$. Furthermore, as $L_r\rightarrow \infty$ the targets can be considered as sufficiently separated such that the following assumption holds:

\textit{Assumption 4 (Well separated targets):} For a given receiving arrays length, $L_r$, and a number of temporal samples, $N$, any two observed targets $q$ and $q'$ are well separated (in space and Doppler) if the corresponding spatio-temporal steering vectors are nearly orthogonal
\be
\left(\hvec_{kl}^{q}\right)^H\hvec_{kl}^{q'} \approx \left\lbrace
\begin{array}{cc}
\frac{L_rE}{L_tM_t}\vert\wvec_k^H\avec_{tk}^q\vert^2 (\zeta_{kl}^q)^2 & q=q'\\
0 & otherwise
\end{array} \right..
\ee
Thus for $Q$ well separated targets
\bea
\label{remot_assmp}
\Hmat_{kl}^H\Hmat_{kl} & \approx & \frac{L_rE}{L_tM_t} diag\left(\vert \wvec_k^H\avec_{tk}^1\vert^2(\zeta_{kl}^1)^2, \vert \wvec_k^H\avec_{tk}^2\vert^2(\zeta_{kl}^2)^2,\right.\nonumber\\
&\ldots &,\left.\vert \wvec_k^H\avec_{tk}^Q\vert^2(\zeta_{kl}^Q)^2\right).
\eea
Using (\ref{remot_assmp}) it can be shown that the variance of the bias term in (\ref{DetAlphaHatAsymp}) approaches zero as $L_r \rightarrow \infty$
\bea
\label{biasalpha}
&var&\left[(\Hmat_{kl}^H\Hmat_{kl})^{-1} \Hmat_{kl}^H\evec_{kl}\right] = 
\sigma^2(\Hmat_{kl}^H\Hmat_{kl})^{-1} \\
&\approx & \sigma^2\frac{L_tM_t}{L_rE} \left[ diag\left(\vert \wvec_k^H\avec_{tk}^1\vert^2(\zeta_{kl}^1)^2, \vert \wvec_k^H\avec_{tk}^2\vert^2(\zeta_{kl}^2)^2,\right.\right. \nonumber\\
&\ldots&,\left.\left.\vert \wvec_k^H\avec_{tk}^Q\vert^2(\zeta_{kl}^Q)^2\right)\right]^{-1} \rightarrow \Zeromat. \nonumber
\eea 
Thus the estimates of $\alphavec_{kl}$ in (\ref{DetAlphaHatAsymp}) are consistent when $L_r \rightarrow \infty$.

Finally, since $F(\hat{\psivec})$ converges to the expected value in (\ref{AsympDet1}) when $L_r$ becomes asymptotically large, the estimate of the noise power $\hat{\sigma}^2(\psivec) = F(\psivec)$ converges to (\ref{DetSigmaHatAsymp}). One can notice by observing (\ref{DetSigmaHatAsymp}) that if the size of the receiving array is sufficiently large such that $L_rN\gg Q$, the estimate of the noise power becomes consistent.

Thus the estimates of the targets' parameters of interest, the noise variance and the targets' reflectivities are consistent when $L_r$ is infinitely large. Unlike adding transmitting and receiving arrays, increasing $L_r$ does not increase the number of nuisance parameters which have to be estimated. Since for infinitely large arrays, all targets can be considered as well-separated, it follows from the standard theory that the ML estimates of $\psivec$ are also asymptotically efficient when $L_r \rightarrow \infty$.

\vspace{6 mm}
To summarize, this section provides a study of the asymptotic properties of the ML estimator under the stochastic and deterministic signal model assumptions for a radar with widely distributed arrays. The ML estimator of $\psivec$ is shown to be consistent and efficient under the stochastic model assumption when the number of transmit-to-receive array paths, $M_tM_r$, is infinitely large. On the other hand, when $M_tM_r$ is finite, increasing the receiving array size, $L_r$, to infinity does not result in the consistent estimates of $\psivec$.

Under the deterministic model assumption, the ML estimator of $\psivec$ is shown to be consistent but not efficient when $M_tM_r\rightarrow \infty$, since the number of nuisance parameters increases as $M_tM_r$ increases. However, when $M_tM_r$ is fixed and $L_r\rightarrow \infty$ the ML estimator for the deterministic model becomes consistent and efficient.

The conducted asymptotic analysis provides insight into the parameter estimation performance of the radar system with widely separated antenna arrays, when $M_tM_r$ and $L_r$ are infinitely large. However, in practice an infinite number of antennas is infeasible, and optimal allocation of a finite number of antennas into a finite number of widely separated arrays remains an open question. In addition, a sufficiently large SNR is required in practice for the derived CRLB to be a good approximate performance measure. In order to further investigate these problems we conduct numerical simulations for finite $M_tM_r$ and $L_r$, and various values of SNR.

\section{Simulation Results}
This section uses the derived CRLB for stochastic and deterministic signal models and Monte Carlo simulations to assess the estimation performance of different configurations of the radar with widely separated antenna arrays. In order to compare different radar systems based on the same amount of transmitted power and occupied bandwidth, we assume that all radar configurations considered in this section have single element transmitting arrays ($L_t = 1$). The results from two sets of simulation scenarios are presented. In the first set the number of transmitting arrays is $M_t=6$, and the total number of receiving elements is $M_rL_r = 512$. In the second set of the simulation scenarios $M_t=2$ and $M_rL_r=128$. In both sets of the simulation scenarios, the target parameter estimation performance is studied for multiple radar configurations by changing the number of receiving arrays $M_r$ and their size $L_r$ while keeping the product $M_rL_r$, and the number of transmitting arrays $M_t$ fixed.

In all the simulation scenarios considered here, the receiver and the transmitter arrays are located equidistantly and symmetrically with respect to the origin \cite{HaimBlum5,HaimBlum6}. Such antenna placements allow for an easy to explain, general methodology for changing the number of arrays whose performance is easy to interpret to facilitate validation of the derived CRLBs. The centers of the transmitting and receiving arrays are
\be
\begin{split}
( x_{tk}, y_{tk} ) = & R \left(\cos \left( ( 2k - 1 ) \pi / M_t \right), \right. \\
& \qquad\left.\sin \left( ( 2k - 1 ) \pi / M_t \right) \right), k = 1,2,\ldots,M_t \nonumber\\
( x_{rl}, y_{rl} ) = & R \left(\cos \left( 2 ( l - 1 ) \pi / M_r \right), \right. \\
& \qquad\left.\sin \left( 2 ( l - 1 ) \pi / M_r \right) \right), l = 1,2,\ldots,M_r 
\end{split}
\ee
where $R$ is a distance from the origin to the transmitter or receiver. In the presented results $R = 1100$m.

\begin{figure*}[htb]
\centerline{\psfig{figure=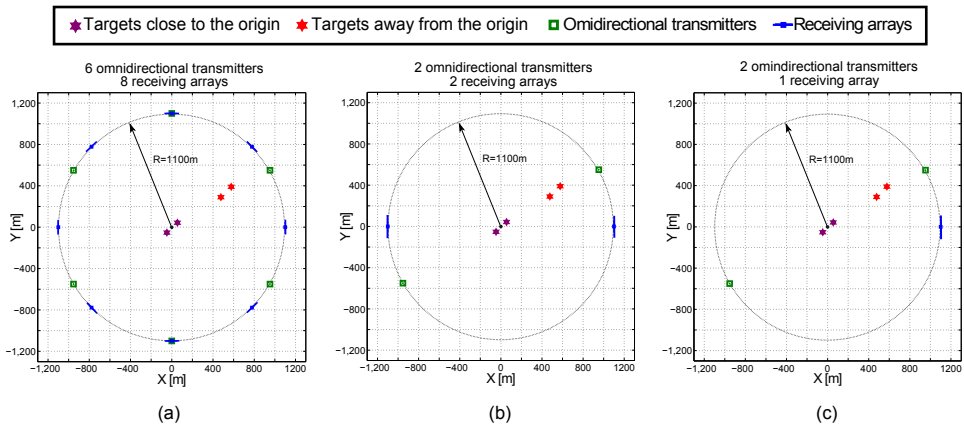,width=5in,height=2.2in}}
\caption{Examples of symmetrical and equdistant antenna placement: (a) $M_t = 6$ and $M_r = 8$; (b) $M_t = 2$ and $M_r = 2$; (c) $M_t = 2$ and $M_r = 1$. The boresighs of all receiving arrays point at the origin (the placements of the array elements are indicated by the lines passing through the corresponding receiver). Two targets positions are considered: 1) the targets are placed close to the origin at $( -40, -50 )$ and $( 60, 50 )$; 2) the targets are located away from the origin at $( 410, 320 )$ and $( 510, 420 )$.} \label{fig:scenario}
\end{figure*}

All receiving antennas are assumed to be uniformly spaced linear arrays (ULA) with $L_r$ elements and half wavelength inter-element spacing. 
%For the assumed carrier frequency of $f_c = 10$GHz, the wavelength is $\lambda = 0.03$m. 
The phase center of the $l$th receiving array is assumed to be in the geometrical center of the array located at $( x_{rl}, y_{rl} )$. The orientation of the $l$th receiving array is chosen such that its boresight direction points towards the origin. Examples of such a symmetrical antenna placement are shown in Fig. \ref{fig:scenario}.

We assume all transmitters use a pulse train waveform of LFM chirps, which in the time-sampled signal domain have the following form
\be
s_k[n] = \sum_{z=0}^{Z-1}s_0\left(n\triangle t - z T_r - \tilde{T}_k\right) \nonumber
\label{LFM}
\ee
\be
s_0(t) = e^{j\pi\frac{f_B}{T_0}\left(t - \frac{1}{2}T_0\right)^2}\left[h(t)-h(t - T_0)\right] \nonumber
\ee
where $Z$ is the number of transmitted pulses, $T_r$ is the pulse repetition interval, $f_B$ is the bandwidth of the chirp, $T_0$ is the pulse duration, and $h(t)$ is the Heaviside step function. Let all transmitted waveforms contain the same number of pulses $Z=3$ with equal pulse repetition interval $T_r = 54$ms, bandwidth of $f_B = 1$MHz, and pulse duration $T_0=20\mu$s. The orthogonality assumption in (\ref{orthog2}) is satisfied by setting the time delays $\tilde{T}_k$ for each transmitter, such that there is no overlap between different transmitted waveforms for the set of time delays of interest. Notice that in such a way the waveform orthogonality can be achieved only if the surveillance area is limited. The derived CRLB and ML results are general and do not depend on the specific type of the transmitted orthogonal waveforms.

The simulation results are obtained for two different positions of $Q=2$ targets. In the first case the targets are close to the origin and approximately equidistant from all transmitting and receiving arrays: $(x_1, y_1) = ( -40, -50 )$ and $( x_2, y_2 ) = ( 60, 50 )$. In the second case the targets' locations are chosen such that they are closer to some transmitting and receiving antennas and more remote from the others: $(x_1, y_1) = ( 410, 320 )$, and $( x_2, y_2 ) = ( 510, 420 )$. The distance between the targets in both cases is $r \approx 141.42$m. Since the MSE assessment of the ML estimator's performance requires computationally intensive Monte Carlo simulations, the simulation scenarios presented here consider the target location estimation only. Thus the vector of unknown parameters is $\psivec = \left[x^1,y^1,x^2,y^2\right]^T$. Notice that the range resolution of the LFM waveform with $1$MHz bandwidth is $\triangle r = c / (2f_b) = 150m > r$, thus the radar cannot reliably resolve the targets in range. We consider such targets as closely spaced.

Let the vector of targets' reflectivities be generated from the complex circular Gaussian distribution with zero-mean and diagonal covariance matrix $\Amat = \sigma_{\alpha}^2\Imat$. Thus the targets are assumed to be uncorrelated. The total SNR is defined as an average of the SNRs over all transmit-to-receiver array pairs and targets
\be
\label{SNRsim}
SNR = \frac{1}{M_tM_rQ}\sum_{k=1}^{M_t}\sum_{l=1}^{M_r}\sum_{q=1}^Q SNR_{kl}^q \nonumber
\ee
where
\be
SNR_{kl}^q = E\frac{\sigma_{\alpha}^2 \left( \zeta_{kl}^q \right)^2}{\sigma^2} \nonumber.
\ee

Recall that the target parameter estimation problem considered in this paper under Assumption 2.1 of the stochastic signal model treats the covariance matrix $\Amat$ as an unknown nuisance parameter. In addition, although the diagonal covaraince matrix is used to generate the signals in the simulation results presented here, no assumptions were made about the shape of the $\Amat$ while deriving the likelihood function in (\ref{likelihood_stoch}).

\begin{figure*}[!t]
\centering
\begin{tabular}{c}
	\subfloat[]{\includegraphics[width=3in]{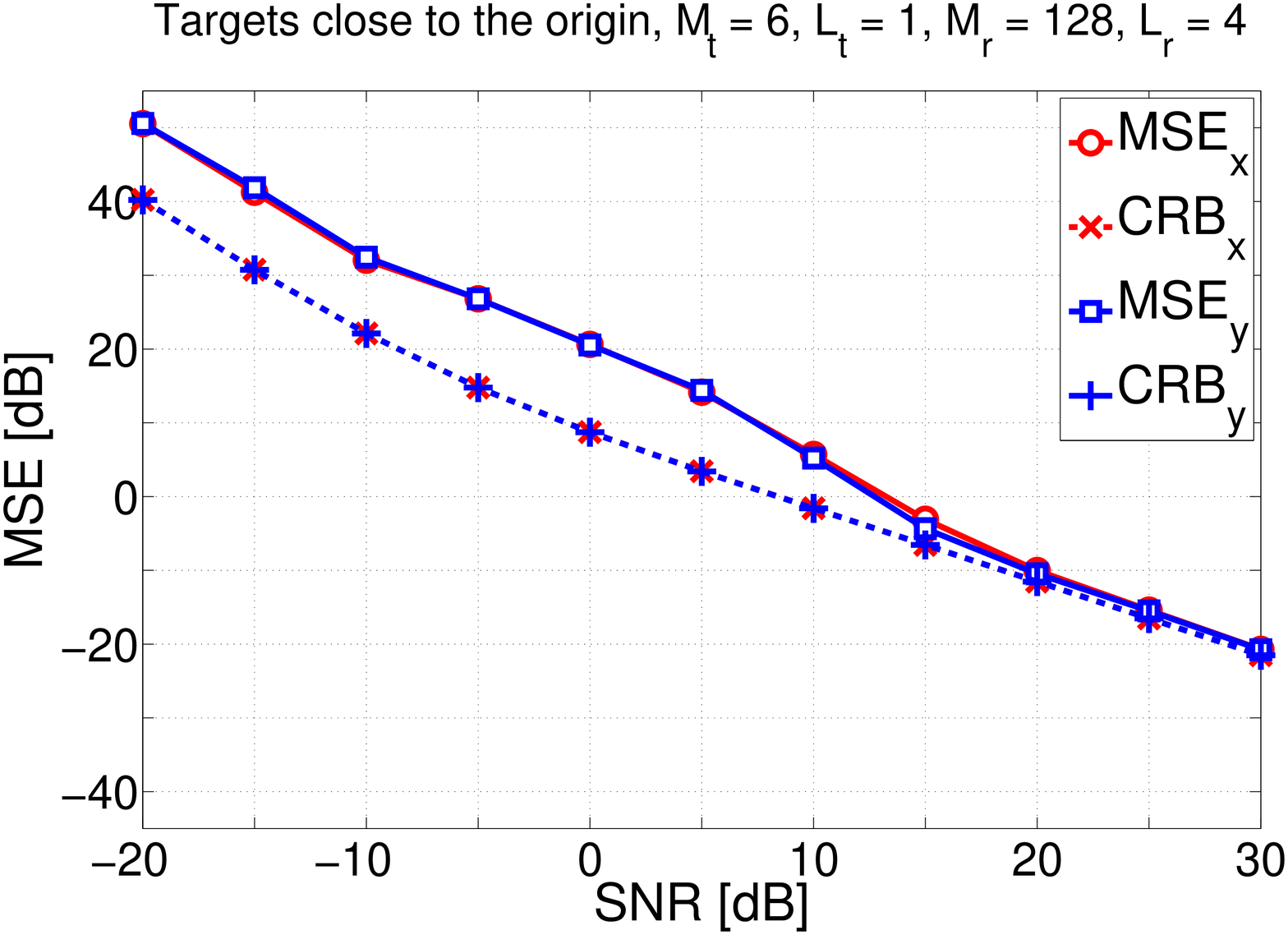}}
	\subfloat[]{\includegraphics[width=3in]{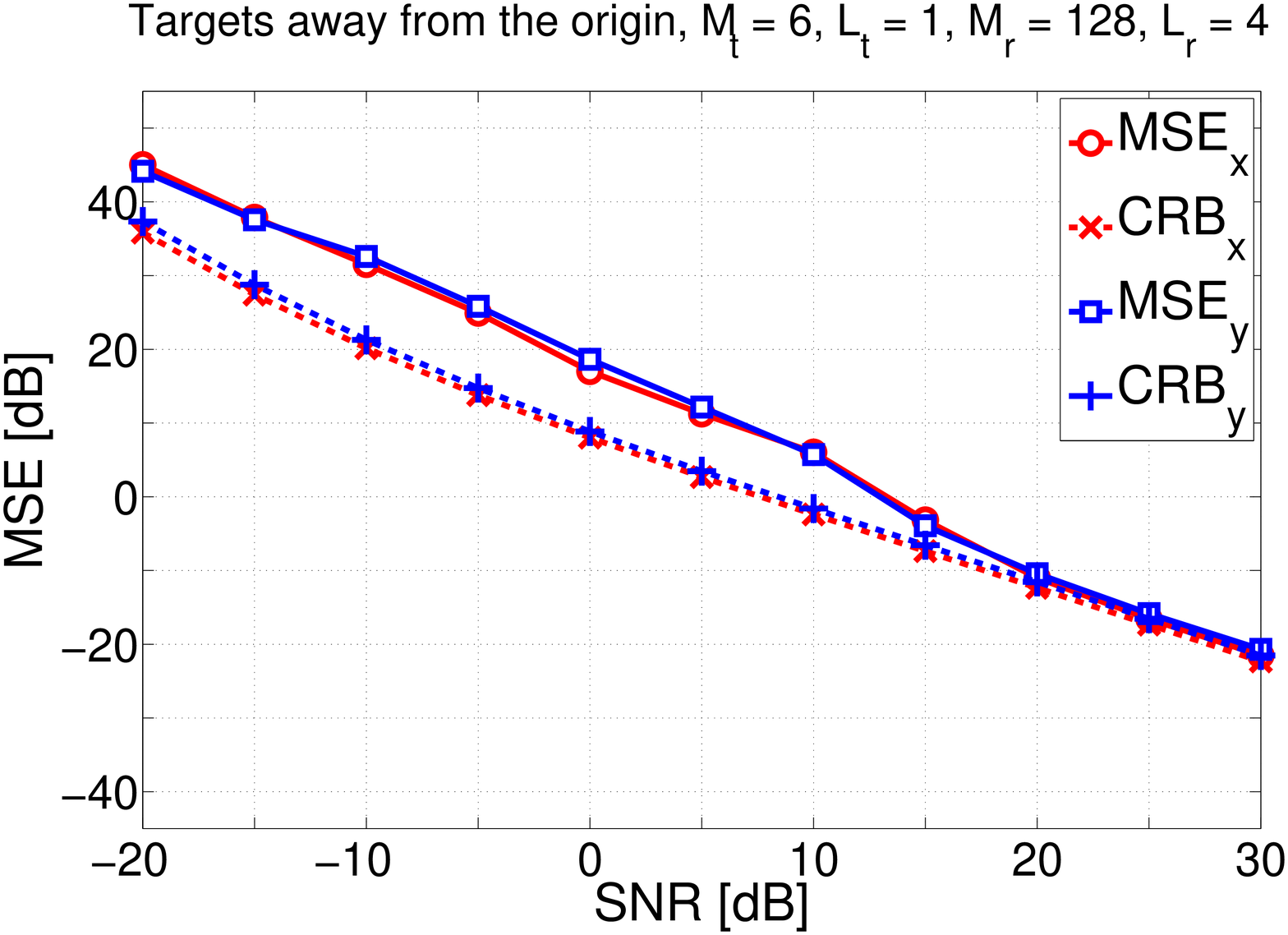}}\\
	
	\subfloat[]{\includegraphics[width=3in]{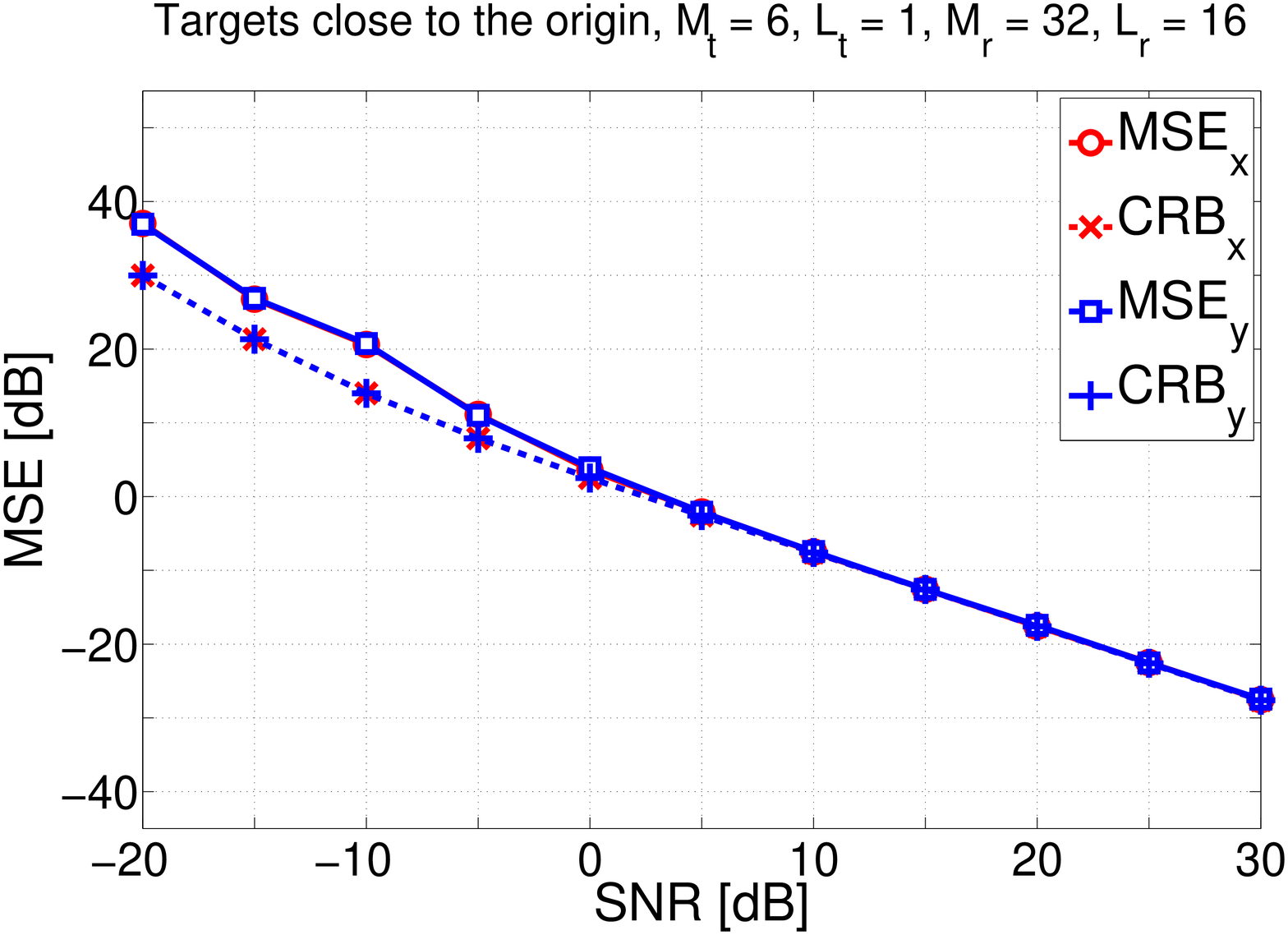}}
	\subfloat[]{\includegraphics[width=3in]{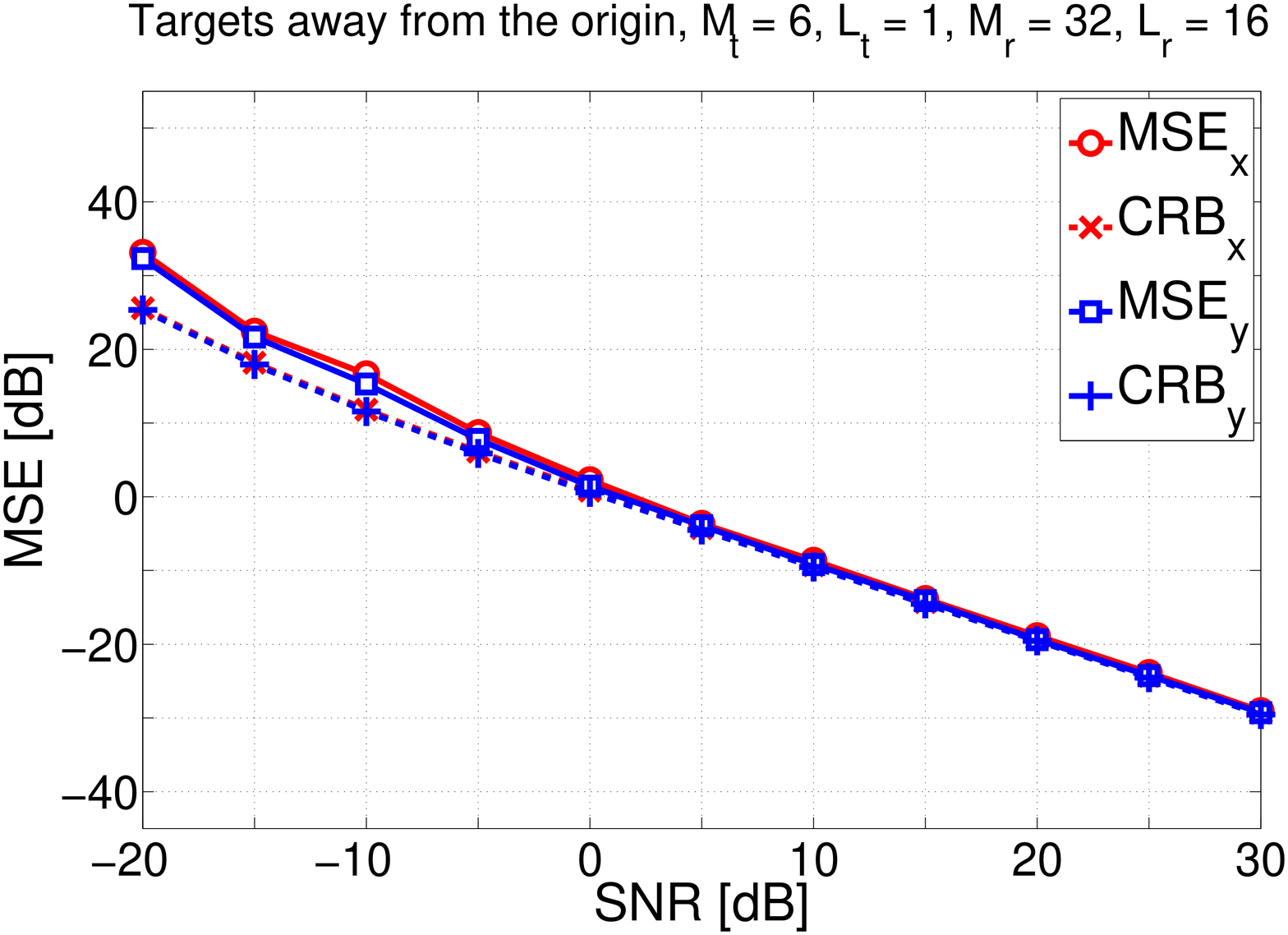}}\\
	
	\subfloat[]{\includegraphics[width=3in]{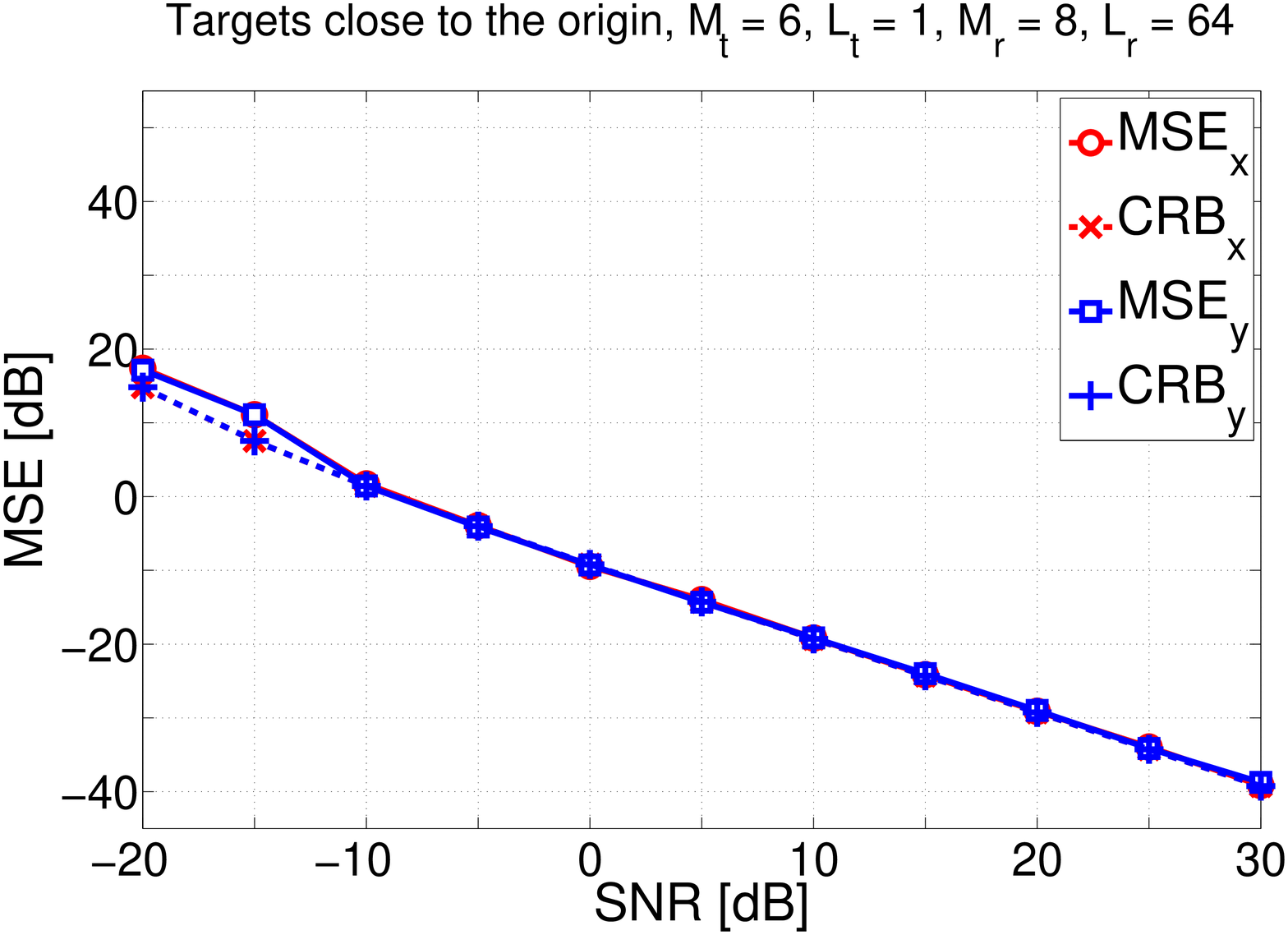}}
	\subfloat[]{\includegraphics[width=3in]{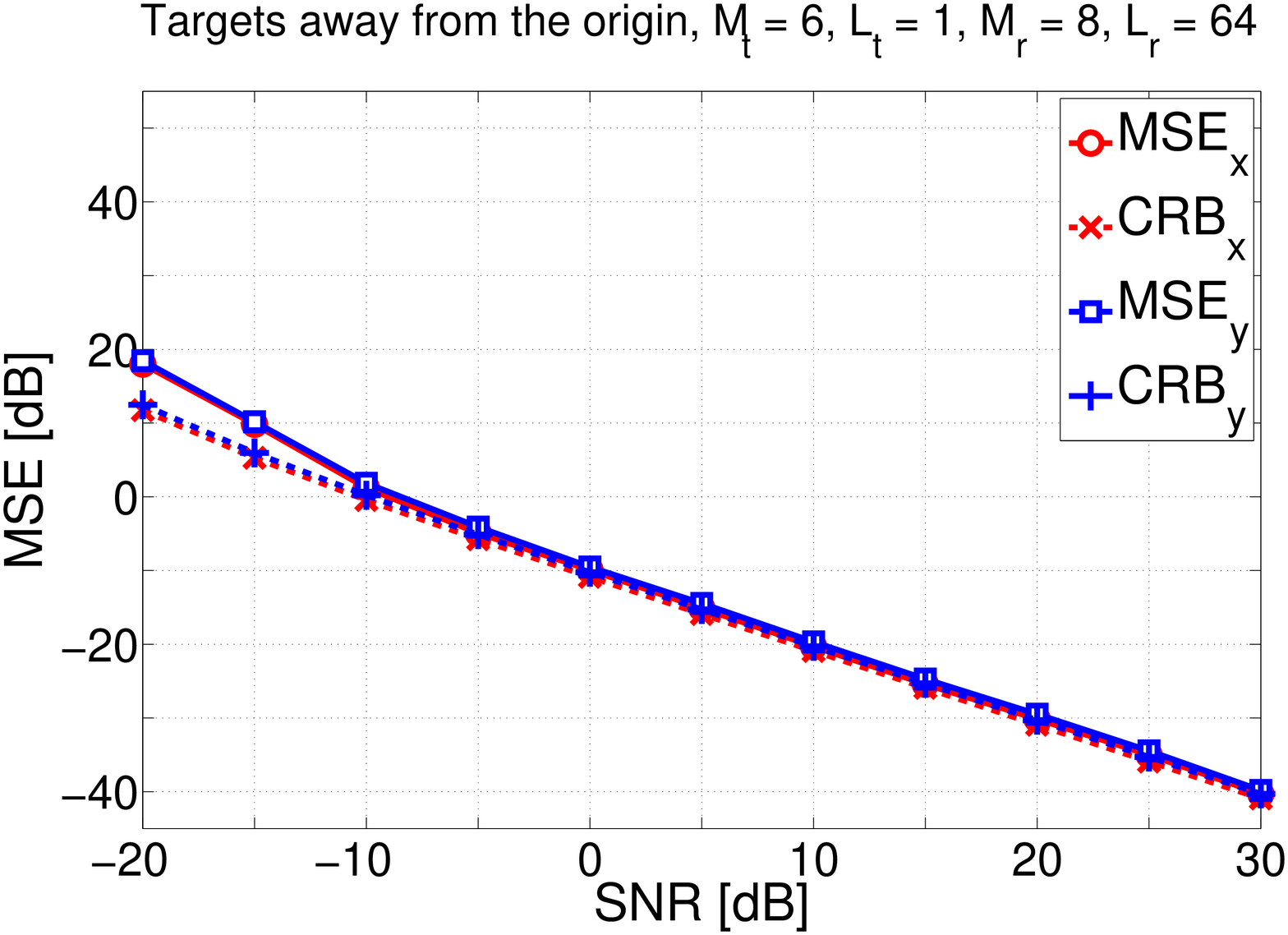}}
\end{tabular}
\caption{The CRLB and the MSE of the ML estimator for the stochastic signal model with two targets. The first target's location estimation results are shown for two closely spaced targets placed close to the origin in (a), (c) and (e), and away from the origin in (b), (d), and (f). Each configuration of the radar with widely separated arrays has $M_t = 6$ single antenna element transmitting arrays and the total number of receiving antenna elements is fixed at $M_rL_r = 512$. The number and the size of the receiving arrays vary: (a) and (b) $M_r = 128$, $L_r = 4$; (c) and (d) $M_r = 32$, $L_r = 16$; (e) and (f) $M_r = 8$, $L_r = 64$.}
\label{fig:CRBsto}
\end{figure*}

\subsection{$M_rL_r = 512$ and $M_t = 6$}
This subsection presents a set of simulation scenarios for different configurations of the radar with multiple widely separated arrays when the total number of receiving antenna elements and the number of transmitting arrays are fixed to $M_rL_r = 512$ and $M_t=6$ respectively. The MSE performance of the ML estimator as a function of SNR for the stochastic signal model is shown in Fig. \ref{fig:CRBsto}. The CRLB is evaluated using the expression in (\ref{FIMstochFianl}), and the ML estimates are obtained by minimizing (\ref{StochLogLike}). The results are shown for the first target's location estimation only, since the results obtained for the second target are identical. Subplots (a), (c), and (e) of Fig. \ref{fig:CRBsto} show the results for the targets located close to the origin, while subplots (b), (d), and (f) show the results for the targets located away from the origin. Subplots (a) and (b) demonstrate the estimation performance of the the radar with the large number of transmit-to-receive array path, $M_tM_r = 768$, but short receiving arrays, $L_r = 4$. The resolution of the receiving arrays, in such a configuration, is not sufficient to separate the targets. Thus, the MSE is predicted well by the CRLB only for large values of SNR. Contrary to this, a configuration with a relatively small value of the product $M_tM_r = 48$, and the long receiving arrays, $L_r = 64$, is shown in (e) and (f). The receiving arrays have sufficient resolution to separate the targets. The MSE starts converging to the CRLB at the SNR$=-10$dB, which is $30$dB lower, than in the case shown in (a) and (b). Subplots (c) and (d) of Fig. \ref{fig:CRBsto} demonstrate the transition between the configuration with the large number of small receiving arrays to the configuration with the long receiving arrays. One can observe from Fig. \ref{fig:CRBsto} that as $L_r$ increases over the limited range shown, for sufficiently large $M_tM_r$ and fixed $M_rL_r$, and targets well surrounded by the separated arrays, the location estimation performance of the radar with widely separated arrays improves in two ways: 1) the CRLB becomes lower, 2) the asymptotic region starts at lower values of the SNR. Notice, that the estimation performance does not change significantly when the targets are moved away from the origin, despite that some of the transmit-to-receive array pairs become more dominant than the others. This can be explained by the large values of $M_tM_r$ considered in these simulation scenarios, which provide a sufficient geometric diversity.

\begin{figure*}[!t]
\centering
\begin{tabular}{c}
	\subfloat[]{\includegraphics[width=3in]{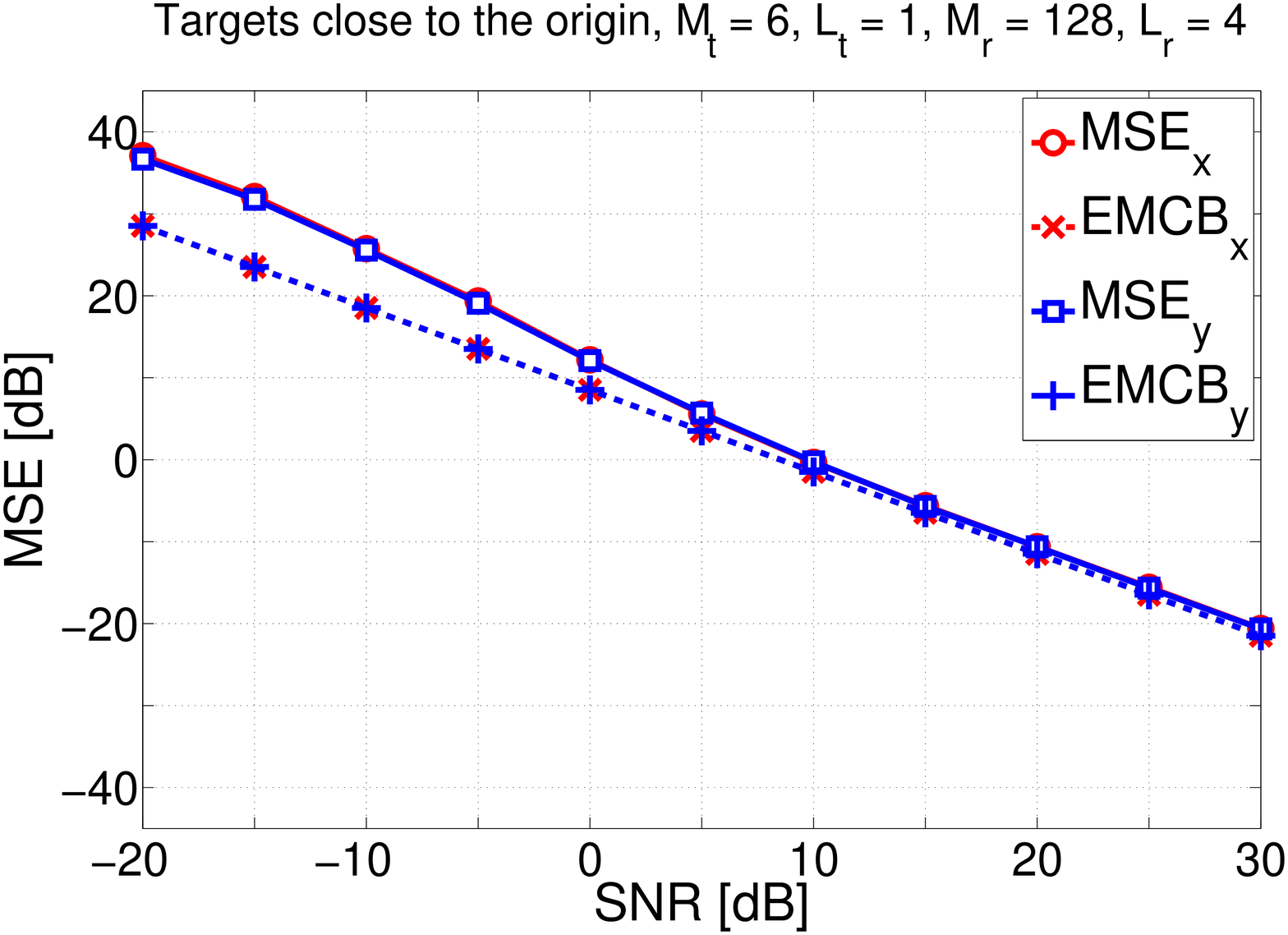}}
	\subfloat[]{\includegraphics[width=3in]{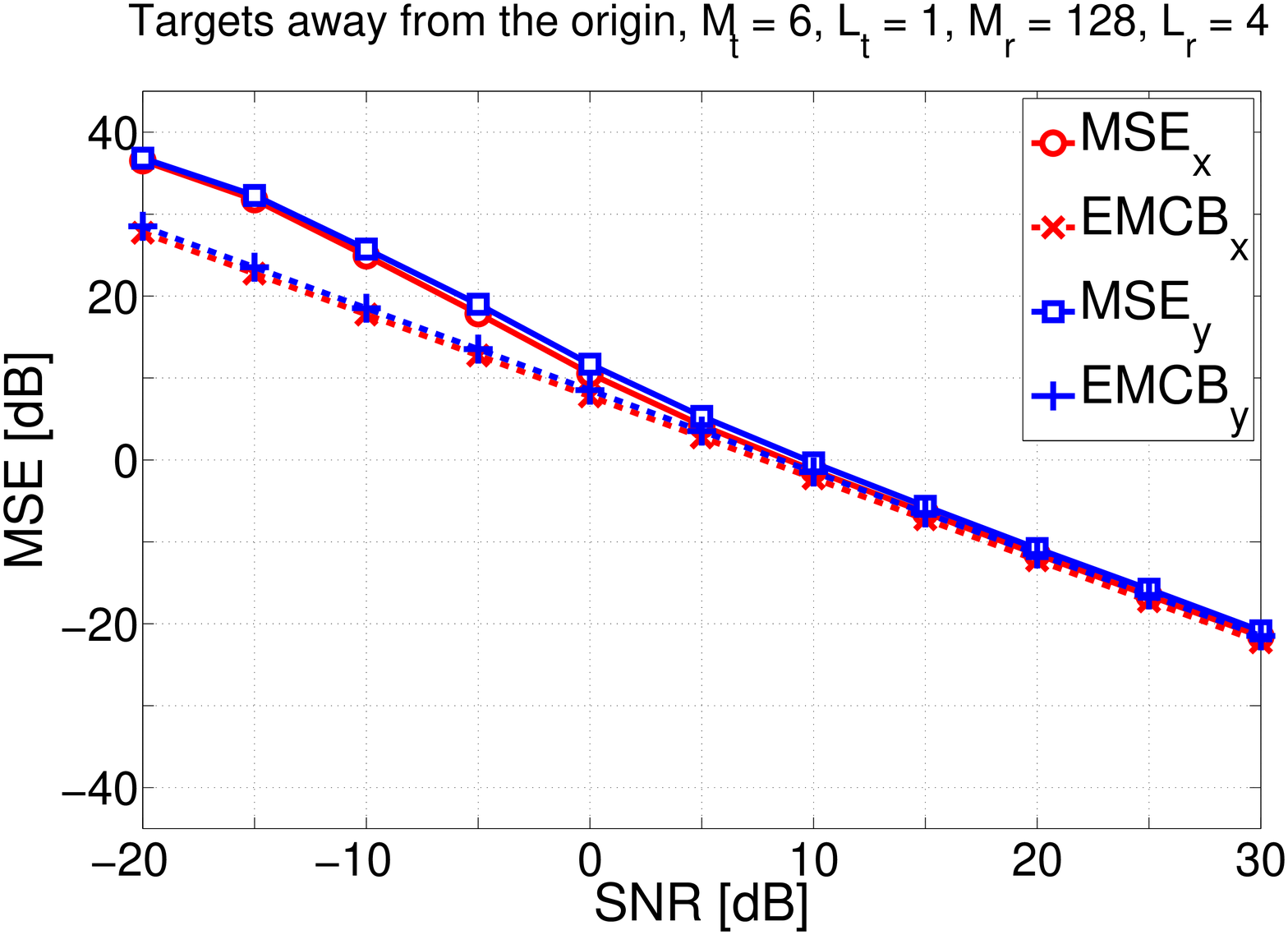}}\\
	
	\subfloat[]{\includegraphics[width=3in]{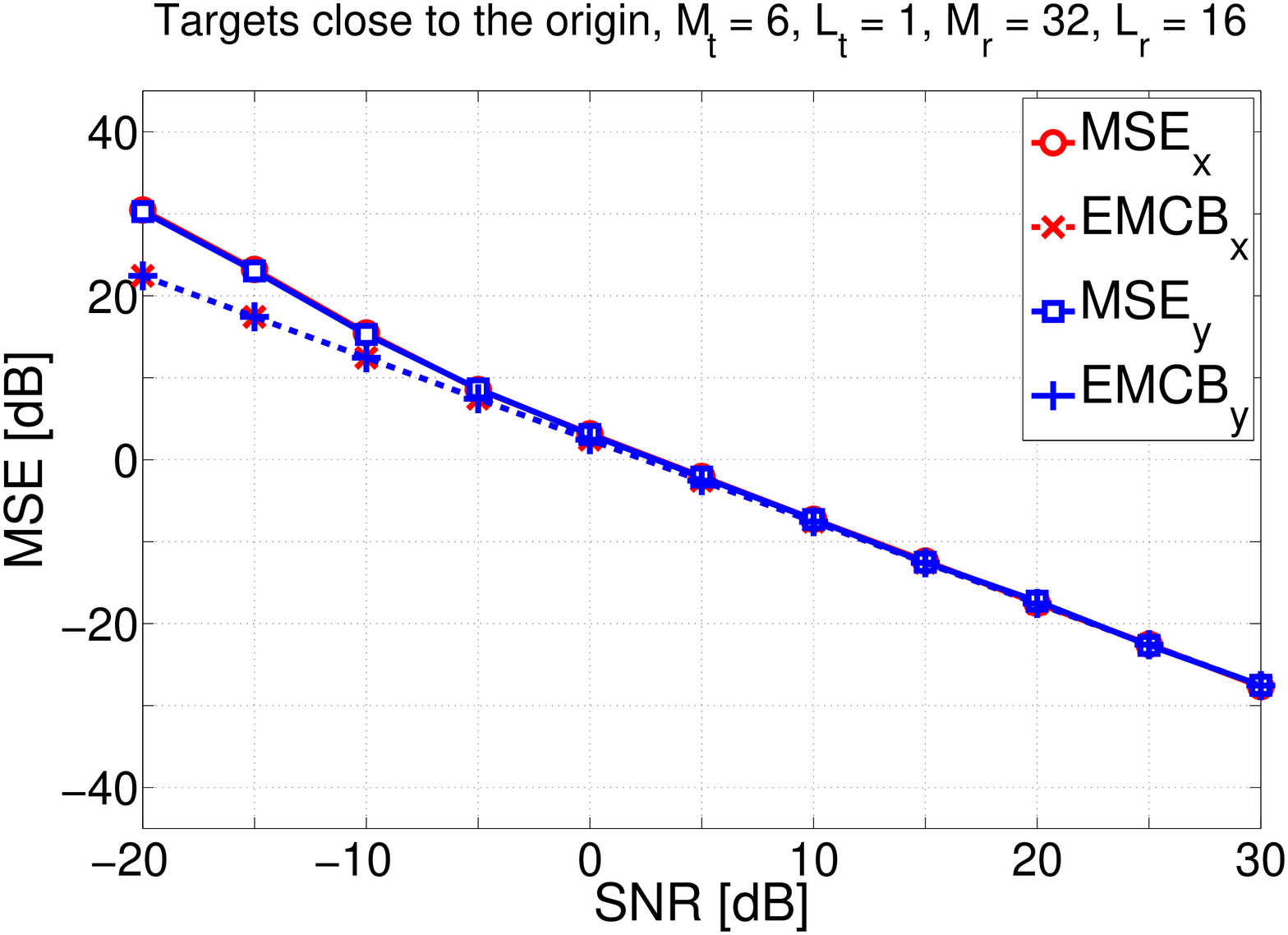}}
	\subfloat[]{\includegraphics[width=3in]{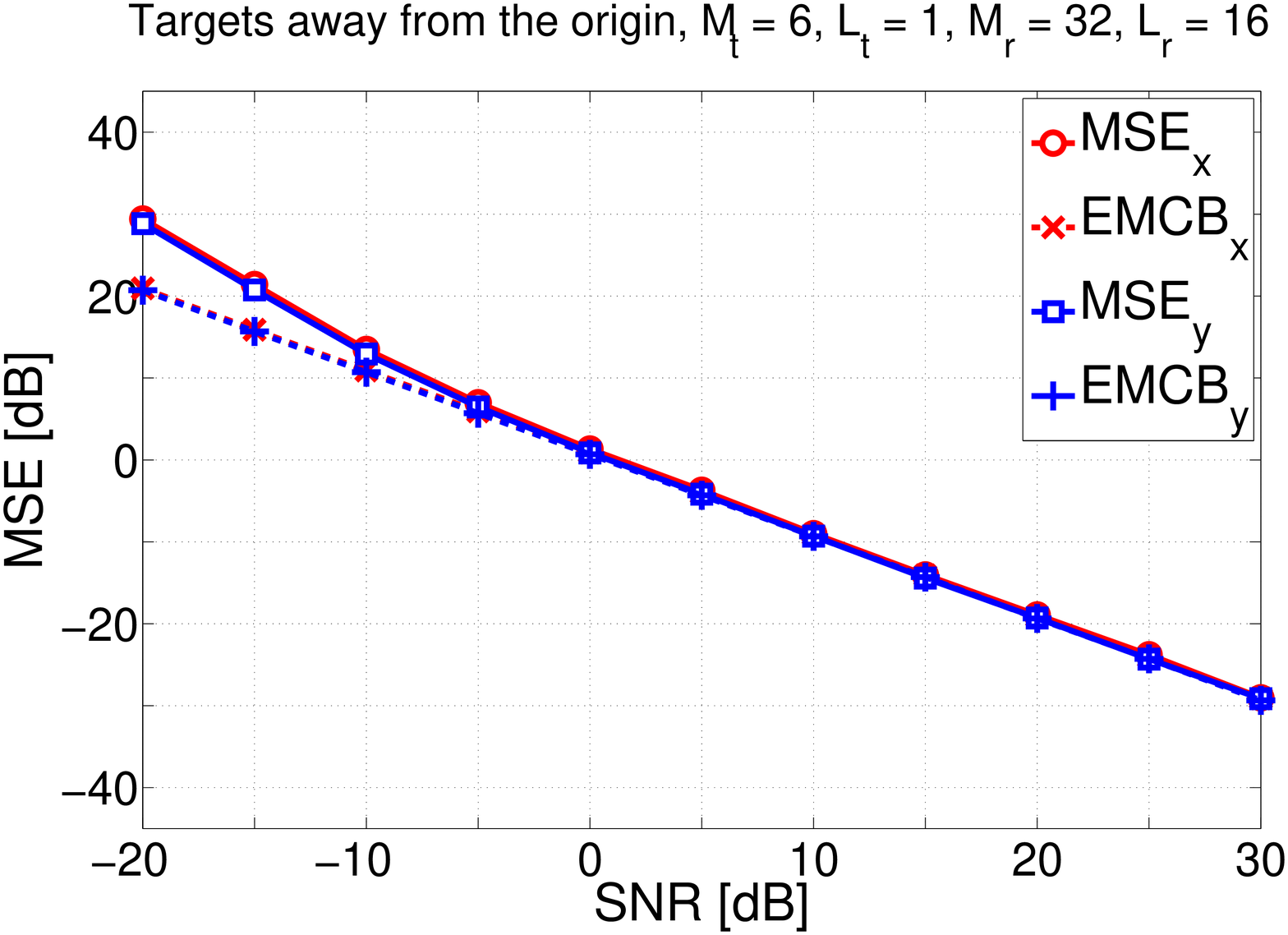}}\\
	
	\subfloat[]{\includegraphics[width=3in]{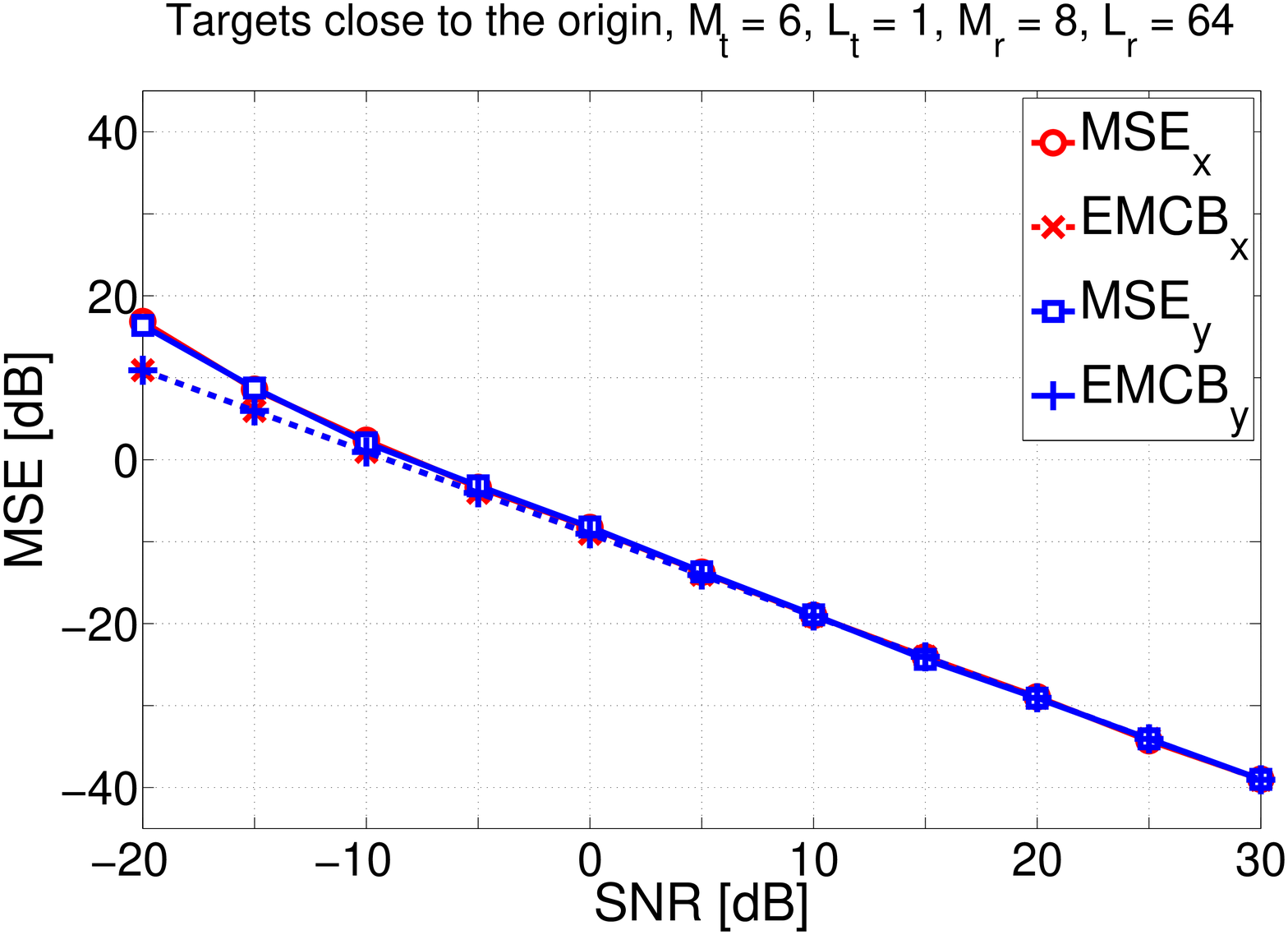}}
	\subfloat[]{\includegraphics[width=3in]{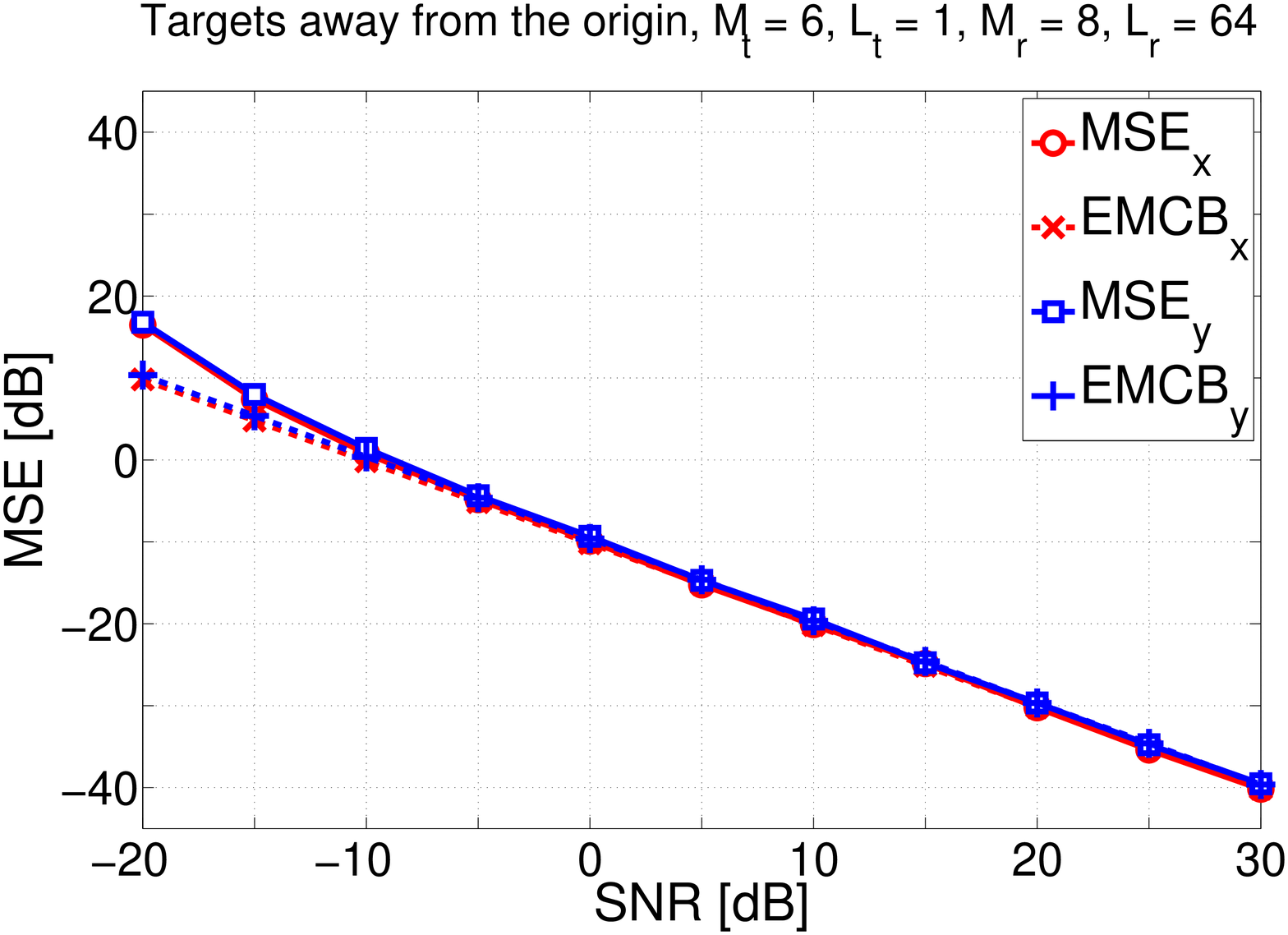}}
\end{tabular}
\caption{The EMCB and the MSE of the ML estimator for the deterministic signal model with two targets. The first target's location estimation results are shown for two closely spaced targets placed close to the origin in (a), (c), and (e), and away from the origin in (b), (d), and (f). Each configuration of the radar with widely separated arrays has $M_t = 6$ single antenna element transmitting arrays and the total number of receiving antenna elements is fixed at $M_rL_r = 512$. The number and the size of the receiving arrays vary: (a) and (b) $M_r = 128$, $L_r = 4$; (c) and (d) $M_r = 32$, $L_r = 16$; (e) and (f) $M_r = 8$, $L_r = 64$.}
\label{fig:CRBdet}
\end{figure*}

Fig. \ref{fig:CRBdet} presents results similar to Fig. \ref{fig:CRBsto} obtained under the deterministic signal model assumption. Here we calculate EMCB instead of the CRLB, as discussed prior to (\ref{emcb}), assuming the targets' reflectivities are generated from the same zero-mean complex Guassian distribution as in the stochastic case. The radar configurations with $M_t = 6$ and $M_rL_r = 512$ are again considered. Since for the deterministic signal model the number of unknowns grows with the number of transmit-to-receiver array paths, $M_tM_r$, it is expected that the radar configuration with smaller $M_tM_r$ and larger receiving arrays will provide a better estimation performance. This can be observed by comparing subplots (a) and (b) to subplots (e) and (f) of Fig. \ref{fig:CRBdet}. Similar to the MSE performance for the stochastic model, for sufficiently large $M_tM_r$, as $L_r$ increases over the range shown, the asymptotic region starts at lower values of the SNR and the overall variance of the estimate decreases. Fig. \ref{fig:CRBdet} shows that even for finite $M_tM_r$ and $L_r$ the EMCB is a good prediction of the MSE when the SNR is high. In addition, Fig. \ref{fig:CRBdet} demonstrates that the results obtained for the targets located close to the origin are similar to the results obtained for the targets located away from origin. Therefore the estimation performance does not significantly depend on the targets locations if $M_tM_r$ is sufficiently large, and the targets are well surrounded by transmitting and receiving antennas.

\begin{figure*}[!t]
\centering
\begin{tabular}{c}
	\subfloat[]{\includegraphics[width=3in]{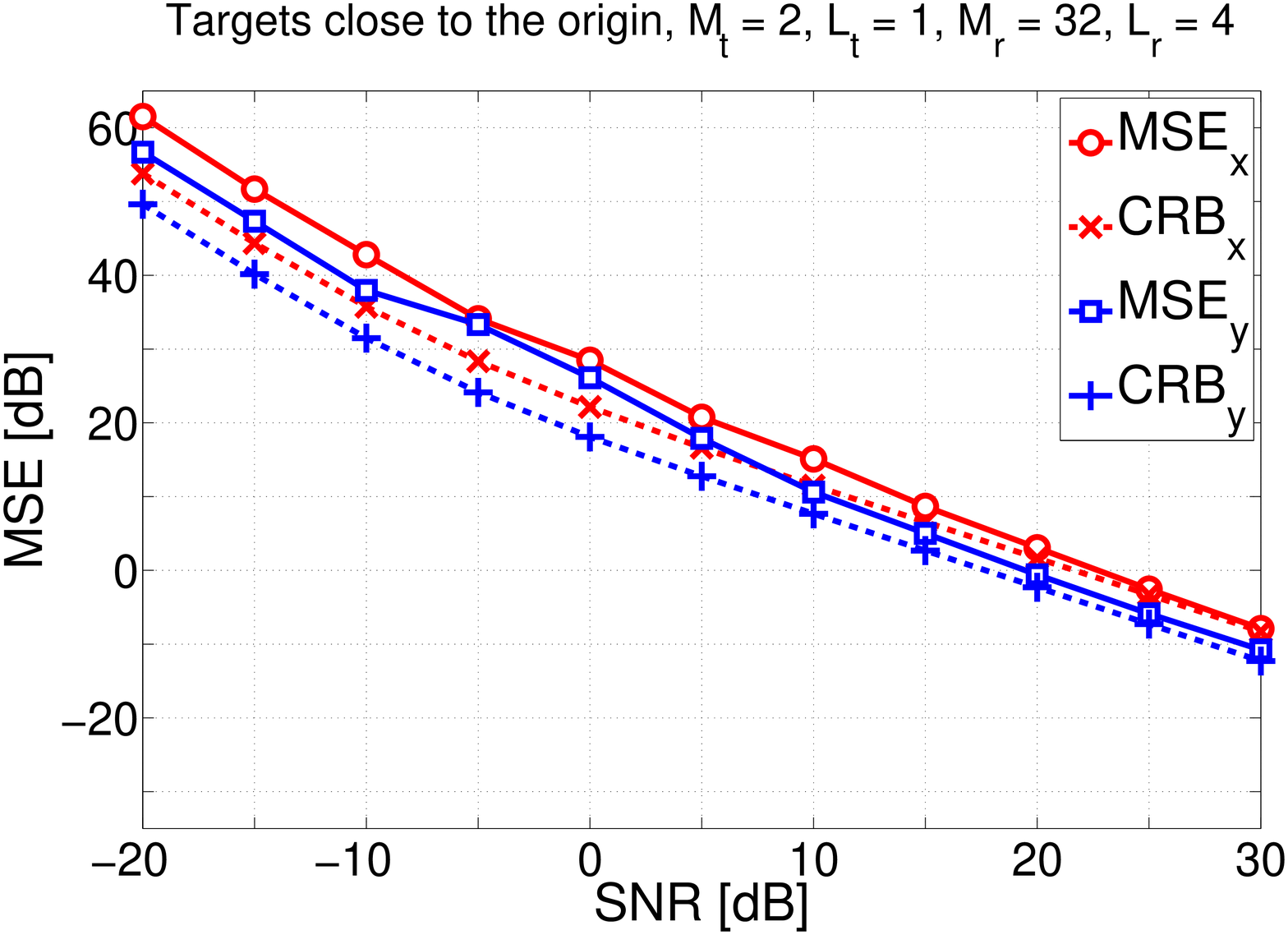}}
	\subfloat[]{\includegraphics[width=3in]{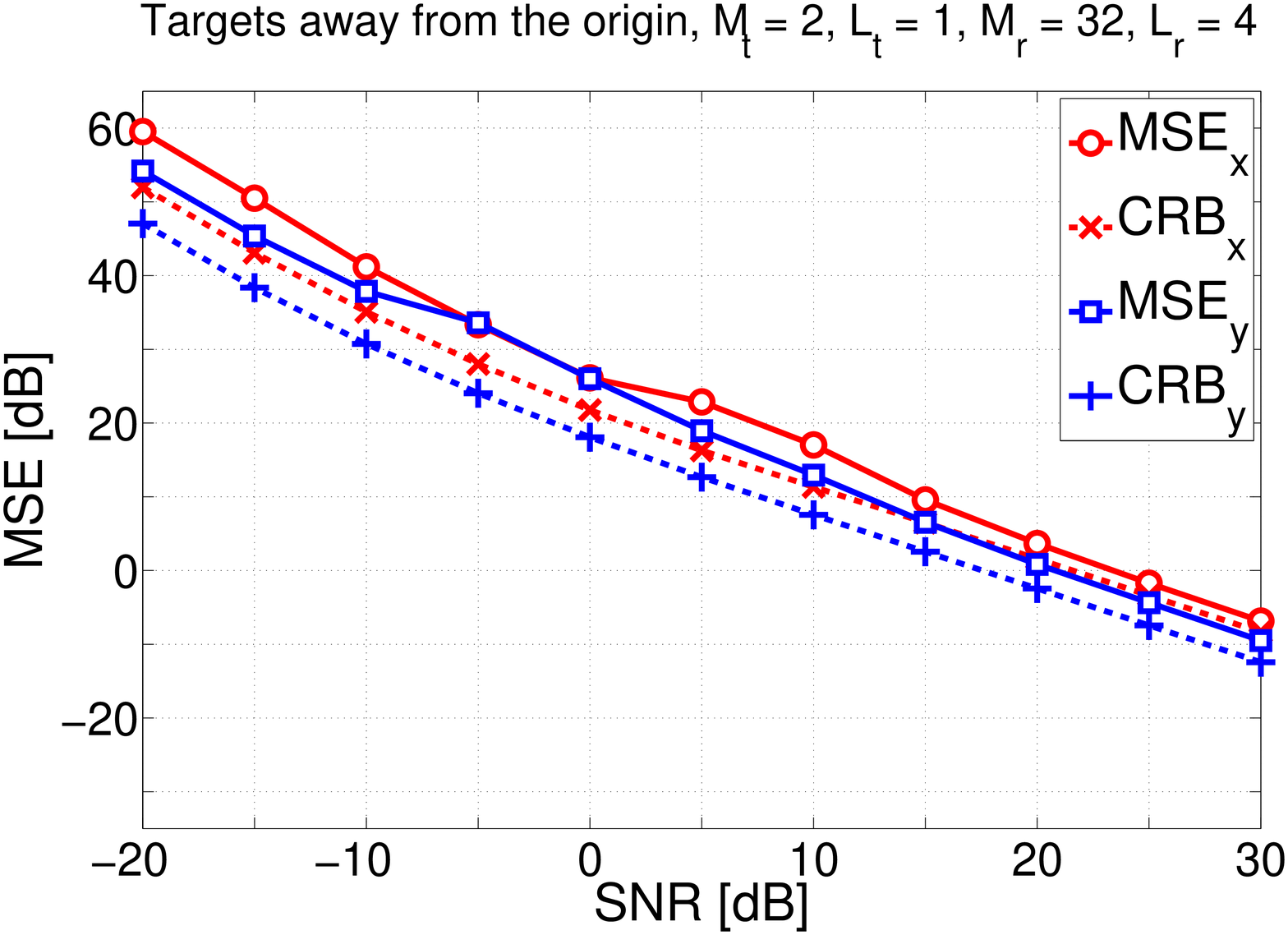}}\\
	
	\subfloat[]{\includegraphics[width=3in]{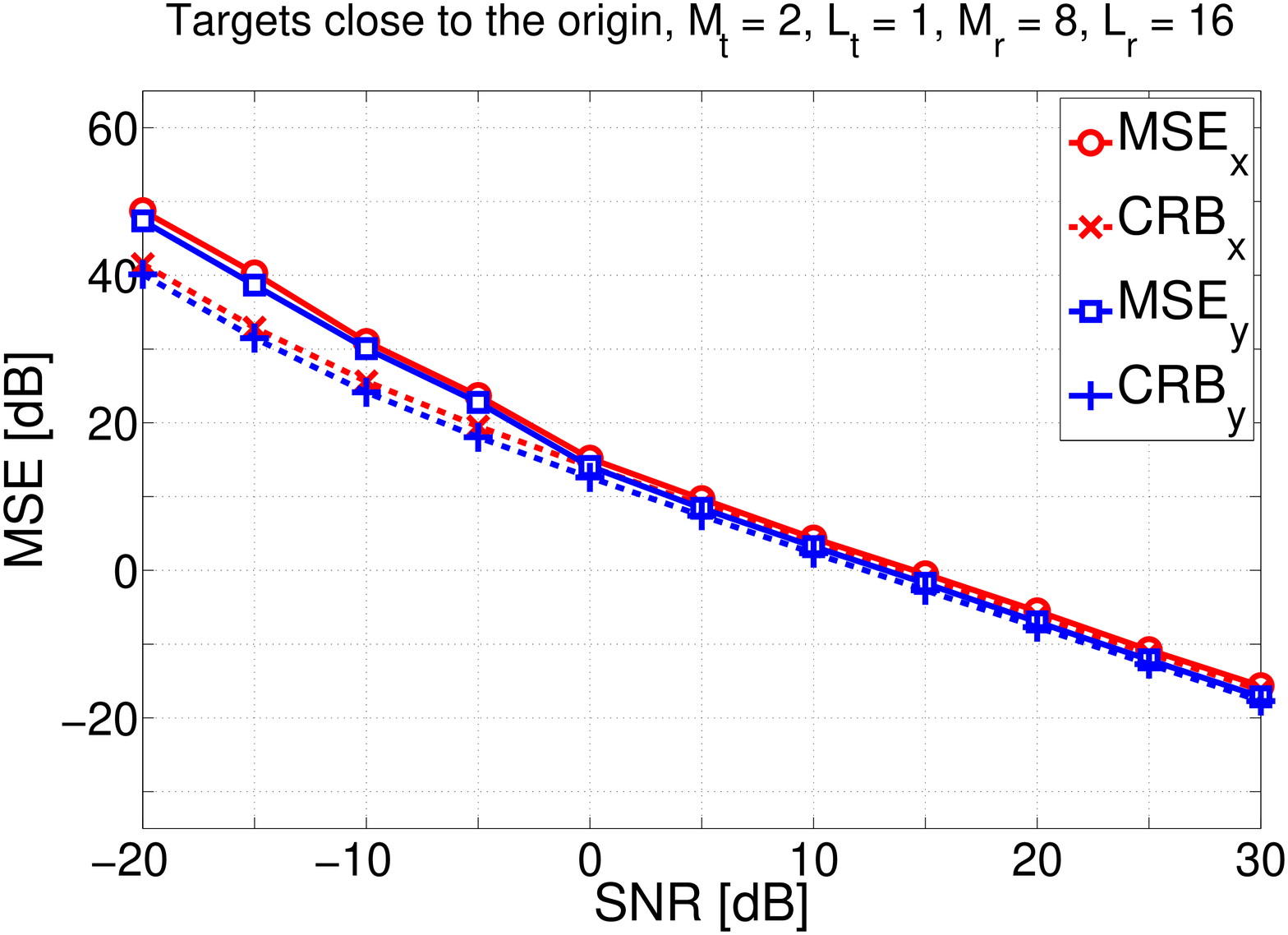}}
	\subfloat[]{\includegraphics[width=3in]{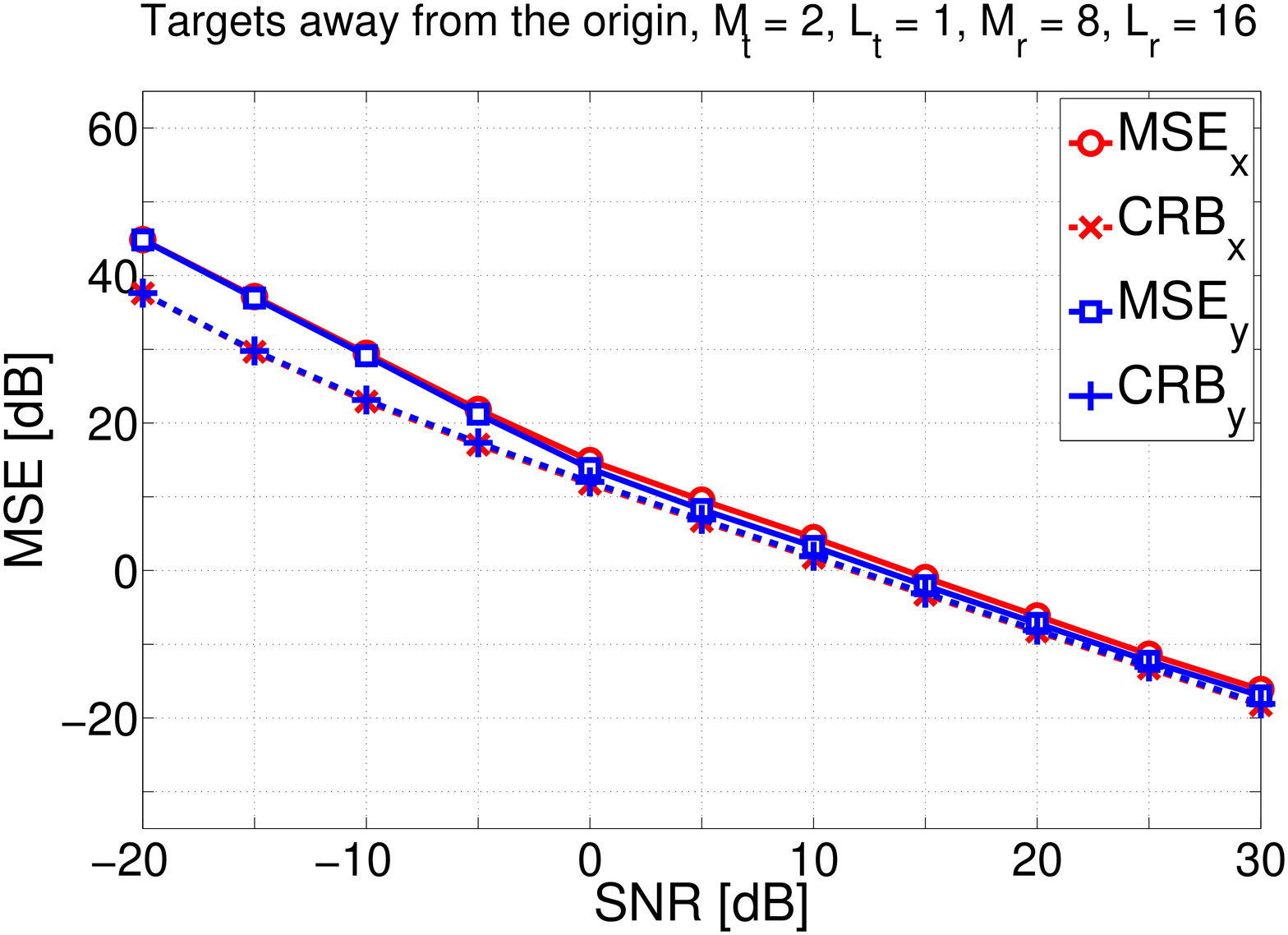}}\\
	
	\subfloat[]{\includegraphics[width=3in]{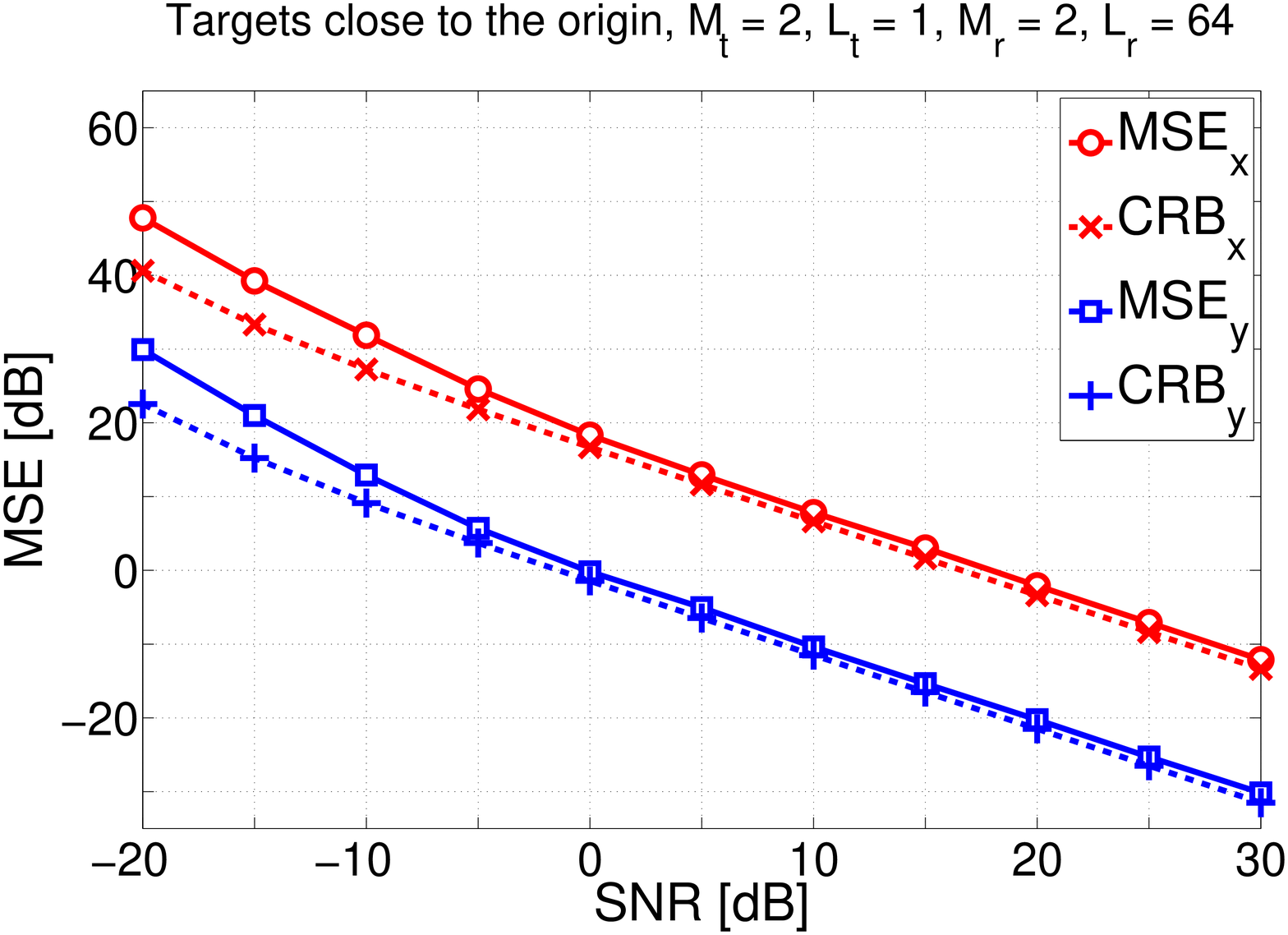}}
	\subfloat[]{\includegraphics[width=3in]{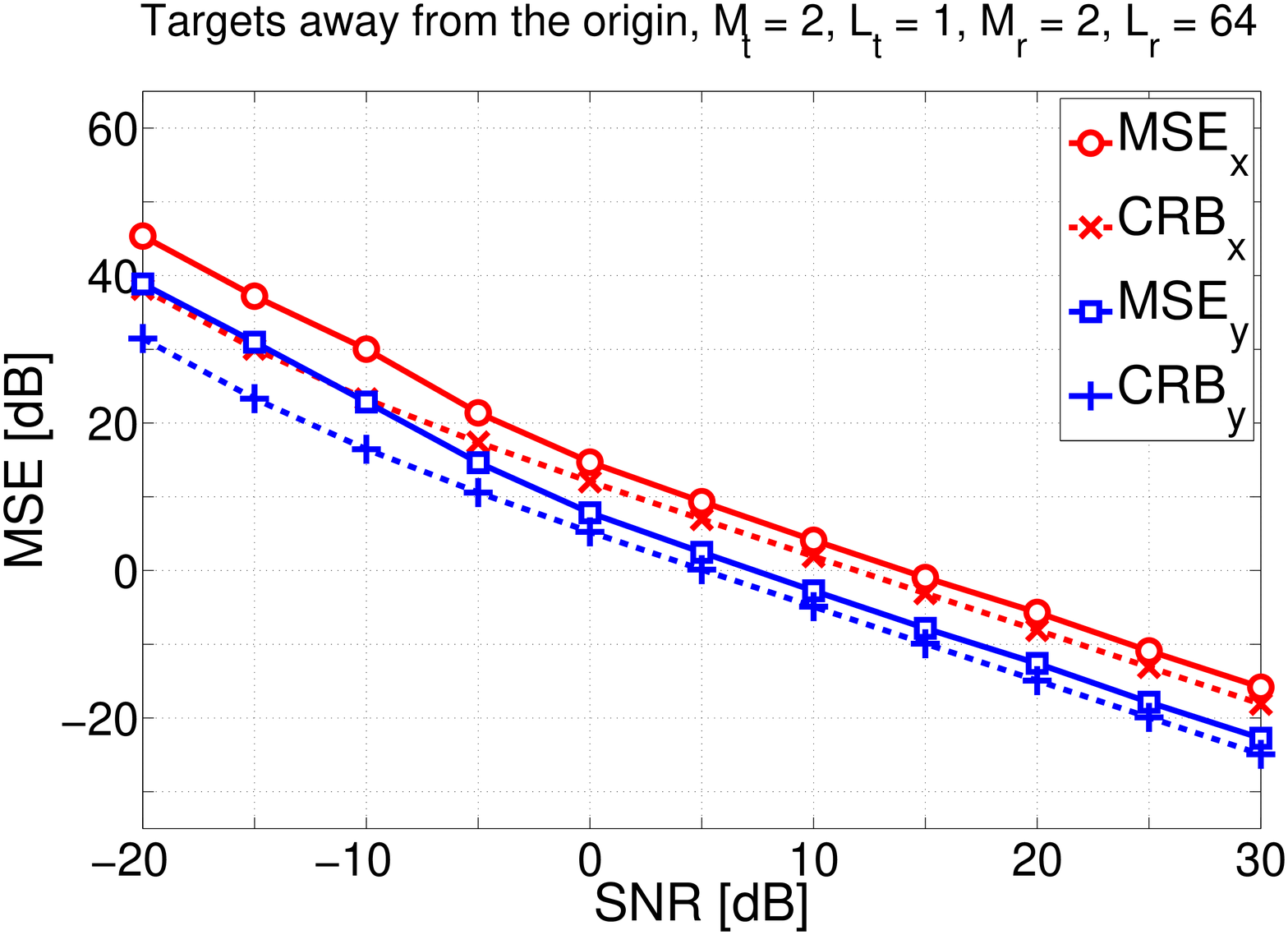}}
\end{tabular}
\caption{The CRLB and the MSE of the ML estimator for the stochastic signal model with two targets. The first target's location estimation results are shown for two closely spaced targets placed close to the origin in (a), (c), and (e), and away from the origin in (b), (d), and (f). Each configuration of the radar with widely separated arrays has $M_t = 2$ single antenna element transmitting arrays and the total number of receiving antenna elements is fixed at $M_rL_r = 128$. The number and the size of the receiving arrays vary: (a) and (b) $M_r = 32$, $L_r = 4$; (c) and (d) $M_r = 8$, $L_r = 16$; (e) and (f) $M_r = 2$, $L_r = 64$.}
\label{fig:sStoch}
\end{figure*}

\subsection{$M_rL_r = 128$ and $M_t = 2$}
The MSE results shown in Fig. \ref{fig:CRBsto} and Fig. \ref{fig:CRBdet} demonstrate a good convergence to the bounds because the number of transmit-to-receive array paths, $M_tM_r$, is always sufficiently large. This subsection considers the estimation performance of the radar configurations with $M_t = 2$ single element transmitting arrays and $M_rL_r=128$ total receiving antenna elements.

Fig. \ref{fig:sStoch} shows the MSE and the CRLB for the stochastic signal model under Assumption 2.1. The target parameter estimation performance for the radar configuration with $M_tM_r=64$ transmit-to-receive array paths is shown in subplots (a) and (b) for the targets located close to and away from the origin respectively. In (a) the CRLB provides a better prediction for the MSE than in (b), however the MSE is similar in both cases. Subplots (e) and (f) show the simulation results for the configuration with $M_tM_r=4$, and $L_r = 64$. The corresponding antenna placement is given in Fig. \ref{fig:scenario}b. Due to the chosen placement and the orientation of the receiving arrays, this radar configuration provides a better resolution along the $y$ axis, and a limited resolution along the $x$ axis, which is consistent with the MSE results in (e) and (f). However, because of a small number of transmit-to-receive array path, $M_tM_r=4$, the estimation performance of this radar configuration is more sensitive to the targets' location when compared to the estimation performance of the configuration with $M_tM_r=64$ in (a) and (b). This can be explained by a loss of the geometric diversity when $M_tM_r$ becomes small. In (c) and (d) the MSE calculated for the radar configuration with $M_t = 2$, $M_r = 8$ and $L_r = 16$ exhibits better convergence to the CRLB, and is more robust to the targets' location. 

%Thus when the number of transmit-to-receive array paths, and the total number of receiving antenna elements is small, the CRLB is most closely achieved when $M_tM_r$ and $L_r$ are balanced.

\begin{figure*}[!t]
\centering
\begin{tabular}{c}
	\subfloat[]{\includegraphics[width=3in]{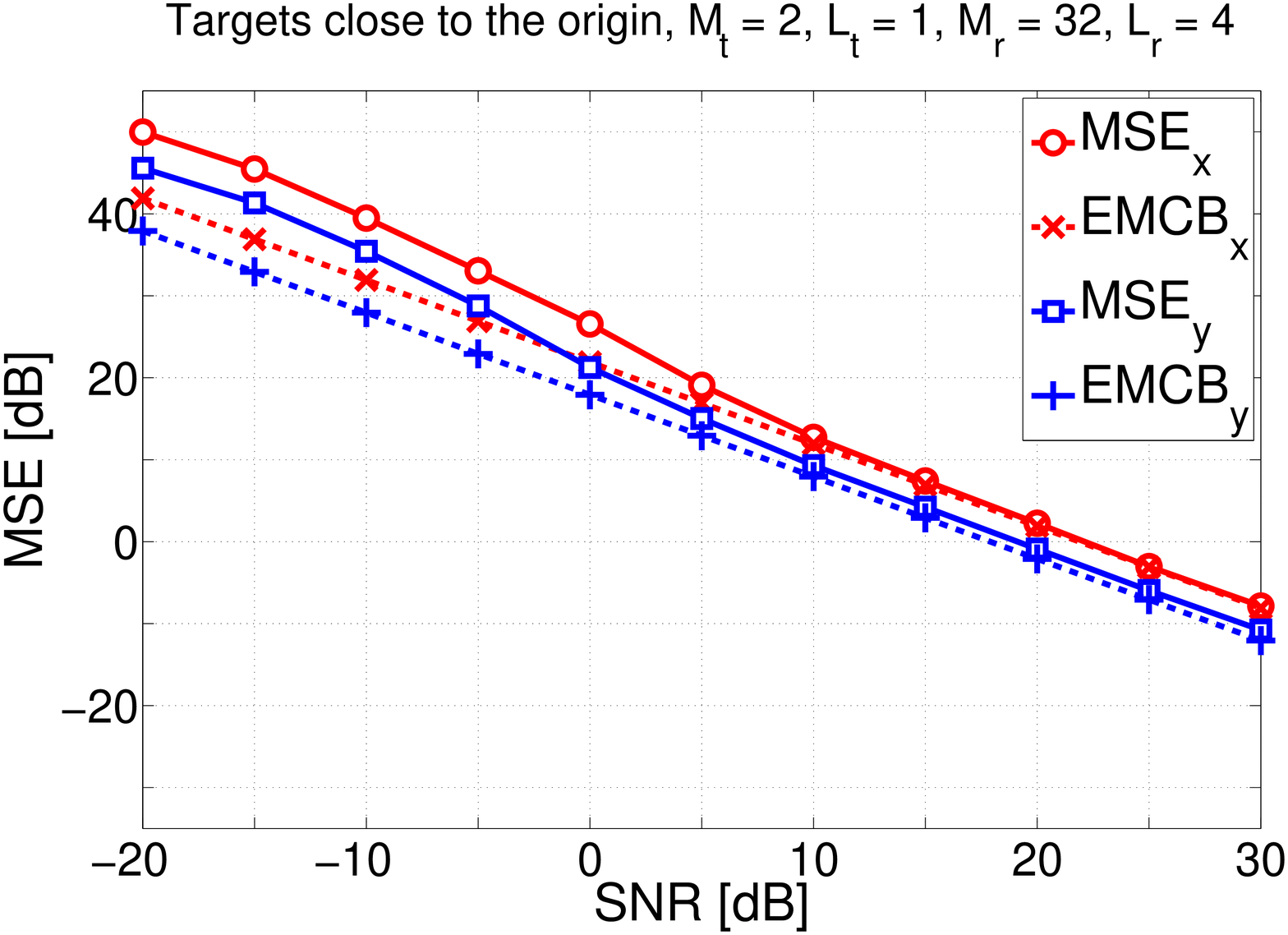}}
	\subfloat[]{\includegraphics[width=3in]{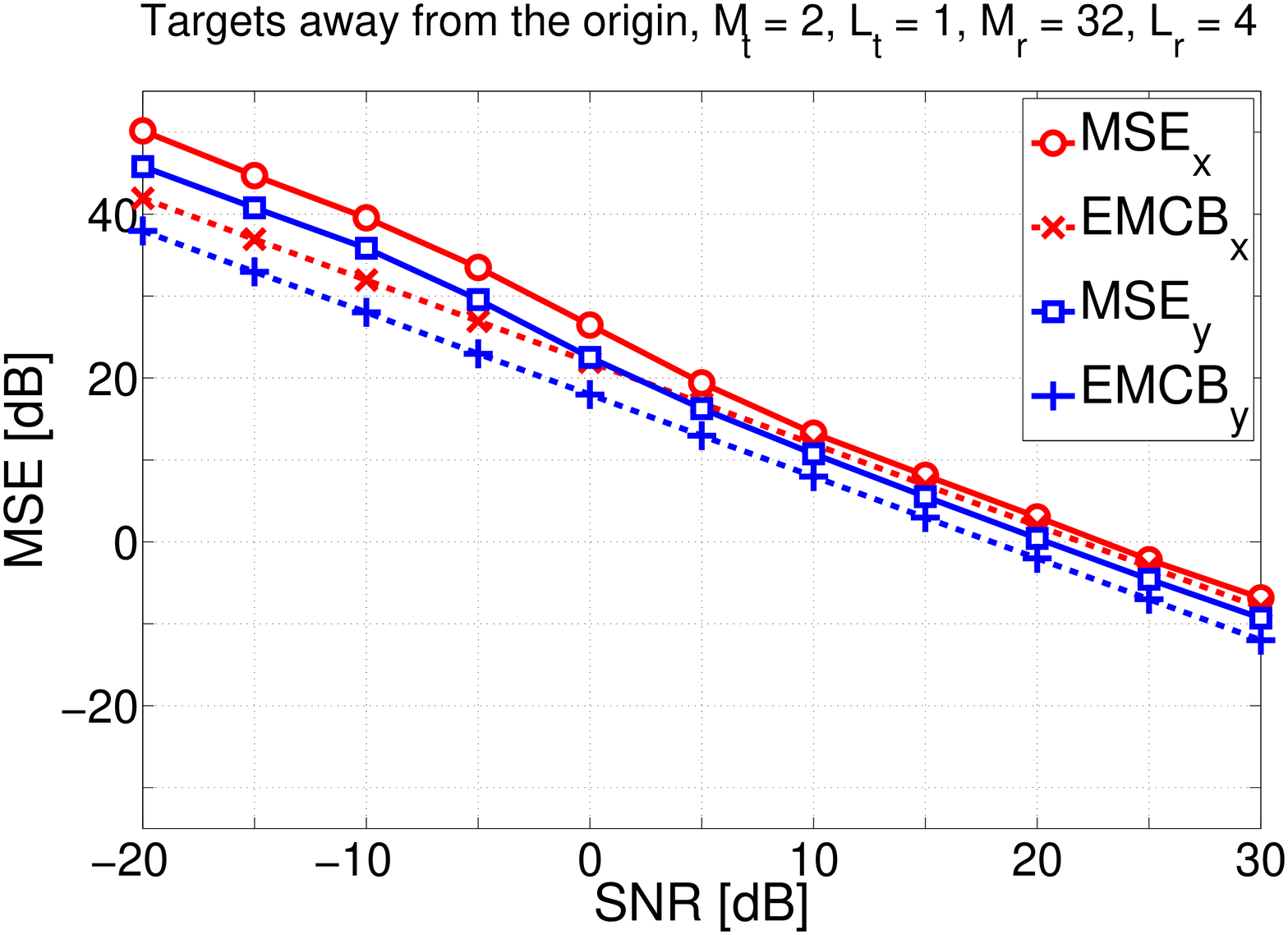}}\\
	
	\subfloat[]{\includegraphics[width=3in]{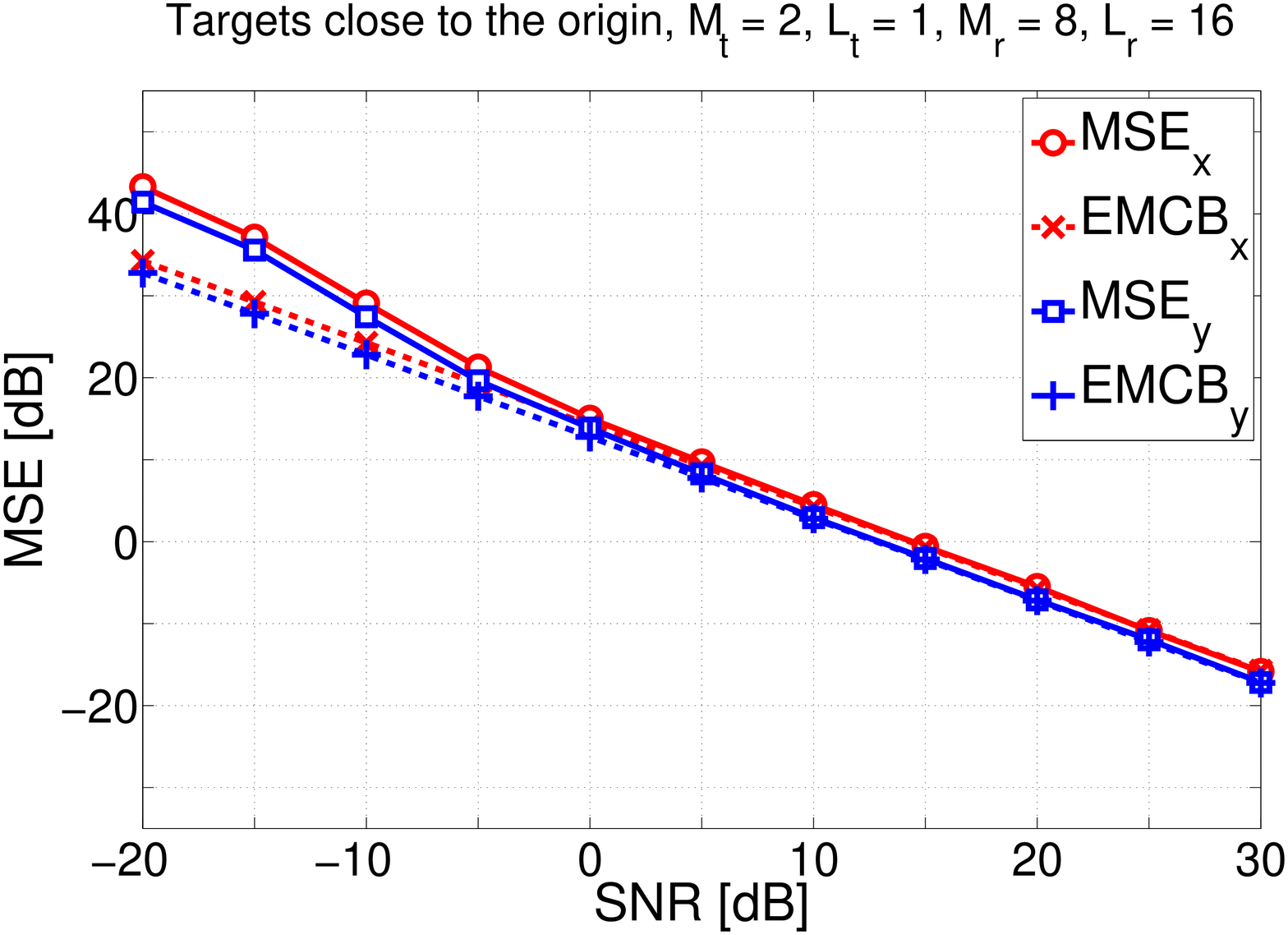}}
	\subfloat[]{\includegraphics[width=3in]{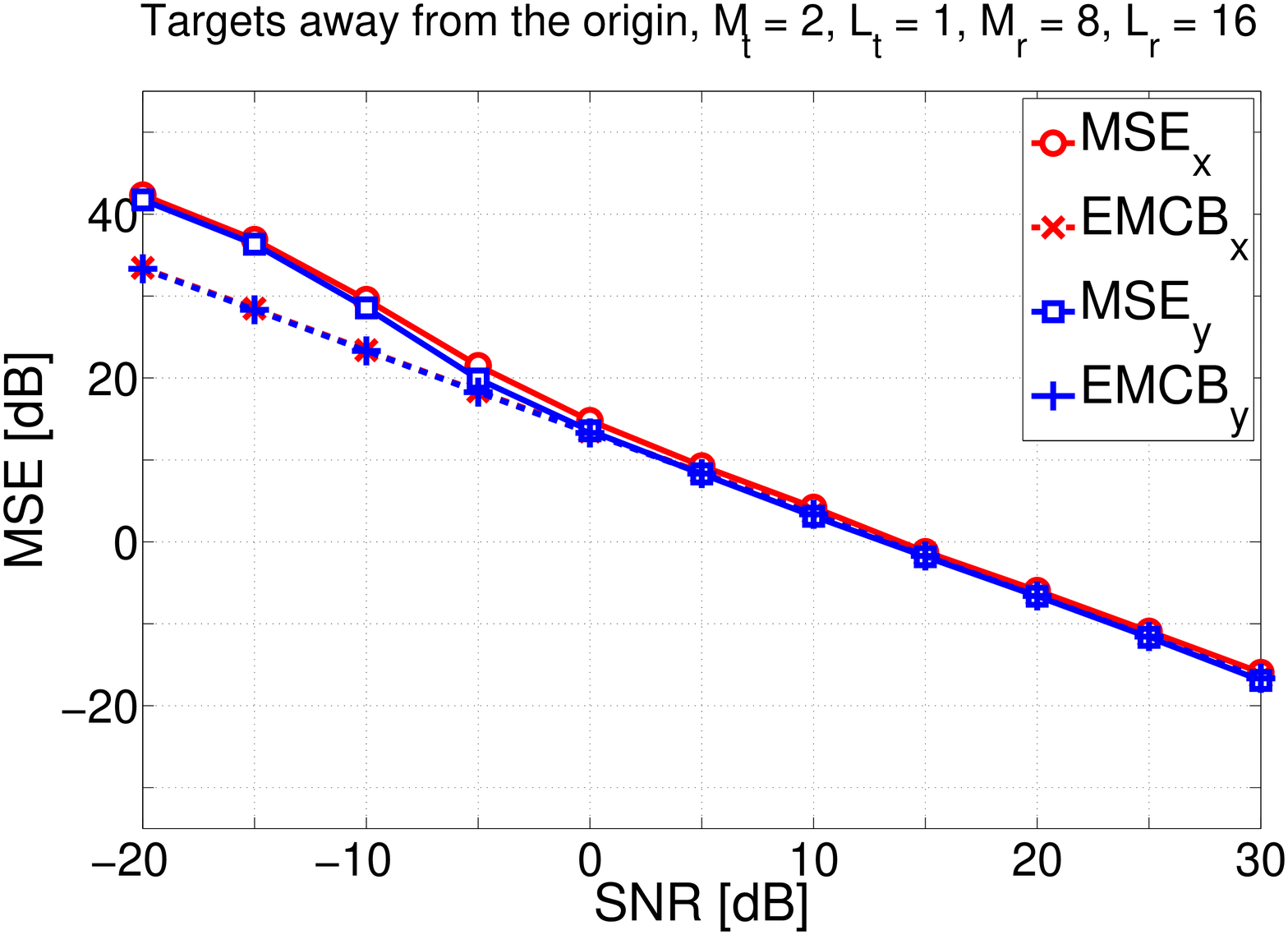}}\\
	
	\subfloat[]{\includegraphics[width=3in]{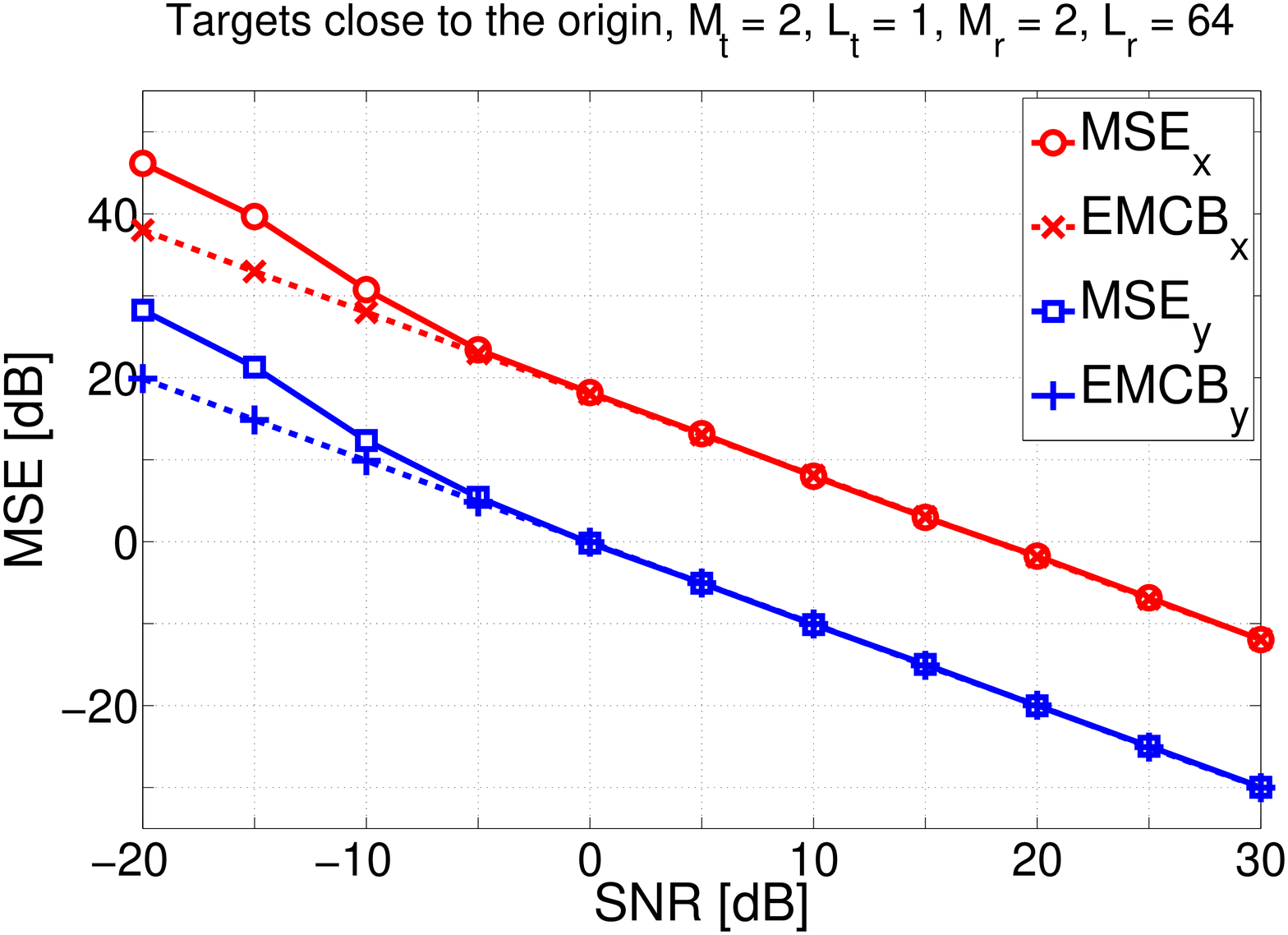}}
	\subfloat[]{\includegraphics[width=3in]{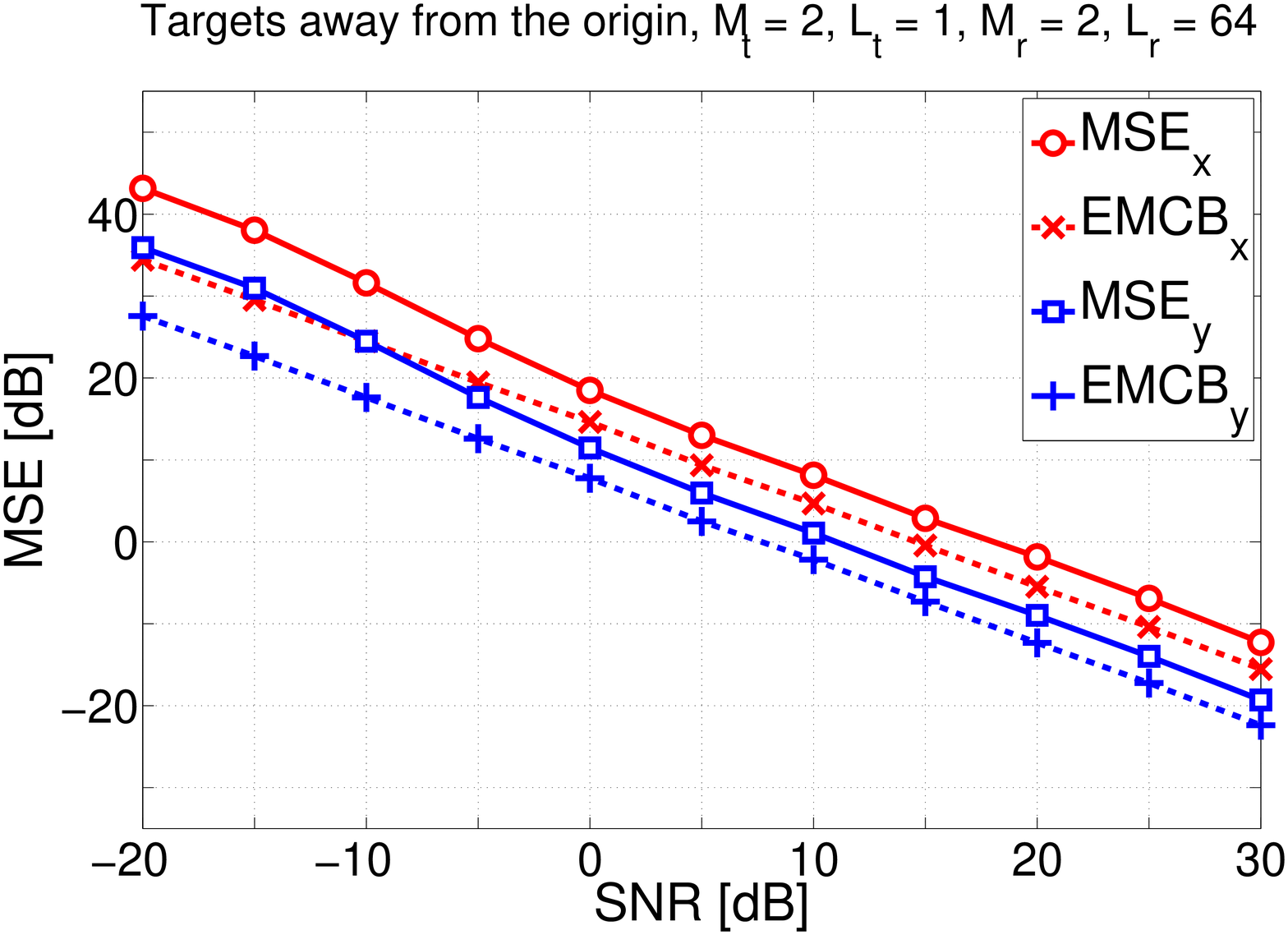}}
\end{tabular}
\caption{The EMCB and the MSE of the ML estimator for the deterministic signal model with two targets. The first target's location estimation results are shown for two closely spaced targets placed close to the origin in (a), (c), and (e), and away from the origin in (b), (d), and (f). Each configuration of the radar with widely separated arrays has $M_t = 2$ single antenna element transmitting arrays and the total number of receiving antenna elements is fixed at $M_rL_r = 128$. The number and the size of the receiving arrays vary: (a) and (b) $M_r = 32$, $L_r = 4$; (c) and (d) $M_r = 8$, $L_r = 16$; (e) and (f) $M_r = 2$, $L_r = 64$.}
\label{fig:sDet}
\end{figure*}

Similar conclusions can be made about the deterministic signal model and the radar configurations with $M_t=2$ and $M_rL_r=128$. The corresponding results of the Monte Carlo simulations are shown in Fig. \ref{fig:sDet}. In (a) and (b) the number of transmit-to-receive array path is $M_tM_r=64$, and the receiving arrays have $L_r=4$ antenna elements. The MSE is well predicted by the EMCB in the high SNR region, and the prediction is better for the targets located close to the origin compared to the targets located away from the origin. Since the number of unknown nuisance parameters (targets' reflectivities) decreases with $M_tM_r$ under the deterministic model assumption, the MSE in (c) and (e) demonstrates better convergence to the EMCB. The antenna placement used to obtain results in (e) and (f) with $M_t=2$, $M_r=2$, and $L_r=64$ is given in Fig. \ref{fig:scenario}b. Similar to the stochastic case, such antenna placement results in a better estimation performance along the $y$ axis than along the $x$ axis, however the estimation performance becomes significantly dependent on the location of the targets with respect to the receiving arrays. 

%Due to the lack of the geometric diversity, when $M_tM_r$ is small, the EMCB provides a poor prediction of the MSE for the targets located away from the origin, even in the asymptotic region. Among the three radar configurations compared in Fig. \ref{fig:sDet}, the configuration with $M_t=2$, $M_r=8$ and $L_r=16$ provides a balance between the MSE and the geometric diversity, such that the MSE does not change significantly with the targets location. Therefore, in terms of the MSE, it is generally more beneficial to have multiple distributed transmitting and receiving arrays such that the targets are observed from multiple angles providing geometric gain in all directions.

\begin{figure*}[!t]
\centering
\begin{tabular}{c}
	\subfloat[]{\includegraphics[width=3in]{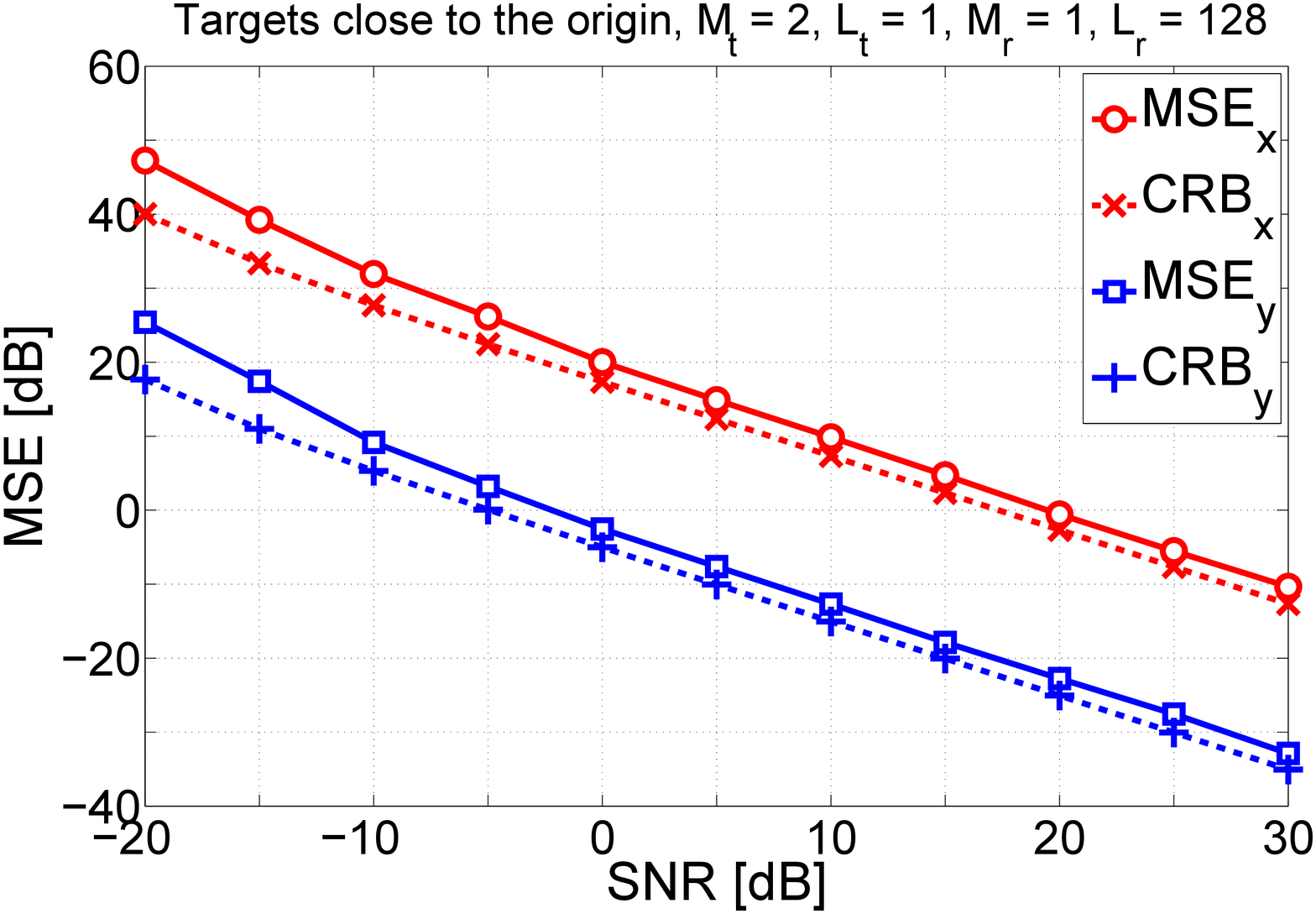}}
	\subfloat[]{\includegraphics[width=3in]{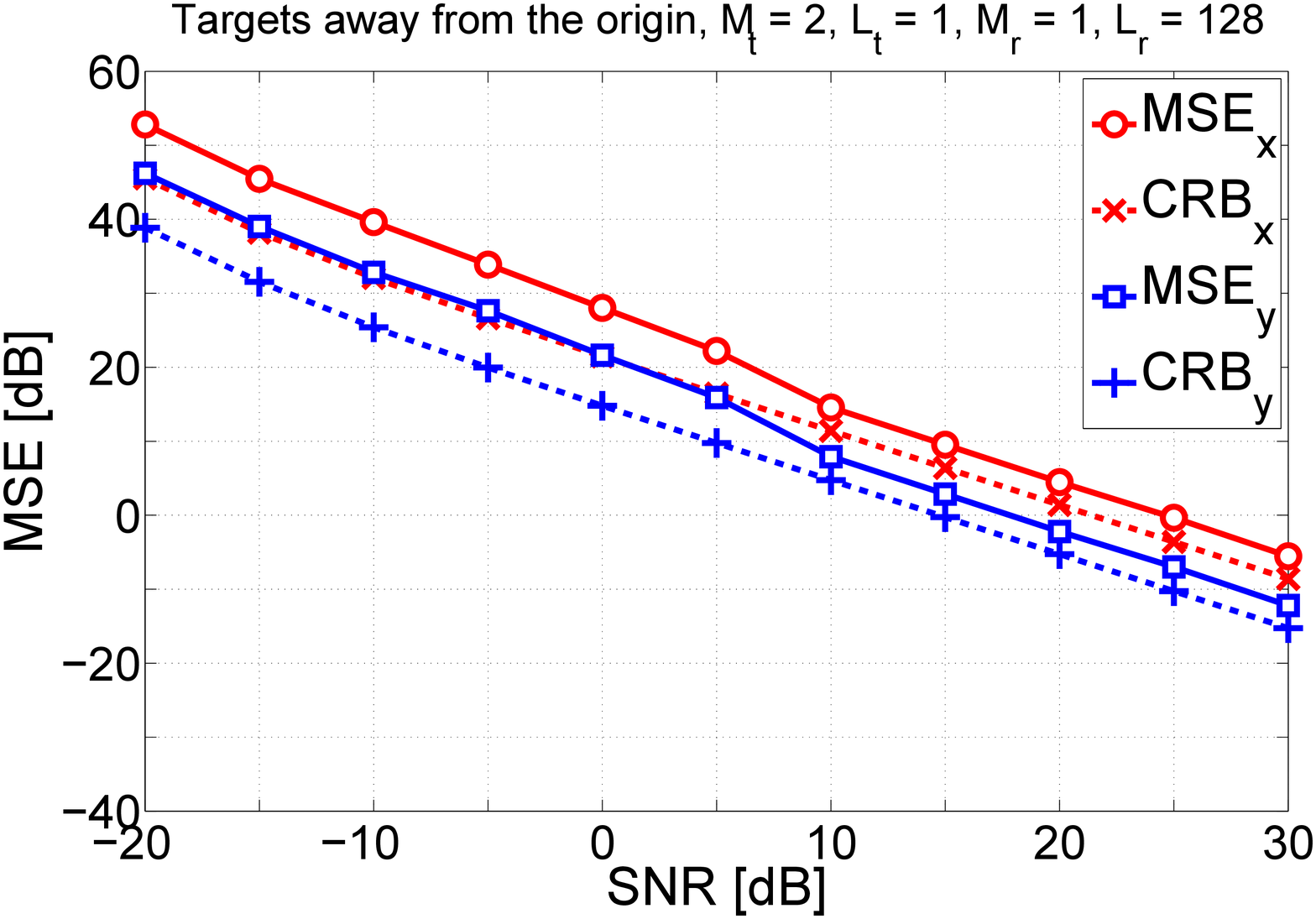}}\\
	
	\subfloat[]{\includegraphics[width=3in]{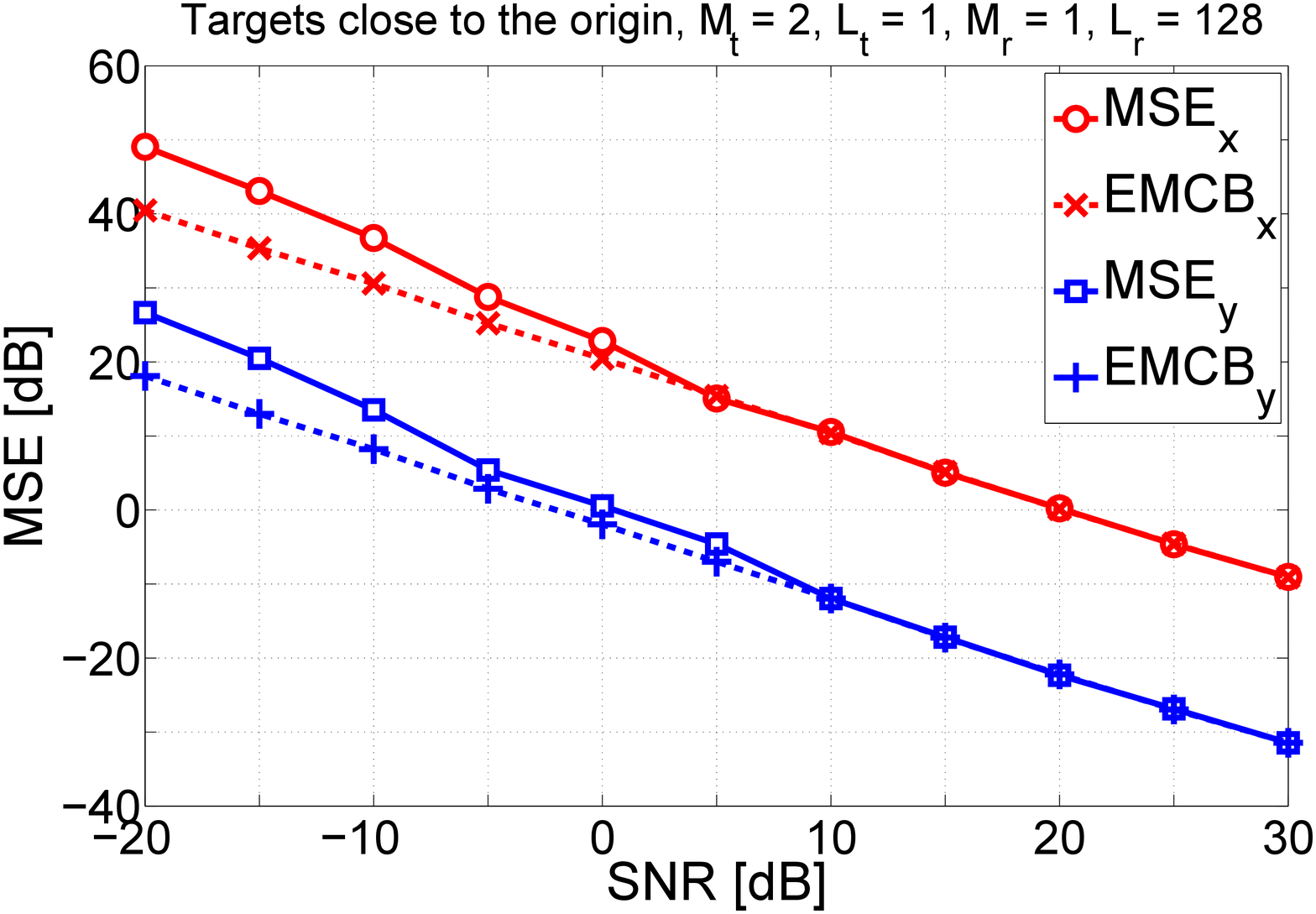}}
	\subfloat[]{\includegraphics[width=3in]{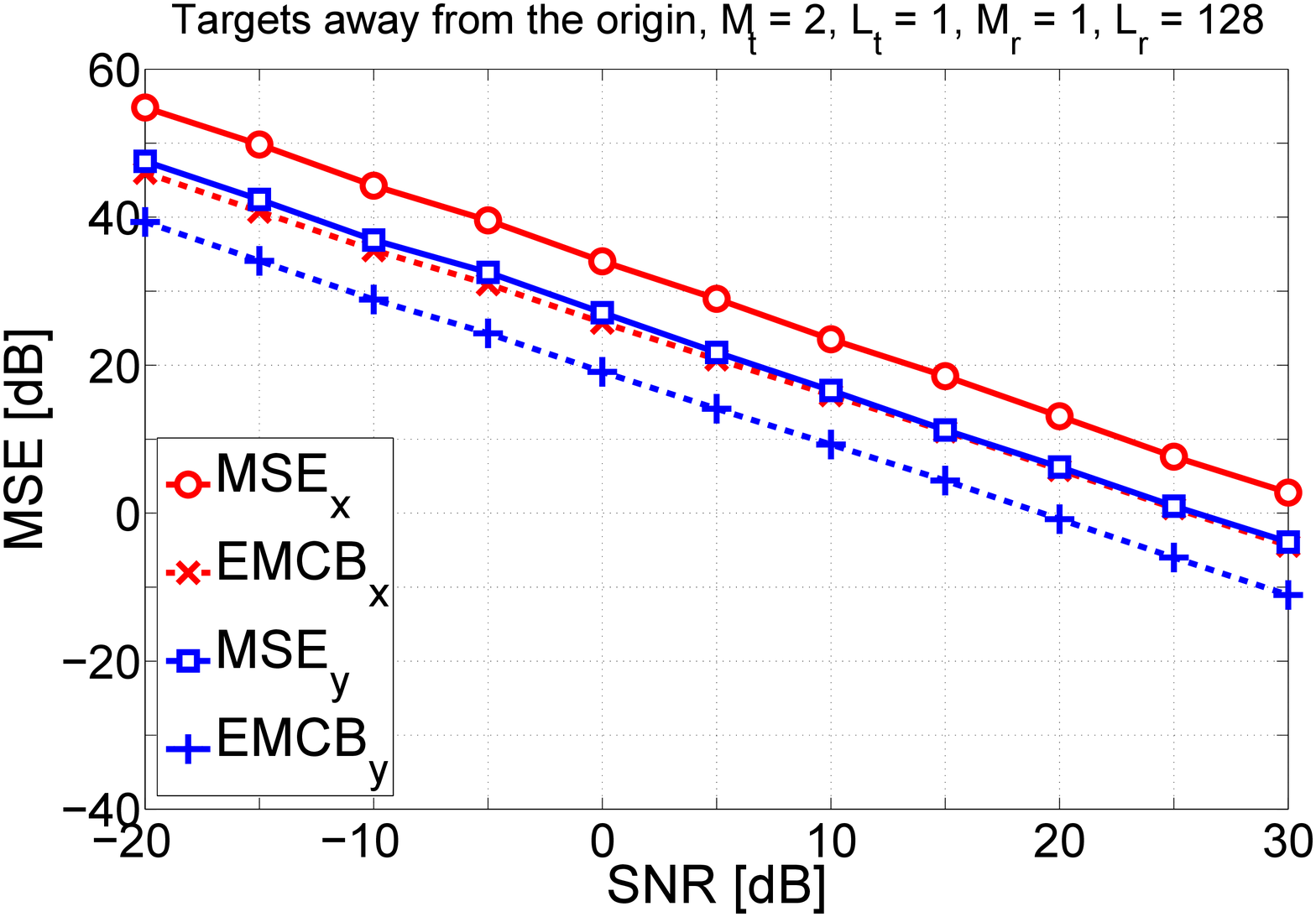}}\\
	
\end{tabular}
\caption{Results of the Monte Carlo simulations for a radar configuration with all receiving antenna elements grouped into one receiving array. The first target's location estimation results are shown for two closely spaced targets: (a) the targets are close to the origin and the stochastic model is assumed; (b) the targets are away from the origin and the stochastic model is assumed; (c) the targets are close to the origin and the deterministic model is assumed; (d) the targets are away from the origin and the deterministic model is assumed. Each configuration of the radar with widely separated arrays has $M_t = 2$ single antenna element transmitting arrays and $M_r = 1$ receiving array with $L_r = 128$ antenna elements.}
\label{fig:longarrays}
\end{figure*}

From the presented results, one might conclude that grouping all receiving antenna elements into one receiving array would provide the smallest MSE. In order to demonstrate that this assumption is generally false, we consider a radar with $M_t=2$ single element transmitting arrays and $M_r=1$ receiving array with $L_r=128$ antenna elements. The antenna placement chosen for the simulations is shown in Fig. \ref{fig:scenario}c. Since such a radar system has only $M_tM_r = 2$ transmit-to-receive array paths, the geometric gain due to observing targets from multiple angles is much smaller compared to the other configurations with $M_tM_r\gg 1$. We provide target location estimation Monte Carlo simulation results in Fig. \ref{fig:longarrays}. In (a) and (b) we show the MSE and the CRLB for the stochastic model, while in (c) and (d) we present the MSE and the EMCB curves for the deterministic model.

Comparing the MSE curves shown in Fig. \ref{fig:longarrays} with those in subplots (e) and (f) in Fig. \ref{fig:sStoch} and Fig. \ref{fig:sDet} indicates that the MSE performance of the ML estimator for the radar with one large receiving array is worse than the performance of the radar with two receiving arrays for both stochastic and deterministic signal models. The MSE is much higher in the single receiving array case, and the CRLB provides a poor prediction of the MSE. Since the receiving array is oriented parallel to the $y$-axis it provides good resolution along the $y$-axis, with very limited resolution along the $x$-axis. Thus, the MSE results for the $y$ coordinates of the targets in Fig. \ref{fig:longarrays} are much better than the corresponding results for the $x$ coordinates. This demonstrates a loss in the geometric gain due to having only a single receiving array. Since the targets have two unknown coordinates, two appropriately oriented arrays are required to obtain accurate estimation performance for both x and y coordinates. 

The simulation results shown in Fig. \ref{fig:longarrays} show that grouping all receiving antenna elements in one array does not generally provide better estimation performance, since the product $M_tM_r=2$ is too small. The MSE depends significantly on the positions of the targets with respect to the receiving array. Therefore, in terms of the MSE, it is generally more beneficial to have multiple distributed transmitting and receiving arrays such that the targets are observed from multiple angles providing geometric gain in all directions.

\section{Conclusions}
This paper studies the parameter estimation performance of a radar with widely separated antenna arrays in multiple target scenarios. The CRLB is derived under the stochastic and the deterministic signal model assumptions. The derived expressions are general and can be used to obtain bounds for different scenarios including a variety of radar system configurations.

The asymptotic properties of the ML estimator are studied under the stochastic and the deterministic signal model assumptions. The stochastic ML estimates are shown to become consistent and efficient if the number of transmit-to-receive array paths, $M_tM_r$, approaches infinity. However, keeping $M_tM_r$ finite and increasing the size of the receiving arrays, $L_r$, does not provide consistency. Under the deterministic signal model assumption the ML estimates are consistent but not efficient when $M_tM_r$ is infinitely large and $L_r$ is fixed, while finite $M_tM_r$, and infinitely large $L_r$ guarantee consistency and efficiency.

The asymptotic study conducted here provides useful insight into the parameter estimation performance of a radar system with distributed arrays. However, the question of the optimal allocation of the antennas into arrays when only a finite number of antenna elements is available cannot be answered by considering the asymptotic performance. This issue is addressed by conducting numerical simulations of scenarios with finite $M_tM_r$ and $L_r$, and a fixed number of available receiving antenna elements, $M_rL_r$.

The simulation results show that the MSE and the threshold SNR decrease as $L_r$ increases for both signal models if $M_tM_r$ is sufficiently large. The MSE for both stochastic and deterministic signal models is shown to be well predicted by the corresponding CRLB and the EMCB in the large SNR region. However, if $M_tM_r$ is too small, the estimation performance becomes strongly dependent on the targets' positions and for certain locations even an infinitely large $L_r$ cannot provide performance close to that with asymptotically large $M_tM_r$. This suggests that when the number of available receiving antenna elements is fixed, the configurations where neither $M_t M_r$ nor $L_r$ are too small provide better target parameter estimation performance. 

The CRLB and the asymptotic analysis of the ML estimator is the first step toward understanding the parameter estimation performance of a radar system with multiple widely separated arrays. This paper introduces a number of questions and shows that further analysis is required in some directions.

\appendices
\section{}
\label{AppendixA}
The derivation of the Cramer-Rao bound for the stochastic model presented in this appendix follows the derivation for a single phased array case developed in \cite{Stoica2}. Using (\ref{vec1}) and (\ref{vec2}), the expression for the $ij$th element of the FIM in (\ref{FIMstoch}) can be rewritten as
\be
\begin{split}
\label{AppendixA0}
\left[\Ical(\gammavec)\right]_{ij} &= \sum_{l=1}^{M_r}\sum_{k=1}^{M_t} \left(vec\left( \frac{d\Rmat_{kl}}{d\gammavec_i}\right)\right)^H\left(\Rmat_{kl}^{-T}\otimes\Rmat_{kl}^{-1}\right)\\ & \cdot vec\left(\frac{d\Rmat_{kl}}{d\gammavec_j}\right).
\end{split}
\ee
Applying (\ref{vec2}) also to (\ref{CovarMat}) we can define
\be
\label{VectorizedR}
\vvec_{kl} = vec\left(\Rmat_{kl}\right) = \left(\Hmat_{kl}^c\otimes \Hmat_{kl}\right)vec\left(\Amat\right)+\sigma^2 vec\left(\Imat_{L_rN}\right).
\ee
Using (\ref{VectorizedR}) in (\ref{AppendixA0}) the $PQ\times PQ$ FIM follows 
\be
\Ical(\gammavec) = \sum_{l=1}^{M_r}\sum_{k=1}^{M_t} \left(\frac{d\vvec_{kl}}{d\gammavec^T}\right)^H\left(\Rmat_{kl}^{-T}\otimes\Rmat_{kl}^{-1}\right)\left(\frac{d\vvec_{kl}}{d\gammavec^T}\right).
\ee
Further, the matrix $\Ical(\gammavec)$ can be partitioned
\be
\label{PartitionedFIM}
\Ical(\gammavec) = \sum_{l=1}^{M_r}\sum_{k=1}^{M_t} \begin{bmatrix}
\Gmat_{kl}^H \\ \Deltamat_{kl}^H
\end{bmatrix} 
\begin{bmatrix}
\Gmat_{kl} & \Deltamat_{kl}
\end{bmatrix} 
\ee
where
\bea
\label{Gmat}
\Gmat_{kl} & = & \left(\Rmat_{kl}^{-T/2}\otimes\Rmat_{kl}^{-1/2}\right)\left(\frac{d\vvec_{kl}}{d\psivec^T}\right)\\
\label{Deltamat}
\Deltamat_{kl} & = & \left(\Rmat_{kl}^{-T/2}\otimes\Rmat_{kl}^{-1/2}\right) \begin{bmatrix} \frac{d\vvec_{kl}}{d\rhovec^T} & \frac{d\vvec_{kl}}{d\sigma^2} \end{bmatrix}.\eea
The CRLB for the vector of the targets' parameters of interest can be found by applying a matrix factorization lemma \cite{Horn}  to (\ref{PartitionedFIM})
\be
\label{AppendixAFIM1}
\begin{split}
&\Cmat_s^{-1}(\psivec) = \sum_{l=1}^{M_r}\sum_{k=1}^{M_t} \Gmat_{kl}^H\Gmat_{kl} \\
&- \Gmat_{kl}^H\Deltamat_{kl}\left(\sum_{l'=1}^{M_r}\sum_{k'=1}^{M_t}\Deltamat_{k'l'}^H\Deltamat_{k'l'}\right)^{-1}\left(\sum_{l'=1}^{M_r}\sum_{k'=1}^{M_t} \Deltamat_{k'l'}^H\Gmat_{k'l'}\right).
\end{split}
\ee
Further we define derivatives of $\vvec_{kl}$ with respect to the elements of $\psivec$ and $\rhovec$, and $\sigma^2$. Let $(\psivec^q)_p$ be the $p$th, $p=1,2,\ldots,P$, parameter of interest of the target $q$, then from (\ref{VectorizedR})
\be
\begin{split}
\frac{d\vvec_{kl}}{d(\psivec^q)_p} &= vec\left(\frac{\Rmat_{kl}}{d(\psivec^q)_p}\right) \\ &= vec\left(\dvec_{kl}^{qp}\left(\cvec^q\right)^H\Hmat_{kl}^H + \Hmat_{kl}\cvec^q\left(\dvec_{kl}^{qp}\right)^H\right)
\end{split}
\ee
where $\cvec^q$ is a $q$th column of $\Amat$, and $\dvec_{kl}^{qp} = \frac{d}{d(\psivec^q)_p}\hvec_{kl}^q$. Thus the $m$th column of the matrix $\Gmat_{kl}$ in (\ref{Gmat}), where $m = P( q - 1 ) + p$, becomes
\bea
\left[\Gmat_{kl}\right]_m &=&  \left(\Rmat_{kl}^{-T/2}\otimes\Rmat_{kl}^{-1/2}\right)\frac{d\vvec_{kl}}{d(\psivec^q)_p}\\& = & vec\left(\Rmat_{kl}^{-1/2}\frac{\Rmat_{kl}}{d(\psivec^q)_p}\Rmat_{kl}^{-1/2}\right)  \nonumber\\
& = & vec\left(\Rmat_{kl}^{-1/2}\left(\Hmat_{kl}\cvec^q\left(\dvec_{kl}^{qp}\right)^H \right.\right.\nonumber\\
& + & \left.\left.\dvec_{kl}^{qp}\left(\cvec^q\right)^H\Hmat_{kl}^H\right)\Rmat_{kl}^{-1/2}\right).\nonumber
\eea
The derivative of (\ref{VectorizedR}) with respect to $\rhovec$ can be simplified after making the following observation: $vec\left(\Amat\right) = \Jmat\rhovec$, where $\Jmat$ is a $Q^2 \times Q^2$ constant block diagonal nonsingular matrix that maps elements of the vector $\rhovec$ into the elements of $vec\left(\Amat\right)$ \cite{Stoica2}. Thus
\be
\label{dvdrho}
\frac{d\vvec_{kl}}{d\rhovec} = \left(\Hmat_{kl}^c\otimes\Hmat_{kl}\right)\Jmat.
\ee 
Finally, the derivative of (\ref{VectorizedR}) with respect to the noise variance is
\be
\label{dvdsigma}
\frac{d\vvec_{kl}}{d \sigma^2} = vec\left(\Imat_{L_rN}\right).
\ee
Using (\ref{dvdrho}) and (\ref{dvdsigma}) in (\ref{Deltamat}), the matrix $\Deltamat_{kl}$ can be written as
\be
\begin{split}
\Delta_{kl} &= \begin{bmatrix}
\left(\Rmat_{kl}^{-T/2}\Hmat_{kl}^c\otimes\Rmat_{kl}^{-1/2}\Hmat_{kl}\right)\Jmat & vec\left(\Rmat_{kl}^{-1}\right)
\end{bmatrix} \\
&= \Fmat_{kl} \bar{\Jmat}
\end{split}
\ee
where
\bea
\Fmat_{kl} & = & \begin{bmatrix}\Rmat_{kl}^{-T/2}\Hmat_{kl}^c\otimes\Rmat_{kl}^{-1/2}\Hmat_{kl} & vec\left(\Rmat_{kl}^{-1}\right) \end{bmatrix}\\
\bar{\Jmat} & = & \begin{bmatrix}
\Jmat & \Zeromat_{Q^2\times 1} \\
\Zeromat_{1 \times Q^2} & 1 
\end{bmatrix}.
\eea
After substituting $\Deltamat_{kl}$ back into (\ref{AppendixAFIM1}) we obtain
\be
\label{AppendixAFIM2}
\begin{split}
\Cmat_s^{-1}(\psivec) & = \sum_{l=1}^{M_r}\sum_{k=1}^{M_t} \Gmat_{kl}^H\Gmat_{kl} \\
&- \Gmat_{kl}^H\Fmat_{kl}\left(\sum_{l'=1}^{M_r}\sum_{k'=1}^{M_t}\Fmat_{k'l'}^H\Fmat_{k'l'}\right)^{-1}\\ &\cdot \left(\sum_{l'=1}^{M_r}\sum_{k'=1}^{M_t} \Fmat_{k'l'}^H\Gmat_{k'l'}\right).
\end{split}
\ee
One can observe that the result in (\ref{AppendixAFIM2}) does not depend on the matrix $\Jmat$, and the explicit form of $\Jmat$ is not important. To our knowledge the expression in (\ref{AppendixAFIM2}) in general cannot be significantly simplified because it requires inversion of the sum of matrices.

\section{}
\label{AppendixB}
The general expression for the deterministic FIM is given in (\ref{FIMdet}). Under Assumption 2.2 the mean and the covariance matrix of the received signal for the $kl$th transmit-to-receive array path are $\muvec_{kl} = \Hmat_{kl}\alphavec_{kl}$ and $\Rmat_{kl} = \sigma^2\Imat_{L_rN}$ respectively. The derivatives of $\muvec_{kl}$ and $\Rmat_{kl}$ with respect to $\psivec$, $\alphavec$, and $\sigma^2$ follow
\be
\label{AppendixBdmudpsi}
\frac{d\muvec_{kl}}{d\psivec^T} = \begin{bmatrix}
\dvec_{kl}^{11}\alpha_{kl}^{1} & \ldots & \dvec_{kl}^{qp}\alpha_{kl}^{q} & \dots & \dvec_{kl}^{PQ}\alpha_{kl}^Q
\end{bmatrix}
\ee
\be
\frac{d\muvec_{kl}}{d \alphavec^T} = \begin{bmatrix} \Zeromat_{L_rN \times 2Q} & \ldots & \Hmat_{kl} & j\Hmat_{kl} & \ldots & \Zeromat_{L_rN \times 2Q} \end{bmatrix}
\ee
\bea
\frac{d\muvec_{kl}}{d \sigma^2} & = & \Zeromat_{L_rN \times 1} \\
\frac{d\Rmat_{kl}}{d \psivec^{qp}} & = & \frac{d\Rmat_{kl}}{d Re\cbl\alpha_{kl}^q\cbr} = \frac{d\Rmat_{kl}}{d Im\cbl\alpha_{kl}^q\cbr} = \Zeromat_{L_rN \times L_rN} \\
\label{AppendixdRmatdsigma}
\frac{d\Rmat_{kl}}{d \sigma^2} & = & \Imat_{L_rN}.
\eea
Notice (\ref{AppendixBdmudpsi}) can be written in a more compact form
\be
\label{AppendixBshort}
\frac{d\muvec_{kl}}{d\psivec^T} = \Dmat_{kl}\Pmat_{kl}
\ee
where
\bea
\Dmat_{kl} & = & \begin{bmatrix} \dvec_{kl}^{11} & \ldots & \dvec_{kl}^{PQ} \end{bmatrix}\\
\Pmat_{kl} & = & diag( \alphavec_{kl} \otimes \onevec_{P\times 1} )
\eea
and $\onevec_{P \times 1}$ is a $P\times 1$ all-ones column vector.
Using (\ref{AppendixBdmudpsi})-(\ref{AppendixdRmatdsigma}) in (\ref{FIMdet}), the FIM under the deterministic model assumption can be written as
\be
\label{AppendixBFIM1}
%\Ical(\gammavec) & = & \frac{2}{\sigma^2}\sum_{l=1}^{M_r}\sum_{k=1}^{M_t}\begin{bmatrix}
%\frac{d\muvec_{kl}^H}{d\psivec}\\
%\frac{d\muvec_{kl}^H}{d\alphavec}\\
%\Zeromat_{1\times L_rN}
%\end{bmatrix}
%\begin{bmatrix}
%\frac{d\muvec_{kl}}{d\psivec^T} & \frac{d\muvec_{kl}}{d\alphavec^T} & \Zeromat_{L_rN\times 1}
%\end{bmatrix} + \begin{bmatrix}
%\Zeromat_{4Q \times 4Q} & \Zeromat \\
%\Zeromat & \frac{L_rN}{2\sigma^2}
%\end{bmatrix} \\
\Ical(\gammavec) = \frac{2}{\sigma^2}\sum_{l=1}^{M_r}\sum_{k=1}^{M_t} Re
\begin{bmatrix}
\frac{d\muvec_{kl}^H}{d\psivec}\frac{d\muvec_{kl}}{d\psivec^T} & \frac{d\muvec_{kl}^H}{d\psivec}\frac{d\muvec_{kl}}{d\alphavec^T} & \Zeromat\\
\frac{d\muvec_{kl}^H}{d\alphavec}\frac{d\muvec_{kl}}{d\psivec^T} & \frac{d\muvec_{kl}^H}{d\alphavec}\frac{d\muvec_{kl}}{d\alphavec^T} & \Zeromat\\
\Zeromat & \Zeromat & \frac{L_rN}{2\sigma^2}
\end{bmatrix}.
\ee
Using a matrix factorization lemma the FIM for the vector of parameters of interest $\psivec$ becomes
\bea
\label{AppendixBFIM2}
\Ical(\psivec) & = & \frac{2}{\sigma^2} \left[\sum_{l=1}^{M_r}\sum_{k=1}^{M_t} Re\cbl \frac{d\muvec_{kl}^H}{d\psivec}\frac{d\muvec_{kl}}{d\psivec^T} \cbr \right.  \\
& - & \left(\sum_{l=1}^{M_r}\sum_{k=1}^{M_t}Re\cbl \frac{d\muvec_{kl}^H}{d\psivec}\frac{d\muvec_{kl}}{d\alphavec^T} \cbr\right) \nonumber\\ &\cdot& \left(\sum_{l=1}^{M_r}\sum_{k=1}^{M_t}Re\cbl \frac{d\muvec_{kl}^H}{d\alphavec}\frac{d\muvec_{kl}}{d\alphavec^T}\cbr \right)^{-1} \nonumber\\ &\cdot & \left.  \left(\sum_{l=1}^{M_r}\sum_{k=1}^{M_t} Re\cbl\frac{d\muvec_{kl}^H}{d\alphavec}\frac{d\muvec_{kl}}{d\psivec^T}\cbr\right)\right] \nonumber.
\eea
It can be verified that 
\be
\label{AppendixBReal1}
Re\cbl\sum_{l=1}^{M_r}\sum_{k=1}^{M_t}\frac{d\muvec_{kl}^H}{d\psivec}\frac{d\muvec_{kl}}{d\psivec^T}\cbr = \begin{bmatrix}
\Ymat_{11} & \cdots & \Zeromat \\
\vdots & \ddots & \vdots\\
\Zeromat & \cdots & \Ymat_{M_tM_r}
\end{bmatrix}
\ee
\bea
Re\cbl\sum_{l=1}^{M_r}\sum_{k=1}^{M_t}\frac{d\muvec_{kl}^H}{d\psivec}\frac{d\muvec_{kl}}{d\alphavec^T}\cbr & = & \begin{bmatrix}
\Tmat_{11} & \ldots & \Tmat_{M_tM_r}
\end{bmatrix}\\
\label{AppendixBReal2}
Re\cbl\sum_{l=1}^{M_r}\sum_{k=1}^{M_t}\frac{d\muvec_{kl}^H}{d\alphavec}\frac{d\muvec_{kl}}{d\psivec^T}\cbr & = & \begin{bmatrix}
\Umat_{11}^T & \ldots & \Umat_{M_tM_r}^T
\end{bmatrix}^T
\eea
where
\bea
\Ymat_{kl} & = & \begin{bmatrix}
Re\cbl\Hmat_{kl}^H\Hmat_{kl}\cbr & - Im\cbl\Hmat_{kl}^H\Hmat_{kl}\cbr\\
Im\cbl\Hmat_{kl}^H\Hmat_{kl}\cbr &  Re\cbl\Hmat_{kl}^H\Hmat_{kl}\cbr\\
\end{bmatrix}\\
\Tmat_{kl} & = & \begin{bmatrix}
Re\cbl\Pmat_{kl}^H\Dmat_{kl}^H\Hmat_{kl}\cbr &  -Im\cbl\Pmat_{kl}^H\Dmat_{kl}^H\Hmat_{kl}\cbr
\end{bmatrix}\\
\Umat_{kl} & = & \begin{bmatrix}
Re\cbl\Hmat_{kl}^H\Dmat_{kl}\Pmat_{kl}\cbr \\  Im\cbl\Hmat_{kl}^H\Dmat_{kl}\Pmat_{kl}\cbr
\end{bmatrix}.
\eea
Substituting (\ref{AppendixBReal1})-(\ref{AppendixBReal2}) and (\ref{AppendixBshort}) into (\ref{AppendixBFIM2}) we obtain
\be
\Ical(\psivec) = \frac{2}{\sigma^2}\sum_{l=1}^{M_r}\sum_{k=1}^{M_t}Re\cbl\Pmat_{kl}^H\Dmat_{kl}^H\Dmat_{kl}\Pmat_{kl}\cbr-\Tmat_{kl}\Ymat_{kl}^{-1}\Umat_{kl}.
\ee
Using identities (\ref{CompMatIdnt1}) and (\ref{CompMatIdnt2}) the FIM can be written as
\bea
\label{AppendixBFIM3}
\Ical(\psivec) & = & \frac{2}{\sigma^2}\sum_{l=1}^{M_r}\sum_{k=1}^{M_t}Re\cbl\Pmat_{kl}^H\Dmat_{kl}^H\Dmat_{kl}\Pmat_{kl}\right. \\
 &-& \left. \Pmat_{kl}^H\Dmat_{kl}^H\Hmat_{kl}\left(\Hmat_{kl}^H\Hmat_{kl}\right)^{-1} \Hmat_{kl}^H\Dmat_{kl}\Pmat_{kl}\cbr \nonumber \\
& = & \frac{2}{\sigma^2}\sum_{l=1}^{M_r}\sum_{k=1}^{M_t}Re\cbl\Pmat_{kl}^H\Dmat_{kl}^H\Pi_{\Hmat_{kl}}^{\perp}\Dmat_{kl}\Pmat_{kl}\cbr \nonumber\\
& = & \frac{2}{\sigma^2}\sum_{l=1}^{M_r}\sum_{k=1}^{M_t}Re\cbl\Dmat_{kl}^H\Pi_{\Hmat_{kl}}^{\perp}\Dmat_{kl} \right. \nonumber\\ &\odot& \left.\left(\alphavec_{kl}\alphavec_{kl}^H \otimes \onevec_{P \times P}\right)^T\cbr \nonumber
\eea
where 
\be
\Pi_{\Hmat_{kl}}^{\perp} = \Imat_{L_rN} - \Hmat_{kl}\left(\Hmat_{kl}^H\Hmat_{kl}\right)^{-1}\Hmat_{kl}^H
\ee
and $\onevec_{P \times P}$ is a $P \times P$ all-ones matrix. An inverse of (\ref{AppendixBFIM3}) results in the CRLB expression in (\ref{FIMdetFinal}).

\section{}
\label{AppendixCA}
This appendix shows that the $kl$th summand in (\ref{StochLogLikeEst}) has a finite variance. Consider a $kl$th summand in (\ref{StochLogLikeEst})
\be
\label{AppCklStoch}
\LL_{s_{kl}} = \frac{1}{M_tM_rL_r}\left( \log\left|\hat{\Rmat}_{kl}\right| + Tr \cbl\hat{\Rmat}_{kl}^{-1} \rvec_{kl} \rvec_{kl}^H\cbr\right).
\ee
Only the trace term in (\ref{AppCklStoch}) is random, thus the variance of $\LL_{s_{kl}}$ is 
\be
\begin{split}
var\left[\LL_{s_{kl}}\right] =& var\left[\frac{1}{M_tM_rL_r}Tr \cbl\hat{\Rmat}_{kl}^{-1} \rvec_{kl} \rvec_{kl}^H\cbr\right]\\
 =& var\left[\frac{1}{M_tM_rL_r}\rvec_{kl}^H \hat{\Rmat}_{kl}^{-1} \rvec_{kl}\right] 
\end{split}
\ee
where to obtain the last identity the cyclic property of the trace operator was used. Using the expression for the variance of the quadratic form in (\ref{QuadFromVar})
\be
\label{AppCStochVar1}
var\left[\LL_{s_{kl}}\right] = \frac{2}{(M_tM_rL_r)^2}Tr\cbl\hat{\Rmat}_{kl}^{-1} \Rmat_{kl}\hat{\Rmat}_{kl}^{-1} \Rmat_{kl}\cbr.
\ee
Further in order to obtain an upper bound on the variance, the matrix trace inequality (\ref{TraceMatrixInequality}) can be applied to the trace in (\ref{AppCStochVar1}) yielding
\be
\label{AppCStochVar2}
var\left[\LL_{s_{kl}}\right] \leq \frac{2}{(M_tM_r)^2} \left( Tr\cbl\hat{\Rmat}_{kl}^{-1} \Rmat_{kl}\cbr\right)^2.
\ee
Let $\hat{\nu}_{kl_{max}}$ and $\nu_{kl_{max}}$ be the largest eigenvalues of the matrices $\hat{\Rmat}_{kl}^{-1}$ and $\Rmat_{kl}$ respectively. Applying the Von Neumanns inequality from (\ref{VonNeumannInq}) to the trace in (\ref{AppCStochVar2}), another bound on the variance of $\LL_{s_{kl}}$ can be obtained
\be
var\left[\LL_{s_{kl}}\right] \leq \frac{2}{(M_tM_rL_r)^2}  \left(N\hat{\nu}_{kl_{max}} \nu_{kl_{max}} \right)^2.
\ee
Since the matrices $\hat{\Rmat}_{kl}$ and $\Rmat_{kl}$ are positive definite the eigenvalues $\hat{\nu}_{kl_{max}}$ and $\nu_{kl_{max}}$ have to be finite, which leads to the variance of $\LL_{s_{kl}}$ being finite.

\section{}
\label{AppStochLr}
This appendix verifies (\ref{LargeLrStochLL}). Using (\ref{Model3}) in (\ref{StochLogLikeEst})
\be
\begin{split}
\LL_s&\left(\hat{\psivec}, \hat{\Amat}, \hat{\sigma}^2\right) = \frac{1}{M_rM_tL_r}\sum_{l=1}^{M_r}\sum_{k=1}^{M_t}\left[\log\left|\hat{\Rmat}_{kl}\right| \right.\\ 
+& Tr\cbl\hat{\Rmat}_{kl}^{-1}\evec_{kl}\evec_{kl}^H\cbr 
+ 2Re\cbl Tr\cbl\hat{\Rmat}_{kl}^{-1}\Hmat_{kl}\alphavec_{kl}\evec_{kl}^H\cbr\cbr \\
+ &\left. Tr\cbl\Hmat_{kl}^H\hat{\Rmat}_{kl}^{-1}\Hmat_{kl}\alphavec_{kl}\alphavec_{kl}^H\cbr\right]
\label{StochLogLikeEstLr1}
\end{split}
\ee
By the definition of a trace operator the second, the third, and the fourth terms of the function in (\ref{StochLogLikeEstLr1}) can be written as summations
\be
\label{StochLogLikeEstLr2}
\begin{split}
\LL_s&\left(\hat{\psivec}, \hat{\Amat}, \hat{\sigma}^2\right) = \frac{1}{M_rM_tL_r}\sum_{l=1}^{M_r}\sum_{k=1}^{M_t}\left[\log\left|\hat{\Rmat}_{kl}\right| \right.\\
& + \sum_{i=1}^{L_rN}\sum_{j=1}^{L_rN}\left[\hat{\Rmat}_{kl}^{-1}\right]_{ij}(\evec_{kl}^H)_{i}(\evec_{kl})_j \\
& + 2\sum_{i=1}^{L_rN}\sum_{j=1}^Q Re\cbl\left[\hat{\Rmat}_{kl}^{-1}\Hmat_{kl}\right]_{ij}(\evec_{kl}^H)_i(\alphavec_{kl})_j\cbr \\
& + \left.\sum_{i=1}^Q\sum_{j=1}^Q \left[\Hmat_{kl}^H\hat{\Rmat}_{kl}^{-1}\Hmat_{kl}\right]_{ij}(\alphavec_{kl}^H)_i(\alphavec_{kl})_j\right]
\end{split}
\ee
The first term in (\ref{StochLogLikeEstLr2}) is deterministic, while the second, the third and the  fourth terms depend on the random targets' reflectivities and the noise. Letting $L_r \rightarrow \infty$ Kolmogorov's strong law of large numbers \cite{Papoulis} can be applied to the second and the third terms in (\ref{StochLogLikeEstLr2}) yielding
\be
\label{term2KSL}
\begin{split}
\sum_{i=1}^{L_rN}&\sum_{j=1}^{L_rN}\left[\hat{\Rmat}_{kl}^{-1}\right]_{ij}(\evec_{kl}^H)_{i}(\evec_{kl})_j\\ 
\xrightarrow{a.s.}&E\left[\sum_{i=1}^{L_rN}\sum_{j=1}^{L_rN} \left[\hat{\Rmat}_{kl}^{-1}\right]_{ij}(\evec_{kl}^H)_{i}(\evec_{kl})_j \right] \\
= & \sum_{i=1}^{L_rN}\sum_{j=1}^{L_rN} \left[\hat{\Rmat}_{kl}^{-1}\right]_{ij} E\left[(\evec_{kl}^H)_{i}(\evec_{kl})_j \right]\\
= & \sum_{i=1}^{L_rN} \left[\hat{\Rmat}_{kl}^{-1}\right]_{ii} \sigma^2 = Tr\cbl\hat{\Rmat}_{kl}^{-1}\cbr\sigma^2 
\end{split}
\ee
and
\be
\label{term3KSL}
\begin{split}
\sum_{i=1}^{L_rN}&\sum_{j=1}^Q Re\cbl\left[\hat{\Rmat}_{kl}^{-1}\Hmat_{kl}\right]_{ij}(\evec_{kl}^H)_i(\alphavec_{kl})_j\cbr \\
\xrightarrow{a.s.}& E\left[\sum_{i=1}^{L_rN}\sum_{j=1}^Q Re\cbl\left[\hat{\Rmat}_{kl}^{-1}\Hmat_{kl}\right]_{ij}(\evec_{kl}^H)_i(\alphavec_{kl})_j\cbr\right] \\
=& \sum_{i=1}^{L_rN}\sum_{j=1}^Q Re\cbl\left[\hat{\Rmat}_{kl}^{-1}\Hmat_{kl}\right]_{ij} E\left[(\evec_{kl}^H)_i(\alphavec_{kl})_j\right]\cbr = 0
\end{split}
\ee
where (\ref{term3KSL}) follows from the independence of the noise and the targets' reflectivities. However Kolmogorov's strong law of large numbers cannot be applied to the fourth term in (\ref{StochLogLikeEstLr2}) since there is no summation over $L_r$. Therefore as $L_r$ approaches infinity, $\LL_s\left(\hat{\psivec}, \hat{\Amat}, \hat{\sigma}^2\right)$ does not approach the expected value in (\ref{mle_stoch_app_asymptotic})
\be
\label{LargeLrTrace3}
\begin{split}
\LL_s & \left(\hat{\psivec}, \hat{\Amat}, \hat{\sigma}^2\right)  \xrightarrow{a.s.} \frac{1}{M_rM_tL_r}\sum_{l=1}^{M_r}\sum_{k=1}^{M_t}\left[\log\left|\hat{\Rmat}_{kl}\right| \right. \\
& \qquad\qquad + \left.\sigma^2 Tr\cbl\hat{\Rmat}_{kl}^{-1}\cbr + \alphavec_{kl}^H\Hmat_{kl}^H\hat{\Rmat}_{kl}^{-1}\Hmat_{kl}\alphavec_{kl}\right] \\
& = \frac{1}{M_rM_tL_r}\sum_{l=1}^{M_r}\sum_{k=1}^{M_t}\left[\log\left|\hat{\Rmat}_{kl}\right| \right.\\
& \qquad\qquad + \left. Tr\cbl\hat{\Rmat}_{kl}^{-1}\left(\Hmat_{kl}\alphavec_{kl}\alphavec_{kl}^H\Hmat_{kl}^H + \sigma^2\Imat\right)\cbr\right]
\end{split}
\ee

\section{}
\label{AppendixDA}
This appendix verifies that the variance of the $kl$th term in (\ref{MLdetEst}) is bounded. Consider a $kl$th summand in (\ref{MLdetEst})
\be
\label{AppenD1}
\begin{split}
F_{kl} &= \frac{1}{L_rNM_tM_r}Tr\cbl\Pi_{\hat{\Hmat}_{kl}}^{\perp} \rvec_{kl}\rvec_{kl}^H\cbr \\
&= \frac{1}{L_rNM_tM_r}\rvec_{kl}^H\Pi_{\hat{\Hmat}_{kl}}^{\perp} \rvec_{kl}
\end{split}
\ee
Under the deterministic signal model assumption the vector $\rvec_{kl}$ has the mean $\muvec_{kl}=\Hmat_{kl}\alphavec_{kl}$ and the covariance matrix $\Rmat_{kl} = \sigma^2\Imat_{L_rN}$. Using the expression for the variance of the quadratic form in (\ref{QuadFromVar2})
\be
\begin{split}
var&[\rvec_{kl}^H\Pi_{\hat{\Hmat}_{kl}}^{\perp} \rvec_{kl}]  = Tr \cbl\Pi_{\hat{\Hmat}_{kl}}^{\perp} \Rmat_{kl} \Pi_{\hat{\Hmat}_{kl}}^{\perp} \Rmat_{kl}\cbr\\
& \qquad\qquad\qquad + 2\muvec_{kl}^H \Pi_{\hat{\Hmat}_{kl}}^{\perp} \Rmat_{kl} \Pi_{\hat{\Hmat}_{kl}}^{\perp}\muvec_{kl}\\
& = \sigma^4 Tr \cbl\Pi_{\hat{\Hmat}_{kl}}^{\perp}\cbr + \sigma^2\muvec_{kl}^H \Pi_{\hat{\Hmat}_{kl}}^{\perp} \muvec_{kl}\\
& = \sigma^4 (L_rN-Q) + \sigma^2Tr\cbl\Pi_{\hat{\Hmat}_{kl}}^{\perp} \muvec_{kl}\muvec_{kl}^H\cbr \label{AppendixDvar}
\end{split}
\ee
where the last identity follows since the trace of the orthogonal projection matrix is equal to its rank $Tr \cbl\Pi_{\hat{\Hmat}_{kl}}^{\perp}\cbr = L_rN-Q$.

In order to show that the variance in (\ref{AppendixDvar}) is bounded we first state a number of inequalities. Since the propagation loss coefficient $\zeta_{kl}^q$ is always positive and smaller than one, and the norm of the temporal steering vector $\bvec_{kl}^q$ according to Assumption 1 is $\frac{E}{M_tL_t}$, the following inequality holds for $Q$ targets and the $kl$th transmit-to-receive array path
\be
\label{AppC1}
\begin{split}
&Tr\cbl\Hmat_{kl}^H\Hmat_{kl}\cbr = \sum_{q=1}^Q (\hvec_{kl}^q)^H\hvec_{kl}^q \\&= \sum_{q=1}^Q (\zeta_{kl}^q)^2\left((\avec_{rl}^q)^H\avec_{rl}^q\right)\left((\bvec_{kl}^q)^H\bvec_{kl}^q\right)\vert\wvec_k^H\avec_{tk}^q\vert^2 \leq QE\frac{L_tL_r}{M_t}.
\end{split}
\ee
If the magnitudes of the targets' reflection ceofficients are bounded such that $\vert\alpha_{kl}^q\vert^2\leq \vert\alpha_{max}\vert^2$, $\forall k,l,q$ then
\be
\label{AppC2}
Tr\cbl\alphavec_{kl}\alphavec_{kl}^H\cbr \leq  Q \vert\alpha_{max}\vert^2.
\ee
Combining (\ref{AppC1}) and (\ref{AppC2}) and using the matrix trace inequality in (\ref{TraceMatrixInequality}) yields:
\be
\label{AppC3}
\begin{split}
Tr\cbl\muvec_{kl}\muvec_{kl}^H\cbr &= Tr\cbl\Hmat_{kl}\alphavec_{kl}\alphavec_{kl}^H\Hmat_{kl}^H\cbr  \\ &\leq Tr\cbl\Hmat_{kl}^H\Hmat_{kl}\cbr Tr\cbl\alphavec_{kl}\alphavec_{kl}^H\cbr \\ &= Q^2E\frac{L_tL_r}{M_t}\vert\alpha_{max}\vert^2.
\end{split}
\ee
Applying the matrix trace inequality in (\ref{TraceMatrixInequality}) to the trace in (\ref{AppendixDvar}) and then using (\ref{AppC3}), the bound on the variance of $F_{kl}$ follows
\be
\begin{split}
var[F_{kl}] &\leq \frac{\sigma^4 (L_rN-Q) + \sigma^2 Tr\cbl\Pi_{\hat{\Hmat}_{kl}}^{\perp}\cbr Tr\cbl\muvec_{kl}\muvec_{kl}^H\cbr}{(L_rNM_tM_r)^2} \\
&\leq \frac{\sigma^2(L_rN-Q)}{(L_rNM_tM_r)^2}\left(\sigma^2 +Q^2E\frac{L_tL_r}{M_t}\vert\alpha_{max}\vert^2 \right) < \infty.
\end{split}
\ee

\section{}
\label{AppendixDB}
Verification that $F(\hat{\psivec})$ in (\ref{MLdetEst}) converges to (\ref{AsympDet1}) as $L_r\rightarrow \infty$. Using (\ref{Model3}) in (\ref{MLdetEst}) the function $F(\hat{\psivec})$ can be written as
\be
\begin{split}
F(\hat{\psivec}) &= \frac{1}{L_rNM_tM_r} \sum_{l=1}^{M_r} \sum_{k=1}^{M_t}\left[\alphavec_{kl}^H\Hmat_{kl}^H \Pi_{\hat{\Hmat}_{kl}}^{\perp}\Hmat_{kl}\alphavec_{kl}\right. \\
&+ \left.\evec_{kl}^H \Pi_{\hat{\Hmat}_{kl}}^{\perp} \evec_{kl} + 2Re\cbl\evec_{kl}^H\Pi_{\hat{\Hmat}_{kl}}^{\perp}\Hmat_{kl}\alphavec_{kl}\cbr\right]
\end{split}
\label{LargeLrTraceDet}
\ee
By the definition of the trace operator
\be
\label{LargeLrTraceDet2}
\begin{split}
F(\hat{\psivec}) & = \frac{1}{L_rNM_tM_r} \sum_{l=1}^{M_r} \sum_{k=1}^{M_t} \left[\vphantom{\cbl\sum_{i=1}^{L_rN}\cbr}\alphavec_{kl}^H\Hmat_{kl}^H\Pi_{\hat{\Hmat}_{kl}}^{\perp}\Hmat_{kl}\alphavec_{kl}\right.\\ 
& + \sum_{i=1}^{L_rN}\sum_{j=1}^{L_rN} \left[\Pi_{\hat{\Hmat}_{kl}}^{\perp}\right]_{ij} (\evec_{kl}^H)_i (\evec_{kl})_j \\
& + \left.2Re\cbl\sum_{i=1}^{L_rN}\sum_{j=1}^Q \left[\Pi_{\hat{\Hmat}_{kl}}^{\perp}\right]_{ij}(\evec_{kl}^H)_i(\alphavec_{kl})_j\cbr\right]
\end{split}
\ee
Note that in (\ref{LargeLrTraceDet2}) only the second and the third terms are random variables since they depend on the noise vector $\evec_{kl}$ and according to the deterministic signal model the targets' reflectivities $\alphavec_{kl}$ are non-random. By letting $L_r \rightarrow \infty$, Kolmogorov's strong law of large numbers can be applied to the second and the third terms in (\ref{LargeLrTraceDet2}) yielding
\be
\begin{split}
\sum_{i=1}^{L_rN}&\sum_{j=1}^{L_rN} \left[\Pi_{\hat{\Hmat}_{kl}}^{\perp}\right]_{ij}  (\evec_{kl}^H)_i (\evec_{kl})_j  \\
\xrightarrow{a.s.} & E\left[ \sum_{i=1}^{L_rN}\sum_{j=1}^{L_rN} \left[\Pi_{\hat{\Hmat}_{kl}}^{\perp}\right]_{ij} (\evec_{kl}^H)_i (\evec_{kl})_j\right] \\
 = & \sum_{i=1}^{L_rN}\sum_{j=1}^{L_rN} \left[\Pi_{\hat{\Hmat}_{kl}}^{\perp}\right]_{ij} E\left[(\evec_{kl}^H)_i (\evec_{kl})_j\right] \\
 = & \sigma^2 Tr\cbl\Pi_{\hat{\Hmat}_{kl}}^{\perp}\cbr = \sigma^2(L_rN-Q) \\
\end{split}
\ee
and
\be
\begin{split}
\sum_{i=1}^{L_rN}&\sum_{j=1}^Q \left[\Pi_{\hat{\Hmat}_{kl}}^{\perp}\right]_{ij}(\evec_{kl}^H)_i(\alphavec_{kl})_j \\
\xrightarrow{a.s.}& E\left[\sum_{i=1}^{L_rN}\sum_{j=1}^Q \left[\Pi_{\hat{\Hmat}_{kl}}^{\perp}\right]_{ij}(\evec_{kl}^H)_i(\alphavec_{kl})_j\right] \\
= & \sum_{i=1}^{L_rN}\sum_{j=1}^Q \left[\Pi_{\hat{\Hmat}_{kl}}^{\perp}\right]_{ij}E\left[(\evec_{kl}^H)_i(\alphavec_{kl})_j\right] = 0
\end{split}
\ee
Thus as $L_r$ approaches infinity $F(\hat{\psivec})$ converges to the expected value
\be
\begin{split}
F(\hat{\psivec}) \xrightarrow{a.s.}& \frac{\sigma^2(L_rN-Q)}{L_rN} \\
+& \frac{1}{L_rNM_tM_r} \sum_{l=1}^{M_r} \sum_{k=1}^{M_t}  \alphavec_{kl}^H\Hmat_{kl}^H\Pi_{\hat{\Hmat}_{kl}}^{\perp}\Hmat_{kl}\alphavec_{kl} 
\end{split}
\ee

\section{Useful Identities and Inequalities}
\label{Inequalities}
\begin{itemize}
\item \textit{Vectorization operator} \cite{Stoica2}
\bea
\label{vec1}
Tr\cbl\Amat\Bmat\cbr = (vec(\Amat^H))^H vec(\Bmat)\\
\label{vec2}
vec(\Amat\Bmat\Cmat) = (\Cmat^T \otimes \Amat) vec(\Bmat).
\eea
\item For a nonsingular complex matrix $\Amat$ and its inverse $\Bmat=\Amat^{-1}$ the following identity was shown to hold in \cite{StoicaNehorai2}
\be
\label{CompMatIdnt1}
\begin{bmatrix}
Re\cbl\Amat\cbr & - Im\cbl\Amat\cbr \\
Im\cbl\Amat\cbr & Re\cbl\Amat\cbr
\end{bmatrix}^{-1} = \begin{bmatrix}
Re\cbl\Bmat\cbr & - Im\cbl\Bmat\cbr \\
Im\cbl\Bmat\cbr & Re\cbl\Bmat\cbr
\end{bmatrix}.
\ee
\item For complex matrices $\Amat$, $\Bmat$ and $\Cmat$ it can be verified that \cite{StoicaNehorai2}
\be
\label{CompMatIdnt2}
\begin{split}
&\begin{bmatrix}
Re\cbl \Amat \cbr & -Im\cbl\Amat\cbr
\end{bmatrix} 
\begin{bmatrix}
Re \cbl \Bmat \cbr & -Im\cbl \Bmat \cbr \\
Im \cbl \Bmat \cbr & Re\cbl \Bmat \cbr
\end{bmatrix}
\begin{bmatrix}
Re \cbl \Cmat \cbr \\
Im \cbl \Cmat \cbr 
\end{bmatrix}\\
& = Re \cbl \Amat\Bmat\Cmat \cbr.
\end{split}
\ee
\item \textit{Expected value and variance of a quadratic form} \cite{Sultan}. For an $n$-dimensional complex Gaussian random vector $\xvec$ with a zero-mean and a covariance matrix $\Sigma$, and a Hermitian matrix $\Amat$
\begin{align}
E\left[\xvec^H\Amat\xvec\right] & = Tr\cbl\Amat\Sigma\cbr\\
E\left[(\xvec^H\Amat\xvec)^2\right] & = Tr\cbl\Amat\Sigma\Amat\Sigma\cbr + \left(Tr\cbl\Amat\Sigma\cbr\right)^2
\end{align}
thus
\be
\label{QuadFromVar}
var\left[\xvec^H\Amat\xvec\right] = 2Tr\cbl\Amat\Sigma\Amat\Sigma\cbr.
\ee
It can be shown that if $\xvec$ has a mean $\muvec$ and covariance matrix $\Sigma$ then
\begin{align}
E\left[\xvec^H\Amat\xvec\right] & = Tr\cbl\Amat\Sigma\cbr + \muvec^H\Amat\muvec\\
var\left[\xvec^H\Amat\xvec\right] & = Tr\cbl\Amat\Sigma\Amat\Sigma\cbr + 2\muvec^H\Amat\Sigma\Amat\muvec \label{QuadFromVar2}
\end{align}

\item \textit{A matrix trace inequality} \cite{Yang}. For the positive semidefinite matrices $\Amat$ and $\Bmat$\\
\be
\label{TraceMatrixInequality}
Tr\cbl \Amat\Bmat \cbr \leq \left( Tr\cbl \Amat \cbr^2  Tr\cbl \Bmat \cbr^2 \right)^{1/2}.
\ee
\item \textit{Von Neumann's trace inequality} \cite{Mirsky}. For $m\times m$ matrices $\Amat$ and $\Bmat$ with singular values $\alpha_1\geq\alpha_2\geq\ldots\geq\alpha_n$ and $\beta_1\geq\beta_2\geq\ldots\geq\beta_n$ respectively\\
\be
\label{VonNeumannInq}
\left|Tr\cbl\Amat\Bmat\cbr\right|\leq\sum_{i=1}^n\alpha_i\beta_i \leq n\alpha_1\beta_1.
\ee
\item For any arbitrary given $n \times n$ positive definite matrix $\Amat$, the inequality below holds for any positive definite $n \times n$ matrix $\Bmat$ \cite{StoicaNehorai3}
\be
\label{StoicasInequality}
\ln\det{\Bmat}+Tr\cbl\Bmat^{-1}\Amat\cbr \geq n + \ln\det{\Amat}.
\ee
\end{itemize}